\documentclass[11pt,reqno,preprint]{article}
\usepackage{jheppub}
\usepackage{epsfig}
\usepackage{amssymb}
\usepackage[utf8]{inputenc}
\usepackage{amsmath}
\usepackage{mathrsfs}
\usepackage{hyperref}
\usepackage{multirow}
\usepackage{subcaption}

\def\be{\begin{equation}}
\def\ee{\end{equation}}
\def\ba{\begin{eqnarray}}
\def\ea{\end{eqnarray}}

\def\a{\alpha}

\def\.{{\cdot}}

\newcommand{\bea}{\setlength\arraycolsep{2pt} \begin{eqnarray}}
\newcommand{\eea}{\end{eqnarray}}

\newcommand{\lsim}{\mathrel{\hbox{\rlap{\lower.55ex \hbox{$\sim$}} \kern-.3em \raise.4ex \hbox{$<$}}}}
\newcommand{\gsim}{\mathrel{\hbox{\rlap{\lower.55ex \hbox{$\sim$}} \kern-.3em \raise.4ex \hbox{$>$}}}}

\def\ft#1#2{{\textstyle{\frac{\scriptstyle #1}{\scriptstyle #2} } }}
\def\fft#1#2{{\frac{#1}{#2}}}

\def\0{{\sst{(0)}}}
\def\1{{\sst{(1)}}}
\def\2{{\sst{(2)}}}
\def\3{{\sst{(3)}}}
\def\4{{\sst{(4)}}}
\def\5{{\sst{(5)}}}
\def\6{{\sst{(6)}}}
\def\7{{\sst{(7)}}}
\def\8{{\sst{(8)}}}
\def\9{{\sst{(9)}}}

\def\sst#1{{\scriptscriptstyle #1}}

\title{Notes on flat-space limit of AdS/CFT}
\author{Yue-Zhou Li}
\affiliation{Department of Physics, McGill University, 3600 Rue University, Montr\'eal, QC Canada H3A 2T8}
\emailAdd{liyuezhou@physics.mcgill.ca}
\abstract{Different frameworks exist to describe the flat-space limit of AdS/CFT, include momentum space, Mellin space, coordinate space, and partial-wave expansion. We explain the origin of momentum space as the smearing kernel in Poincare AdS, while the origin of latter three is the smearing kernel in global AdS. In Mellin space, we find a Mellin formula that unifies massless and massive flat-space limit, which can be transformed to coordinate space and partial-wave expansion. Furthermore, we also manage to transform momentum space to smearing kernel in global AdS, connecting all existed frameworks. Finally, we go beyond scalar and verify that $\langle VV\mathcal{O}\rangle$ maps to photon-photon-massive amplitudes.
}
\begin{document}

\maketitle

\def\x#1{x_{#1}^2}
\def\xsq#1{x_{#1}^4}
\def\y#1{y_{#1}^2}
\def\zb{\overline{z}}
\def\a{\alpha}
\def\ab{\overline{\alpha}}
\def\j{J}
\def\zbar{\bar{z}}
\def\eps{\epsilon}
\def\GN{G}
\def\mubar{\bar{\mu}}
\def\gammaE{\gamma_{\rm E}}

\section{Introduction}
\label{intro}

Including negative cosmological constant, gravity theory coupled to other local fields can be formulated as weakly coupled quantum field theory (QFT) by perturbatively expanding the curvatures around the Anti-de Sitter (AdS) background. Although the resulting QFT lives on AdS, we are still able to apply the standard techniques, which utilize the propagators in AdS to calculate the ``AdS amplitudes'' for local quantum fields. As interpreted by  the AdS/CFT correspondence, these AdS amplitudes are corresponding to correlation functions of large-$N$ expanded conformal field theory (CFT) on the AdS boundary \cite{Maldacena:1997re,Witten:1998qj,Gubser:1998bc}.

Naively, at the level of effective Lagrangian, we can take the large AdS radius limit $\ell\rightarrow\infty$, QFTs on AdS then make no difference from flat-space. We can also easily observe the limit $\ell\rightarrow\infty$ reduces AdS background to a flat-space. It is, however, rather nontrivial to incorporate the AdS amplitudes into this flat-space limit, where we expect that AdS amplitudes degrade and give rise to S-matrix or scattering amplitudes of QFT in flat-space. Employing AdS/CFT, the flat-space limit of AdS then suggests that boundary CFT correlation function shall encode the flat-space S-matrix \footnote{It is worth noting that the flat-space limit of AdS/CFT is different from flat holography proposal, e.g., \cite{Bagchi:2016bcd}. In the flat-space limit of AdS/CFT, we expect CFT encodes one higher dimensional S-matrix, but the S-matrix can not fully encode CFT. While by flat holography, flat-space physics and CFT should be able to be transformed back and forth between each other}.

\begin{figure}[t]
\centering \hspace{0mm}\def\svgwidth{100mm}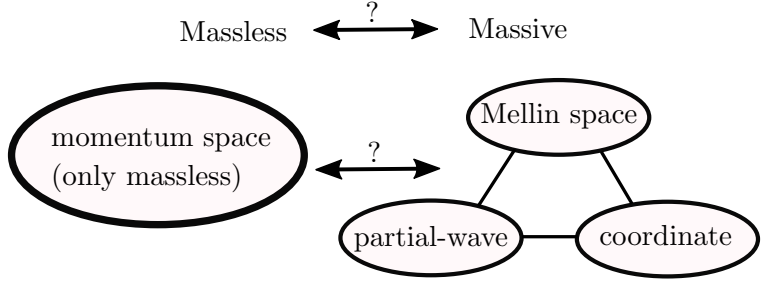
\caption{The existed frameworks describing the flat-space limit of AdS/CFT, where the question mark denotes the undiscovered relation.}
\label{known frameworks}
\end{figure}

The idea on the flat-space limit of AdS/CFT enjoys a long history \cite{Polchinski:1999ry,Giddings:1999jq,Gary:2009ae,Gary:2009mi,Heemskerk:2009pn,Fitzpatrick:2010zm,Fitzpatrick:2011jn}, and more quantitative and precise maps were established in the recent decade \cite{Okuda:2010ym,Penedones:2010ue,Fitzpatrick:2011hu,Maldacena:2015iua,Komatsu:2020sag,Paulos:2016fap,Raju:2012zr}. However, in the literature, there exist several frameworks which work in different representations of CFT: momentum space \cite{Raju:2012zr}, Mellin space \cite{Penedones:2010ue,Fitzpatrick:2011hu,Paulos:2016fap}, coordinate space \cite{Okuda:2010ym,Maldacena:2015iua,Komatsu:2020sag}, and partial-wave expansion (conformal block expansion) \cite{Maldacena:2015iua,Paulos:2016fap}, as summarised in Figure \ref{known frameworks}. The latter three representations are natural to consider conformal bootstrap \cite{Caron-Huot:2020adz}, so our focus will be mostly on the latter three frameworks, for which the formulas describing massless scattering and massive scattering (defined for external legs) are sharply different. The massless particles are described by operators with finite conformal dimension, while massive particles are described by operators with infinite conformal dimension $\Delta\sim\ell\rightarrow\infty$ \footnote{For the framework in momentum space, as far as we know, only the massless formula was proposed \cite{Raju:2012zr}}. The details shall be reviewed in subsection \ref{sec known frameworks} and here we simply provide a chronological history: the massless formula in coordinate space for four-point case was first proposed in \cite{Okuda:2010ym} and was reformulated by the proposal of Mellin space \cite{Penedones:2010ue}, which is later known as the bulk-point limit \cite{Maldacena:2015iua}, and a contact example of the partial-wave coefficients was provided in \cite{Maldacena:2015iua}; the massive Mellin space formula and the phase-shift formula (which is basically the coefficient of the partial-wave) was later proposed in \cite{Paulos:2016fap}, and the massive formula in the coordinate space was recently conjectured in \cite{Komatsu:2020sag}.

Two natural questions that we aim to answer in this paper are:
\begin{itemize}
\item What is the origin of these seemingly different frameworks of the flat-space limit?

\item Why do the formulas describing massless scattering and massive scattering look different and how do we unify them?
\end{itemize}
Considering the Mellin space, coordinate space and partial-wave expansion can be translated to each other, we expect they share the same origin. The origin follows the spirit of the HKLL formula \cite{Hamilton:2005ju,Hamilton:2006az}, which represents the flat-space S-matrix in terms of boundary correlation function via smearing over the boundary against a scattering smearing kernel. Such scattering smearing kernel for massless scattering was constructed in \cite{Fitzpatrick:2011jn} and was applied to rigorously derive the massless Mellin formula later \cite{Fitzpatrick:2011hu}. A scattering smearing kernel that is generally valid for both massless and massive cases was proposed in \cite{Hijano:2019qmi}, which slightly overlaps with this paper. We find, crucially, only the scattering smearing kernel constructed from global AdS can be served as the origin of the flat-space limit in Mellin space, coordinate space, and partial-wave expansion; on the other hand, when we construct the scattering smearing kernel from Poincare AdS, we find it simply performs the Fourier-transform and thus gives rise to the framework of flat-space limit in momentum space. According to subregion duality \cite{Bousso:2012sj,Czech:2012bh,Bousso:2012mh} which states subregion of CFT is encoded in the corresponding subregion of AdS (usually the causal wedge \cite{Bousso:2012sj} or more generally entanglement wedge \cite{Harlow:2016vwg}), we expect that the Poincare scattering smearing kernel can be transformed to the global smearing kernel, simply because the Poincare patch is a part of the global AdS. We indeed find that the global scattering can be obtained from Poincare scattering, which also suggests a momentum-coordinate duality for CFT at large momentum and conformal dimensions.

Notably, scattering smearing kernels never treat massless and massive scattering distinguishingly, we should be able to unify the massless flat-space limit and massive flat-space limit. In this paper, we find a Mellin formula applying to all masses, which can be easily translated to other frameworks for both massless and massive cases. Typically, in terms of CFT language, the massive scattering is more like a ``limit'' of massless one, because nonzero masses provide additional large parameters $\Delta\sim\ell\rightarrow\infty$ that further dominate the scattering smearing kernel.

\begin{figure}[t]
\centering \hspace{0mm}\def\svgwidth{110mm}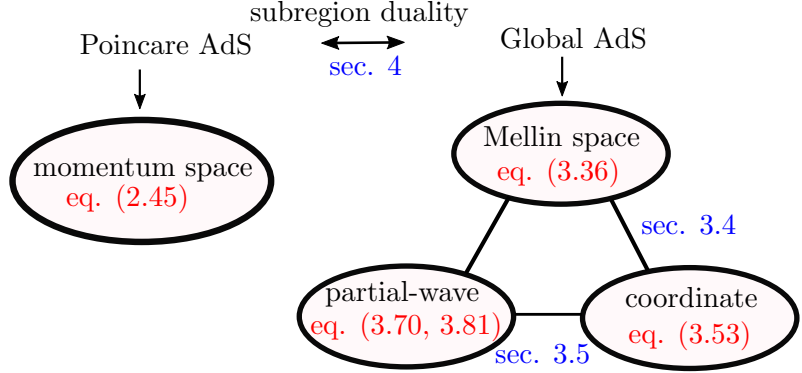
\caption{Massless and massive unified frameworks of the flat-space limit, where the origins are clarified.}
\label{our frameworks}
\end{figure}

The outline of our finding is illustrated in Figure \ref{our frameworks}.
This paper is organized as follows. In section \ref{scattering smearing}, we take the flat-space limit for bulk reconstruction in both global AdS and Poincare AdS to construct scattering smearing kernels that represent flat-space S-matrix in terms of CFT correlator. The Poincare scattering smearing kernel automatically Fourier-transforms the CFT correlator and gives rise to flat-space limit in momentum space. In section \ref{Mellin etc}, we review the existed flat-space limit, include Mellin space, coordinate space, and partial-wave expansion. We start with the global scattering smearing kernel and find saddle-points that dominate the smearing integral. Using the saddle-points, we find a Mellin formula that applies to both massless scattering and massive scattering. We then show this Mellin formula gives rise to the flat-space limit in coordinate space, and then to the partial-wave/phase-shift formula. In section \ref{sec momentum-coordinate duality}, use the notion of subregion duality, we propose a momentum-coordinate duality, which relates the flat-space limit in momentum space to global scattering smearing kernel. In section \ref{sec spinning flat-space}, we propose a flat-space parameterization of embedding coordinate for spinning operators. We apply our proposal to $\langle VV\mathcal{O}\rangle$ three-point function where $V$ is conserved current, we verify the momentum-coordinate duality as well as a map to flat-space amplitude.

In appendix \ref{app Euclidean momentum}, we analytically continue the flat-space limit in momentum space to Euclidean CFT, which effectively turns AdS into dS. In appendix \ref{Normalization}, we show how to fix the normalization of scattering smearing kernel. In appendix \ref{Mellin-derivation}, we provide more details on derivation of Mellin flat-space limit. In appendix \ref{app contact witten diagram}, we compute four-point scalar contact Witten diagram (no derivative) and verify it is equivalent to momentum conservation delta function in the flat-space limit. In appendix \ref{massive conformal block}, we introduce a new conformal frame, which helps us solve the conformal block at limit $\Delta,\Delta_i\rightarrow\infty$. We double-check our conformal block by working explicitly in $d=2,4$.

\section{Quantization and scattering smearing kernel}
\label{scattering smearing}

\subsection{Global quantization and the flat-space limit}
\label{global}
\begin{figure}[t]
\centering \hspace{0mm}\def\svgwidth{45mm}
\begingroup%
  \makeatletter%
  \providecommand\color[2][]{%
    \errmessage{(Inkscape) Color is used for the text in Inkscape, but the package 'color.sty' is not loaded}%
    \renewcommand\color[2][]{}%
  }%
  \providecommand\transparent[1]{%
    \errmessage{(Inkscape) Transparency is used (non-zero) for the text in Inkscape, but the package 'transparent.sty' is not loaded}%
    \renewcommand\transparent[1]{}%
  }%
  \providecommand\rotatebox[2]{#2}%
  \newcommand*\fsize{\dimexpr\f@size pt\relax}%
  \newcommand*\lineheight[1]{\fontsize{\fsize}{#1\fsize}\selectfont}%
  \ifx\svgwidth\undefined%
    \setlength{\unitlength}{200.0277839bp}%
    \ifx\svgscale\undefined%
      \relax%
    \else%
      \setlength{\unitlength}{\unitlength * \real{\svgscale}}%
    \fi%
  \else%
    \setlength{\unitlength}{\svgwidth}%
  \fi%
  \global\let\svgwidth\undefined%
  \global\let\svgscale\undefined%
  \makeatother%
  \begin{picture}(1,0.94791127)%
    \lineheight{1}%
    \setlength\tabcolsep{0pt}%
    \put(0,0){\includegraphics[width=\unitlength,page=1]{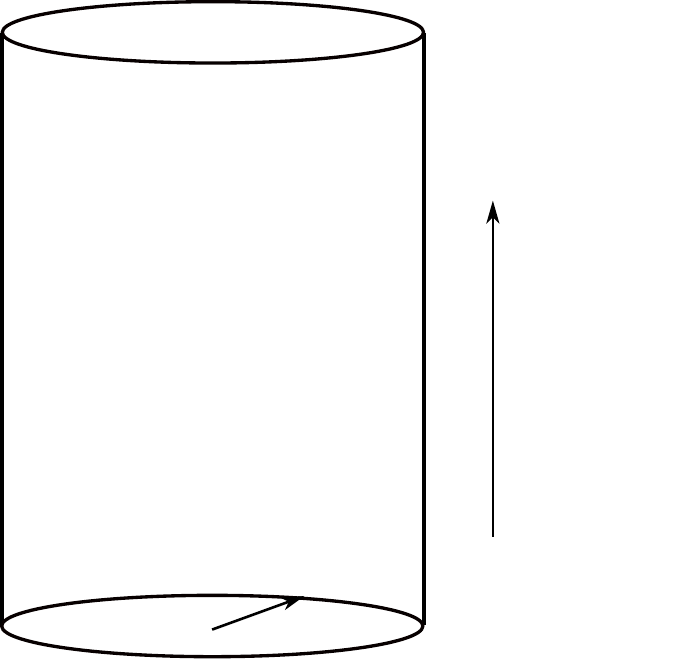}}%
    \put(0.7378366,0.39403735){\color[rgb]{0,0,0}\makebox(0,0)[lt]{\lineheight{1.25}\smash{\begin{tabular}[t]{l}$\tau$\end{tabular}}}}%
    \put(0.38705299,0.02874166){\color[rgb]{0,0,0}\makebox(0,0)[lt]{\lineheight{1.25}\smash{\begin{tabular}[t]{l}$\rho$\end{tabular}}}}%
    \put(0,0){\includegraphics[width=\unitlength,page=2]{global_AdS.pdf}}%
  \end{picture}%
\endgroup%

\caption{Cylinder diagram of global AdS.}
\label{fig global AdS}
\end{figure}

We first consider global Euclidean AdS coordinate
\be
ds^2=\fft{\ell^2}{\cos\rho^2}(d\tau^2+d\rho^2+\sin\rho^2d\Omega_{d-1}^2)\,,
\ee
where its boundary is located at $\rho=\pi/2$. The advantage of global AdS is that it provides a $R\times S_{d-1}$ background for boundary CFT, i.e.,
\be
ds_{\rm CFT}^2=d\tau^2+d\Omega_{d-1}^2\,,
\ee
which is natural for radial quantization in CFT. This global coordinate is depicted in Fig \ref{fig global AdS}.
Moreover, to make contact with flat Minkowski space where physical scattering processes happen, we may start with Lorentzian AdS. To do this, we simply wick rotate $\tau$
\be
ds^2=\fft{\ell^2}{\cos\rho^2}(-d\tau^2+d\rho^2+\sin\rho^2d\Omega_{d-1}^2)\,,
\ee
for which the AdS and CFT embedding coordinate $X$ and $P$ are parameterized by
\be
X=\fft{\ell}{\cos\rho}(\cos\tau,-i\sin\tau,\sin\rho\, \hat{r})\,,\quad P=(\cos\tau,-i\sin\tau,\hat{r})\,,\label{embedding}
\ee
respectively.

Let us consider a free scalar with mass $m$ in global AdS, which can be quantized by \cite{Fitzpatrick:2010zm}
\bea
&& \phi=\sum_{n,J,m_i}e^{i E_{nJ} \tau}R_{n,J}(\rho)Y_{J m_i}(\hat{\rho})a_{nJm_i}+{\rm c.c}\,,
\cr &&
 \cr && R_{n,J}(\rho)=\fft{1}{N_{\Delta J}}\sin^J\rho \cos^\Delta \rho\,_2F_1\big(-n,\Delta+J+n,J+\fft{d}{2},\sin\rho^2\big)\,,\label{globalmode}
\eea
where the energy eigenvalues are discretized as $E_{nJ}=\Delta+J+2n$, and
\be
m^2\ell^2=\Delta(\Delta-d)\,.
\ee
This spectra correspond to a primary operator $\mathcal{O}$ and all its descendent family $\partial^{2n}\partial_{\mu_1}\cdots \partial_{\mu_J}\mathcal{O}$. The normalization factor $N_{\Delta J}$ can be found by usual quantization procedure
\be
[\phi(\vec{x},\tau),\pi(\vec{y},\tau)]=i\fft{\delta(\vec{x}-\vec{y})}{\sqrt{-g}}\,,\quad [a_{n Jm_i},a^\dagger_{n' J'm'_i}]=\delta_{nn'}\delta_{JJ'}\delta_{m_im_i'}\,,
\ee
which yields \cite{Fitzpatrick:2010zm}
\be
N_{\Delta J}=\sqrt{\fft{n!\Gamma(J+\fft{d}{2})^2\Gamma(\Delta+n-\fft{d-2}{2})\ell^{d-1}}{\Gamma(n+J+\fft{d}{2})\Gamma(\Delta+n+J)}}\,.
\ee
Since we are starting with global AdS, we may call this quantization ``global quantization''.

Now with this preliminary of global quantization, we can move to discuss the flat-space limit. At first, we shall discuss how to take the flat-space limit for coordinates. Our notation of flat-space is
\be
ds^2=-dt^2+dr^2+r^2d\Omega_{d-1}^2\,.
\ee
We can see now taking the flat-space limit for coordinates is quite trivial, we can take the coordinate transformation
\be
\ell\tan\rho=r\,,\quad \tau\ell=t\,,\label{global limit }
\ee
and then send $\ell\rightarrow\infty$. It immediately follows that to make the Fourier factor $e^{iE\tau}$ in (\ref{globalmode}) valid with flat-space limit, the energy must scale as $\ell$, i.e., $E=\omega \ell$, where we denote $\omega$ as the energy in flat-space. This fact also indicates that $n\sim \ell$ for massless particles, more specifically we have $\omega=2n/\ell$. Note also in the context of AdS/CFT, we should be aware of $m\sim\Delta/\ell$. Thus any primary scalar operators with finite conformal dimensions $\Delta$ corresponds to massless particles in the flat-space limit \cite{Okuda:2010ym}, and it is necessary to consider scalar operators with large conformal dimensions scaling linear in $\ell$ to probe massive particles in flat-space \cite{Paulos:2016fap}.

Before we discuss the flat-space limit of quantization, we shall briefly review the quantization of scalar fields in flat-space in spherical coordinates. To avoid confusion, we denote $\varphi$ as scalars in flat-space. We have
\be
\varphi=\sum_{J, m_i}\int d\omega (a_{\omega J m_i}e^{i \omega t}R_{|\vec{p}|,J}(r)Y_{J m_i}(\hat{r})+{\rm c.c})\,,
\ee
where $Y_{Jm_i}$ is the spherical harmonics on $S^{d-1}$ (in which $m_i$ denotes all ``magnetic'' angular momenta), and the radial function $R_{|\vec{p}|,J}(r)$ is given by
\be
R_{|\vec{p}|,J}(r)=\fft{1}{\sqrt{2}}r^{\fft{2-d}{2}}J_{\fft{d-2}{2}+J}(|\vec{p}|r)\,.
\ee
The quantization condition is also straightforward
\be
[\varphi(\vec{x},t),\pi_{\varphi}(\vec{y},t)]=i\fft{\delta(\vec{x}-\vec{y})}{\sqrt{-g}}\,,\quad [a_{\omega Jm_i},a^\dagger_{\omega'J'm_i'}]=
\delta(\omega-\omega')\delta_{JJ'}\delta_{m_im_i'}\,.
\ee
Now we can easily take the flat-space limit for radial function and we can observe that
\be
R_{n,J}(\rho)\big|_{\ell\rightarrow\infty}=\sqrt{\fft{2}{\ell}}R_{|\vec{p}|, J}(r)\,.
\ee
It is also not hard to probe the flat-space limit for creation and annihilation operators by comparing the canonical quantization condition for those operators, i.e.,
\bea
&& [a_{njm_i},a^\dagger_{n'j'm_i'}]\big|_{\ell\rightarrow\infty}=\delta(n-n')
\delta_{JJ'}\delta_{m_im_i'}=\delta(\fft{(\omega-\omega')\ell}{2})\delta_{JJ'}\delta_{m_im_i'}
\cr &&
=\fft{2}{\ell}\delta(\omega-\omega')\delta_{JJ'}\delta_{m_im_i'}=\fft{2}{\ell}[a_{\omega Jm_i},a^\dagger_{\omega' J'm_i'}]\,.
\eea
It thus immediately follows
\be
a_{njm_i}\big|_{\ell\rightarrow\infty}=\sqrt{\fft{2}{\ell}}e^{i\eta}a_{\omega Jm_i}\,,
\ee
with an arbitrary phase factor $\eta$ that is to be fixed by convenience later. Trivially, the Fourier factor is simply $
e^{iE\tau}=e^{i\omega t}$, and the flat-space limit of measure in summation over all energy spectra is also consistent
\be
\sum_n\rightarrow \int d\omega \fft{\ell}{2}\,.
\ee
By including above factors, we are led to
\be
\phi\big|_{\ell\rightarrow \infty}\simeq\varphi\,.\label{flatlimsc}
\ee
In other words, the flat-limit of the quantized scalars in global AdS is equivalent to the quantized scalars in flat-space.

Using the quantization in global AdS, the corresponding primary operator $\mathcal{O}$ that is dual to $\phi$ can be quantized via
\be
\mathcal{O}=\sum_{n,j,m_i}(e^{iE_{nj}\tau}Y_{jm_i}(\hat{\rho})a_{njm}+{\rm c.c})N^{\mathcal{O}}_{\Delta,n,j}\,,\label{quantize operator global}
\ee
where the normalization can be fixed by normalizing the two-point function \cite{Fitzpatrick:2010zm}
\be
N^{\mathcal{O}}_{\Delta,n,j}=\sqrt{\fft{\Gamma(1+\Delta-\fft{d}{2}+n)\Gamma(\Delta+J+n)}{\Gamma(1+n)\Gamma(\fft{d}{2}+J+n)}}
\fft{1}{\Gamma(1+\Delta-\fft{d}{2})}\,.
\ee
It then follows that we can represent creation operator by $\mathcal{O}$ via
\be
a^\dagger_{njm_i}=\int_{-\fft{\pi}{2}-\tau_0}^{\fft{\pi}{2}-\tau_0} \fft{d\tau}{\pi}d\Omega_{d-1}e^{i E_{nj}\tau}\fft{Y_{jm_i}(\hat{\rho})}{N^{\mathcal{O}}_{\Delta,n,j}}\mathcal{O}(\tau,\hat{\rho})\,,
\ee
where $\tau_0$ is the (finite) reference time which can be chosen for convenience and doesn't affect the integral. This reflects the $\tau$ translation symmetry.
Take the flat-space limit on both sides of above formula, we obtain
\be
a^\dagger_{\omega Jm_i}=\int_{-\fft{\pi}{2}\ell-\tau_0}^{\fft{\pi}{2}\ell-\tau_0}\fft{dtd\Omega_{d-1}}{\sqrt{2\pi^2\ell}}e^{i\omega t}Y_{jm_i}(\hat{\rho})
2^{\Delta-\fft{d}{2}}(|\vec{p}|\ell)^{\fft{d}{2}-\Delta}\xi_{\omega\Delta}\Gamma(1+\Delta-\fft{d}{2})\times e^{-i\eta}\mathcal{O}(\tau,\hat{\rho})\,,
\ee
where we define
\be
\xi_{\omega\Delta}=\big(\fft{\omega\ell-\Delta}{\omega \ell+\Delta}\big)^{\fft{\omega\ell}{2}}e^{\Delta}=\exp[\fft{\omega\ell}{2}
\log\big(\fft{\omega\ell-\Delta}{\omega \ell+\Delta}\big)+\Delta]\,,
\ee
which, as an exponent factor, is well-defined for both massive and massless cases. We can readily verify that $\xi_{\omega\Delta}$ is simply $1$ at $\ell\rightarrow\infty$ limit for massless particles.

Using this formula, we can construct the smearing kernel $K_a(t,\hat{r})$ that represents scattering states $|p\rangle$ in terms of primary operator in CFT \cite{Fitzpatrick:2011jn}
\be
|p\rangle=\int dtd\Omega_{d-1} K_a(t,\hat{r}) \mathcal{O}(\tau,\hat{r})|0\rangle\,,
\ee
To find the smearing kernel, we can decompose the momentum eigenstate $|p\rangle$ into angular momentum eigenstate
\be
|p\rangle=\sum_{J,m_i}\langle J,m_i|p\rangle |J,m_i\rangle\,, \quad \langle J,m_i|p\rangle= i^J 2^{\fft{d+1}{2}}\pi^{\fft{d}{2}}|\vec{p}|^{\fft{2-d}{2}}Y_{Jm_i}(\hat{p})\,,
\ee
from which we can derive the smearing kernel
\bea
K_a(t,\hat{r})&=&e^{i\omega t}\sum_{Jm_i}\ell^{\fft{d-1}{2}-\Delta}\xi_{\omega,\Delta}|\vec{p}|^{1-\Delta}\times 2^\Delta\pi^{\fft{d-2}{2}}Y_{Jm_i}(\hat{r})
Y_{Jm_i}(\hat{p})\Gamma(1+\Delta-\fft{d}{2})
\cr &=& e^{i\omega t}\ell^{\fft{d-1}{2}-\Delta}\xi_{\omega,\Delta}|\vec{p}|^{1-\Delta}\times
2^\Delta\pi^{\fft{d-2}{2}}\Gamma(1+\Delta-\fft{d}{2})\delta(\hat{p}-\hat{r})\,,\label{smearsc}
\eea
in which we choose $\eta=-J\pi/2$ to cancel the funny $i^J$ factor. Note this smearing kernel is obtained for a free scalar theory. Nevertheless, we assume it also works whenever the plane-wave state is asymptotically free, which is exactly the scattering states defined at infinite past or future. We can then apply this smearing kernel to establish a formula relating flat-space ($n$-particle) S-matrix to CFT $n$-point function (or AdS amplitudes)
\bea
S&=&\,_{+\infty}\langle p_1p_2\cdots p_k|p_{k+1}\cdots p_n\rangle_{-\infty}=\mathbb{I}+i \delta^{(d+1)}(p_{\rm tot}) T(p_i)
\cr &&
 \cr &=& \lim_{\ell\rightarrow\infty} \int \big(\prod_i dt_i  e^{i\omega_i t_i}\ell^{\fft{d-1}{2}-\Delta_i}\xi_{\omega_i\Delta_i}|\vec{p}_i|^{1-\Delta}2^{\Delta_i}
 \pi^{\fft{d-2}{2}}\Gamma(1+\Delta_i-\fft{d}{2})\big)\langle \mathcal{O}_1\cdots\mathcal{O}_n\rangle\,,
 \label{full global smearing}
\eea
where $\mathbb{I}$ denotes the disconnected part of S-matrix and $T$ the scattering amplitudes, and in the second line we analytically continue the momenta such that all momenta are in-states before employing the smearing kernel (\ref{smearsc}). The interpretation of eq.~\eqref{full global smearing} shall be briefly discussed before we move on. A pure CFT does know nothing about $\ell$ without the notion of AdS/CFT. One job that AdS/CFT (with large $\ell$ limit of AdS) does is to provide a specific kernel $K_{\rm s}$ in eq.~\eqref{full global smearing}. Then we can study a particular CFT correlator in a single CFT and notice that the smeared version (smear over $\tau$) of the CFT correlator with a large $\ell$ limit of the kernel will approximate the flat-space S-matrix, where $\Delta/\ell$ estimates the masses. However, from the dynamics, to define a flat-space QFT with gravity, we have to take a family of AdS and follow the sequence that $\ell$ grows. The estimation of flat-space S-matrix by using eq.~\eqref{full global smearing} becomes more and more accurate if we have a family of CFTs supported with large $N$ limit and sparse gap $\Delta_{\rm gap}$. Thus to extract S-matrix accurately by using eq.~\eqref{full global smearing}, one should consider a family of CFTs. We shall call
\be
K_{\rm s}= \big(\prod_i e^{i\omega_i t_i}\ell^{\fft{d-1}{2}-\Delta_i}\xi_{\omega_i\Delta_i}|\vec{p}_i|^{1-\Delta_i}2^{\Delta_i}
 \pi^{\fft{d-2}{2}}\Gamma(1+\Delta_i-\fft{d}{2})\big)\,,\label{scatterkernel}
\ee
the global scattering smearing kernel. This global scattering smearing kernel generalizes the massless smearing written down in \cite{Fitzpatrick:2011hu}, and was also recently obtained by requiring the consistency with HKLL formula \cite{Hijano:2019qmi} (where they take $\Delta\sim m\ell\rightarrow\infty$ to simplify the prefactor). Note that the integration range in $t$ is different from \cite{Fitzpatrick:2011hu} for massless case. In \cite{Fitzpatrick:2011hu}, the scattering smearing kernel integrates time within $t\in(-\pi/2\ell-\delta t,-\pi/2\ell+\delta t)$, because it was argued that the flat-space physics emerges from the wave packets starting around $\tau=-\pi/2$ \cite{Gary:2009mi}, and $\delta t$ exists to make sure the in and out wave packets don't overlap. Here we construct the scattering smearing kernel from the exact free theory and thus the integration range runs over the reasonable range of $\tau$, i.e., $(-\pi/2-\tau_0,\pi/2-\tau_0)$. In the next section, we prove that there is indeed $\tau=-\pi/2$ (for reference point $\tau_0>0$) dominates the scattering smearing kernel and thus effectively gives $t\in(-\pi/2\ell-\delta t,-\pi/2\ell+\delta t)$.

\subsection{Poincare quantization and the momentum space}
\label{Poincare}
\begin{figure}[t]
\centering \hspace{0mm}\def\svgwidth{100mm}
\begingroup%
  \makeatletter%
  \providecommand\color[2][]{%
    \errmessage{(Inkscape) Color is used for the text in Inkscape, but the package 'color.sty' is not loaded}%
    \renewcommand\color[2][]{}%
  }%
  \providecommand\transparent[1]{%
    \errmessage{(Inkscape) Transparency is used (non-zero) for the text in Inkscape, but the package 'transparent.sty' is not loaded}%
    \renewcommand\transparent[1]{}%
  }%
  \providecommand\rotatebox[2]{#2}%
  \newcommand*\fsize{\dimexpr\f@size pt\relax}%
  \newcommand*\lineheight[1]{\fontsize{\fsize}{#1\fsize}\selectfont}%
  \ifx\svgwidth\undefined%
    \setlength{\unitlength}{471.8699794bp}%
    \ifx\svgscale\undefined%
      \relax%
    \else%
      \setlength{\unitlength}{\unitlength * \real{\svgscale}}%
    \fi%
  \else%
    \setlength{\unitlength}{\svgwidth}%
  \fi%
  \global\let\svgwidth\undefined%
  \global\let\svgscale\undefined%
  \makeatother%
  \begin{picture}(1,0.40182381)%
    \lineheight{1}%
    \setlength\tabcolsep{0pt}%
    \put(0,0){\includegraphics[width=\unitlength,page=1]{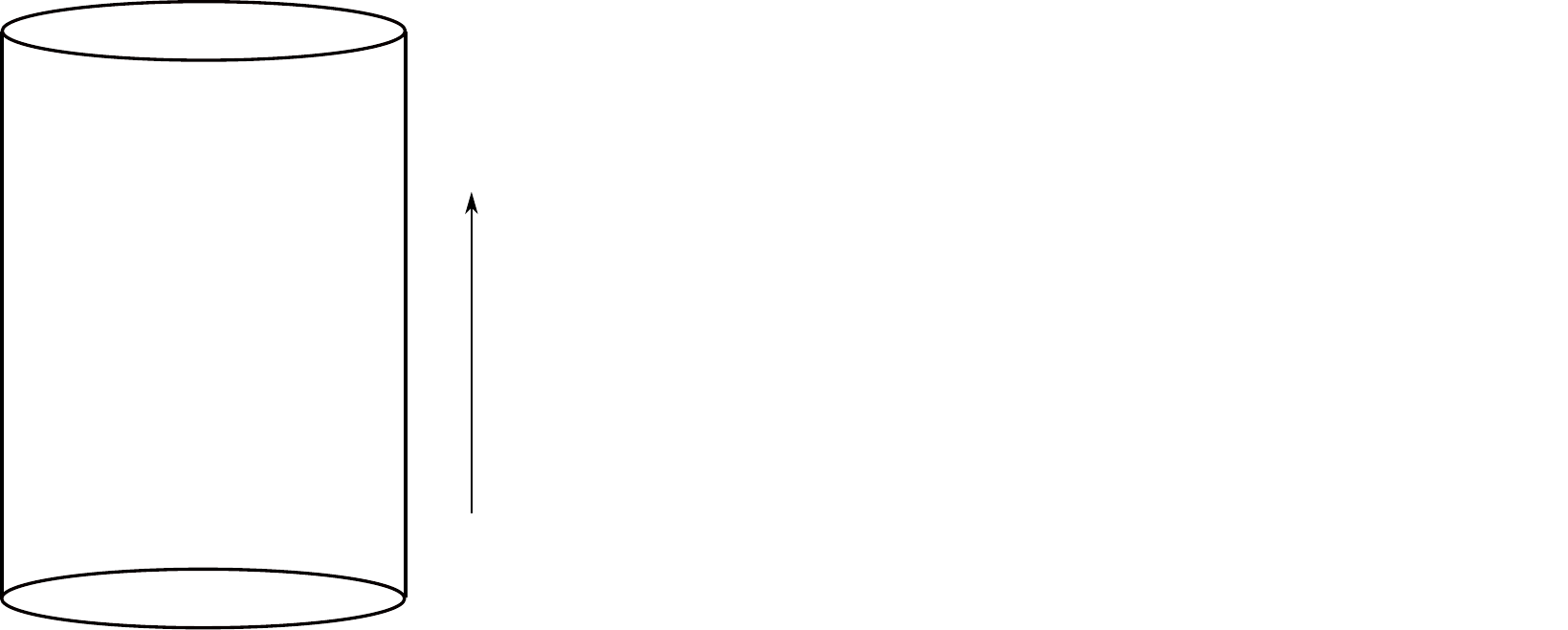}}%
    \put(0.31277224,0.16703418){\color[rgb]{0,0,0}\makebox(0,0)[lt]{\lineheight{1.25}\smash{\begin{tabular}[t]{l}$\tau$\end{tabular}}}}%
    \put(0,0){\includegraphics[width=\unitlength,page=2]{Poincare_AdS.pdf}}%
    \put(0.05482157,0.26717516){\color[rgb]{0,0,0}\makebox(0,0)[lt]{\lineheight{1.25}\smash{\begin{tabular}[t]{l}$\mathcal{B}$\end{tabular}}}}%
    \put(0,0){\includegraphics[width=\unitlength,page=3]{Poincare_AdS.pdf}}%
    \put(0.82525488,0.22091734){\color[rgb]{0,0,0}\makebox(0,0)[lt]{\lineheight{1.25}\smash{\begin{tabular}[t]{l}$\mathcal{B}$\end{tabular}}}}%
    \put(0.66135373,0.16008808){\color[rgb]{0,0,0}\makebox(0,0)[lt]{\lineheight{1.25}\smash{\begin{tabular}[t]{l}$T$\end{tabular}}}}%
    \put(0.71939355,0.15648037){\color[rgb]{0,0,0}\makebox(0,0)[lt]{\lineheight{1.25}\smash{\begin{tabular}[t]{l}$Y$\end{tabular}}}}%
    \put(0.71602535,0.21561996){\color[rgb]{0,0,0}\makebox(0,0)[lt]{\lineheight{1.25}\smash{\begin{tabular}[t]{l}$z$\end{tabular}}}}%
  \end{picture}%
\endgroup%

\caption{Poincare AdS only covers a wedge of global AdS. On LHS, the lines marked $\mathcal{B}$ meet the global AdS boundary. $\mathcal{B}$ is the boundary of Poincare AdS where CFT lives. On the RHS, we depict a local figure near $\mathcal{B}$.}
\label{fig Poincare AdS}
\end{figure}

We can also consider quantization in Poincare coordinates
\be
ds^2=\fft{\ell^2}{z^2}(dz^2-dT^2+\sum_{i=1}^{d-1}dY_i^2)\,,
\ee
which can be depicted as Fig \ref{fig Poincare AdS}.
It is straightforward to work with the quantization in this coordinate, which gives
\be
\phi=\fft{1}{\sqrt{2}\ell^{\fft{d-1}{2}}}\int_{E>|K|} \fft{dE d^{d-1}K}{(2\pi)^{\fft{d-1}{2}}} \big(a_{EK}e^{-iE T+i\vec{K}\cdot Y}z^{\fft{d}{2}}J_{\Delta-\fft{d}{2}}(z|{\bf K}|)+{\rm c.c}\big)\,,\label{poincare}
\ee
where we denote $|{\bf K}|=\sqrt{E^2-K^2}>0$, and the overall factor is determined by canonical quantization condition
\be
[\phi(Y),\pi_\phi(Y')]=i\fft{\delta^{(d)}(Y-Y')}{\sqrt{-g}}\,,\quad [a_{EK},a^\dagger_{E'K'}]=\delta(E-E')
\delta^{(d-1)}(K-K')\,.
\ee
Note this quantization is only valid for $E>K$ where the momentum is time-like, which is the necessary condition for the field to have its CFT dual. For the space-like spectrum $E<K$, it is equivalent to consider Euclidean AdS, and this quantization crashes because of the divergence at Poincare horizon $z\rightarrow 0$. Instead of the Bessel function of the first kind, the quantization for spatial momentum should be expanded by the modified Bessel function of the second kind $K_\nu$ which does, however, not have the appropriate fall-off to admit operator dual. We shall emphasize it does not contradict the Euclidean AdS/CFT, it only indicates that in Euclidean space the quantization of CFT operators is not compatible with the bulk quantization described above if we persist AdS. Nevertheless, \cite{Raju:2012zr} established a flat-space limit in the momentum space for spatial momentum, and the price is to have an imaginary momentum in the bulk. We show in appendix \ref{app Euclidean momentum} their limit is equivalent to ours but wick rotates $z\rightarrow i z$, which in effect analytically continues AdS to dS.

\begin{figure}[t]
\centering \hspace{0mm}\def\svgwidth{50mm}
\begingroup%
  \makeatletter%
  \providecommand\color[2][]{%
    \errmessage{(Inkscape) Color is used for the text in Inkscape, but the package 'color.sty' is not loaded}%
    \renewcommand\color[2][]{}%
  }%
  \providecommand\transparent[1]{%
    \errmessage{(Inkscape) Transparency is used (non-zero) for the text in Inkscape, but the package 'transparent.sty' is not loaded}%
    \renewcommand\transparent[1]{}%
  }%
  \providecommand\rotatebox[2]{#2}%
  \newcommand*\fsize{\dimexpr\f@size pt\relax}%
  \newcommand*\lineheight[1]{\fontsize{\fsize}{#1\fsize}\selectfont}%
  \ifx\svgwidth\undefined%
    \setlength{\unitlength}{195.83455748bp}%
    \ifx\svgscale\undefined%
      \relax%
    \else%
      \setlength{\unitlength}{\unitlength * \real{\svgscale}}%
    \fi%
  \else%
    \setlength{\unitlength}{\svgwidth}%
  \fi%
  \global\let\svgwidth\undefined%
  \global\let\svgscale\undefined%
  \makeatother%
  \begin{picture}(1,1.2071038)%
    \lineheight{1}%
    \setlength\tabcolsep{0pt}%
    \put(0,0){\includegraphics[width=\unitlength,page=1]{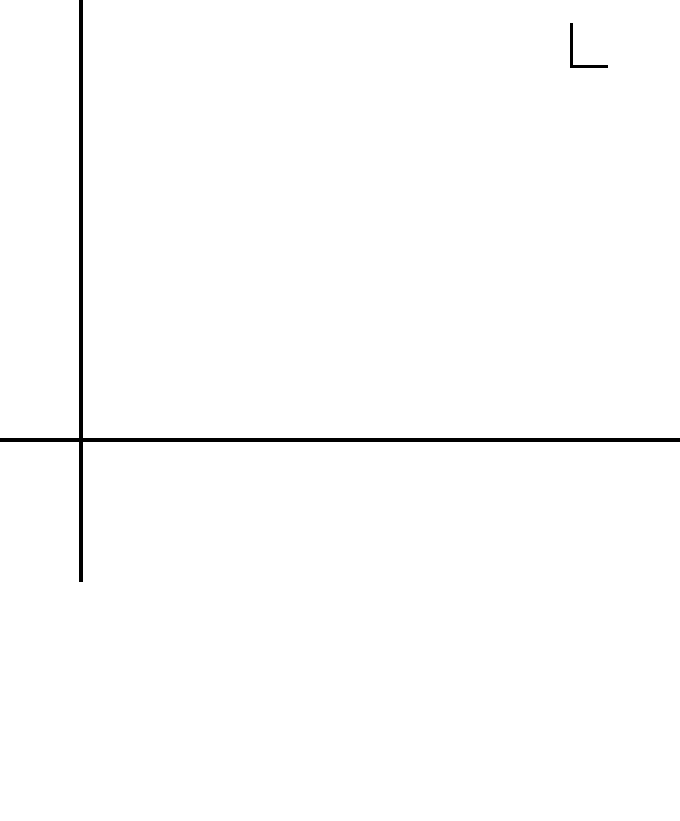}}%
    \put(0.86233961,1.14442143){\color[rgb]{0,0,0}\makebox(0,0)[lt]{\lineheight{1.25}\smash{\begin{tabular}[t]{l}$z$\end{tabular}}}}%
    \put(0,0){\includegraphics[width=\unitlength,page=2]{BesselJ.pdf}}%
    \put(0.65498324,0.4439967){\color[rgb]{0,0,0}\makebox(0,0)[lt]{\lineheight{1.25}\smash{\begin{tabular}[t]{l}$3$\end{tabular}}}}%
    \put(0.44150077,0.4439967){\color[rgb]{0,0,0}\makebox(0,0)[lt]{\lineheight{1.25}\smash{\begin{tabular}[t]{l}$2$\end{tabular}}}}%
    \put(0,0){\includegraphics[width=\unitlength,page=3]{BesselJ.pdf}}%
    \put(0.2102282,0.43510177){\color[rgb]{0,0,0}\makebox(0,0)[lt]{\lineheight{1.25}\smash{\begin{tabular}[t]{l}$1$\end{tabular}}}}%
    \put(0,0){\includegraphics[width=\unitlength,page=4]{BesselJ.pdf}}%
  \end{picture}%
\endgroup%

\caption{The original integral contour of $z$, as depicted as dotted line, picks up poles denoted as cross at positive integers, which sums to Bessel function. The contour is deformed to pass through the saddle-points in the desired limit.}
\label{fig BesselJ}
\end{figure}

The scalar field in flat-space is standardly quantized via
\be
\varphi=\int \fft{d^{d}k}{(2\pi)^{d}2\omega}\big(a_{k}e^{-i\omega t+\vec{k}\cdot x}+a_k^\dagger e^{i\omega t-i\vec{k}\cdot x}\big)\,,\label{flat-simp}
\ee
where
\be
[\varphi(x),\pi_{\varphi}(x')]=i\delta^{(d)}(x-x')\,,\quad [a_k,a_{k'}^\dagger]=(2\pi)^d 2\omega \delta^{(d)}(k-k')\,.
\ee
Our first goal is thus to understand that how the flat-space limit brings (\ref{poincare}) to (\ref{flat-simp}). For this purpose, we change the variables
\be
z=e^{\fft{x_{d}}{\ell}}\,,
\ee
such that the limit $\ell\rightarrow \infty$ would nicely give rise to Minkowski space
\be
ds^2=-dt^2+\sum_{i=1}^d dx_i^2\,,\quad t=\ell T\,,\quad x_{i<d}=\ell Y_i\,.\label{flat space from Poincare}
\ee
To fully understand the flat-space limit of quantization, we have to clarify $\ell\rightarrow\infty$ limit of mode functions. As before, the Fourier phase factor is trivial, we just need to take the energy and the momenta in AdS scaling as $\ell$, i.e., $E=\omega\ell\,, K=k\ell$. Probing the large $\ell$ limit of Bessel functions is more technically difficult. We shall first explicitly write down the series representation of Bessel function
\be
J_{\nu}(x)=(\fft{1}{2}x)^\nu \sum_{n=0}^\infty (-1)^n \fft{\big(\fft{1}{4}x^2\big)^n}{\Gamma(\nu+n+1)\Gamma(n+1)}\,,\label{sereisofJ}
\ee
and we should be interested in its limit at $\nu,x \rightarrow\infty$ with $\nu/x$ fixed. The strategy is to rewrite this series in terms of a complex integral
\be
J_\nu(x)=\int_{C} \fft{dz}{2\pi i}\fft{\big(\fft{1}{2}x\big)^{2z+\nu}}{\Gamma(\nu+z+1)\Gamma(z+1)}
\fft{e^{iz\pi}}{e^{2i z\pi}-1}\,.
\ee
When we deform the contour to pick up poles located at $z\in\mathbb{Z}^+$, the series representation (\ref{sereisofJ}) is produced. The trick to find its limit is to notice that the limit exponentiates the integrand, and thus we can deform the integral contour to pick up the saddle-points, which gives
\be
J_{\nu}(x)\Big|_{\nu,x\rightarrow\infty,\nu/x\, \text{fixed}}=\fft{e^{-\fft{3i\pi}{4}-i\chi}x^{-i\chi}(\nu-i\chi)^{\fft{1}{2}
(i\chi-\nu)}(\nu+i\chi)^{\fft{1}{2}(i\chi+\nu)}}{\sqrt{2\pi}(e^{i\pi\nu-\pi\chi}-1)\chi^{\fft{1}{2}}}+{\rm c.c}\,,
\ee
where $\chi=\sqrt{x^2-\nu^2}$. The process is depicted in Fig \ref{fig BesselJ}.
This trick is actually the main tool of this paper, and we will use it to derive the flat-space limit formula in following sections. After simple algebra, we find
\bea
J_{\Delta-\fft{d}{2}}(|{\bf K}|z)|_{\ell\rightarrow\infty}=\alpha_{k_d}e^{ik_dx_d}+\alpha^\dagger_{k_d}e^{-ik_dx_d}\,,
\eea
where $k_d=\sqrt{|{\bf k}|^2-m^2}$ and
\be
\alpha_{k_d}=\fft{e^{i\ell k_d-i\fft{\pi}{4}}(m+ik_d)^{-\fft{\Delta}{2}}
(m-ik_d)^{\fft{\Delta}{2}}}{\sqrt{2\pi\ell}k_d^{\fft{1}{2}}}
\,.
\ee
Then it is readily to evaluate
\be
\phi|_{\ell\rightarrow\infty}=\fft{\ell^{\fft{d+1}{2}}}{\sqrt{2}}\int \fft{dk_d d^{d-1}k}{(2\pi)^{\fft{d-1}{2}}} \fft{k_d}{\omega}(a_{EK}\alpha_{k_d}e^{-i\omega t+i\vec{k}\cdot x}+{\rm c.c})\,,
\ee
where the covariant momentum now is
\be
p^{(d+1)}=(\omega,k)=(\omega,k_{i<d},k_d)=(p^{(d)},k_d)\,,
\ee
which satisfies the on-shell condition trivially. We have used on-shell condition to replace $d\omega$ by $dk_d$ with a Jacobian factor $k_d/\omega$, it is then easy to observe that $\alpha_{k_d}=(2\pi\ell k_d)^{-\fft{1}{2}}e^{i\tilde{\alpha}_{k_d}-i\fft{\pi}{4}}$, where $\tilde{\alpha}_{k_d}$ is purely real in the Lorentzian signature and denotes the nontrivial phase. We thus obtain the limit for annihilation (or creation) operator
\be
a_{E K}|_{\ell\rightarrow\infty}=\fft{1}{\sqrt{2\ell^{d-1}(2\pi)^{d-1}}}\alpha_{k_d}^\dagger e^{i(\eta+\fft{\pi}{4})}a_k\,,
\ee
which suggests the same formula (\ref{flatlimsc}). We can then readily obtain the smearing kernel in Poincare coordinate (we simply choose $\eta=-\pi/4$ to cancel the pure number in the phase)
\be
|p\rangle=2^{1-\fft{d}{2}+\Delta}\ell^{-\Delta}\sqrt{\fft{\Gamma(1+\Delta-\fft{d}{2})}{\Gamma(\fft{d}{2}-\Delta)}}\fft{k_d^{\fft{1}{2}}}
{|{\bf k}|^{\Delta-\fft{d}{2}}}e^{-i\tilde{\alpha}_{k_d}}\int d^dx e^{ip^{(d)}\cdot x}\mathcal{O}(T,Y)|0\rangle\,.
\ee
We can thus conclude
\be
S=\lim_{\ell\rightarrow\infty}\int \big(\prod_id^dx_i 2^{1-\fft{d}{2}+\Delta_i}\ell^{-\Delta_i}\sqrt{\fft{\Gamma(1+\Delta_i-\fft{d}{2})}{\Gamma(\fft{d}{2}-\Delta_i)}}\fft{k_{id}^{\fft{1}{2}}}
{|{\bf k}_i|^{\Delta_i-\fft{d}{2}}}e^{-i\tilde{\alpha}_{k_d}}e^{ip^{(d)}_i\cdot x_i}\big)\langle\mathcal{O}_1\cdots \mathcal{O}_n\rangle_{\rm L}\,,\label{scattering kernel Poincare}
\ee
where the subscript ${\rm L}$ denotes the Lorentzian correlator. In other words, written in Poincare patch, the flat-space S-matrix is simply the Fourier-transform of correlators, up to prefactors with robust dependence on the momentum. This formula reminds us the flat-space limit in momentum space of AdS proposed in \cite{Raju:2012zr} for massless particles, which is actually related to ours by wick rotations to Euclidean CFT and is also shared by dS flat-space limit. We explain the details in appendix \ref{app Euclidean momentum}, and here we simply quote the formula
\be
S=\lim_{\ell\rightarrow\infty}\int \big(\prod_id^dx_i 2^{1-\fft{d}{2}+\Delta_i}\ell^{-\Delta_i}\sqrt{\fft{\Gamma(1+\Delta_i-\fft{d}{2})}{\Gamma(\fft{d}{2}-\Delta_i)}}\fft{\omega_i^{\fft{1}{2}}}
{|p_i|^{\Delta_i-\fft{d}{2}}}e^{-i\tilde{\alpha}_{\omega}}e^{ip_i\cdot x_i}\big)\langle\mathcal{O}_1\cdots \mathcal{O}_n\rangle_{\rm E}\,,\label{scattering kernel Poincare Euclidean}
\ee
where $p$ is spatial and satisfies $-\omega^2+p^2=-m^2$.

\subsection{HKLL $+$ LSZ $=$ scattering smearing kernel}
\label{HKLL LSZ sec}
\begin{figure}[t]
\centering \hspace{0mm}\def\svgwidth{30mm}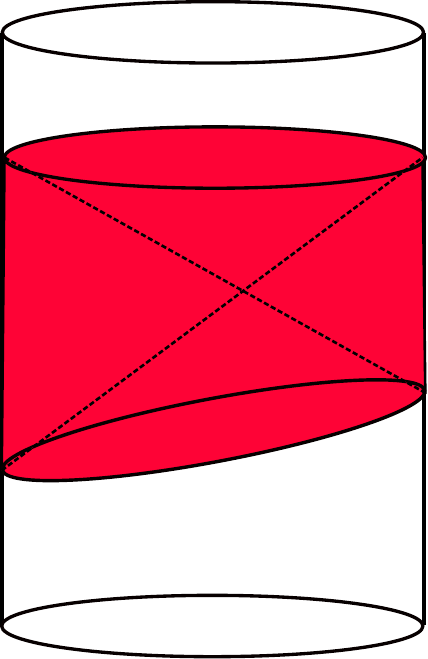
\caption{The red strip of boundary can reconstruct the bulk fields living in the region enclosed by the red strip.}
\label{fig bulk reconstruction}
\end{figure}

In preceding sections, we constructed the scattering smearing kernel for both global AdS and Poincare AdS by quantization procedures. The quantization and mode sum approach is also used to construct the HKLL formula which reconstructs the bulk fields from boundary CFT operators \cite{Hamilton:2005ju,Hamilton:2006az}
\be
\phi(X)=\int d^dP K(X;P)\mathcal{O}(P)\,,\label{HKLLfree}
\ee
where $X$ is bulk coordinate and $P$ boundary coordinate. An illustrative example is depicted in Fig \ref{fig bulk reconstruction}.
Eq.~\eqref{HKLLfree} is the HKLL formula encoding only the free theory. In order to reconstruct bulk fields with interactions, the HKLL formula should include more terms perturbatively in couplings. Nevertheless, the free theory version above is enough for our purpose as we consider perturbative QFT: the Feynman rules consist of only the free fields supplemented by the form of interaction vertices, while the exact propagator is not necessary. We can expect that the flat-space limit of HKLL formula simply represents flat-space fields in terms of CFT operators. In flat-space, S-matrix can be constructed from correlator of fields through LSZ reduction. For scalars, it reads
\be
S=\int \big(\prod_{i=1}^n d^{d+1}x_i e^{ip_i\cdot x_i} (p_i^2+m_i^2)\big)\langle {\rm T}\phi(x_1)\cdots \phi(x_n)\rangle\,,
\ee
where ${\rm T}$ refers to time ordering. Thus it is natural that scattering smearing kernel could be constructed by simply combining HKLL formula and LSZ reduction, in a way that we have
\be
S=\lim_{\ell\rightarrow\infty}
\int \big(\prod_{i=1}^n d^{d+1}x_id^{d}x_i^\prime e^{ip_i\cdot x_i} (p_i^2+m_i^2)K(x_i;x_i^\prime)\big)\langle \mathcal{O}(x_1^\prime)\cdots \mathcal{O}(x_n^\prime)\rangle\,.\label{HKLL+LSZ}
\ee
In this subsection, we provide strong evidence that this procedure indeed works for both global smearing and Poincare smearing. For simplicity, we consider HKLL formula in even bulk dimensions, which is then free of logarithmic term. In odd bulk dimensions, although HKLL formula contains a further logarithmic term, we can argue that such a logarithmic term just gives an factor that is naturally absorbed in the normalization.

In both global and Poincare AdS, the smearing function $K$ in HKLL formula eq.~\eqref{HKLLfree} is written as \cite{Hamilton:2006az}
\bea
K(x,\rho;x^\prime)=\fft{(-1)^{\fft{d-1}{2}}2^{\Delta-d-1}\Gamma(\Delta-\fft{d}{2}+1)}{\pi^{\fft{d}{2}}\Gamma(\Delta-d+1)}
\sigma(x,x^\prime)^{\Delta-d}\,,\label{global HKLL}
\eea
where $\sigma(x,x^\prime)$ is the geodesic length connecting bulk points $x$ and boundary points $x^\prime$, which reads, respectively for global and Poincare AdS
\be
\sigma_{\rm global}=\cos(\tau-\tau^\prime)-\sin\rho\, \hat{r}\cdot\hat{r^\prime}\,,\quad \sigma_{\rm Poincare}=
z^2+|Y-Y^\prime|^2-|T-T^\prime|^2\,.
\ee
To derive the scattering smearing kernel, we rewrite $\sigma^{\Delta-d}$ as $\exp[(\Delta-d)\log\sigma]$, then we can first integrate over $x_i$ in eq.~\eqref{HKLL+LSZ} by picking up the saddle-points of time at large $\ell$ limit.

Let's first discuss the global smearing, where we have integrands for each $x_i$ as follows
\bea
\int dt_id^{d}x_i \exp[-i\omega_i t_i+i |p_i|\hat{p_i}\cdot x_i+(\Delta-d)\log\sigma_{{\rm global}}]\,.
\eea
We simply slip off the normalization factor in HKLL formula (\ref{global HKLL}). We can use eq.~\eqref{global limit } and find that there is a saddle-point for time $t_i$
\be
t_i^\ast=(\arctan(-i\fft{\omega_i}{m_i})+\tau_i^\prime)\ell\,.
\ee
Expanding the exponents around this saddle-point and integrating $t_i$ yields
\bea
&& \int d^dx_i e^{-i\omega_i \tau_i^\prime\ell-i(|p_i|\hat{p_i}
\cdot x_i-\sqrt{\omega_i^2-m_i^2}\hat{r_i^\prime}\cdot x_i)}\times \sqrt{\ell}\,i^{-\Delta_i+\omega_i\ell+d}m_i^{\Delta_i-d+\fft{1}{2}} \fft{(\omega_i\ell-\Delta_i)^{\fft{\omega_i\ell}{2}}}
{(\omega_i\ell+\Delta_i)^{\fft{\omega_i\ell}{2}}}(\omega_i^2-m_i^2)^{\fft{d-\Delta_i}{2}-1}\,.
\cr &&
\eea
Note that we should not take the on-shell condition $\omega^2-m^2=|p|^2$ at this moment, since there is literally not such constraint in AdS, rather we expect
\be
|p|\sim \sqrt{\omega^2-m^2}+\fft{\#}{\ell}\,.
\ee
On the other hand, keeping $p^2+m^2\neq 0$ is helpful for keeping track of how one-particle factor $p_i^2+m_i^2$ in eq.~\eqref{HKLL+LSZ} get canceled. In fact, we can observe that there is a Dirac delta function of the on-shell condition coming from the remaining Fourier factor when we integrate over $x_i$, which can cancel one-particle factor. More precisely, we have
\be
\int d^dx_i e^{-i(|p_i|\hat{p_i}
\cdot x_i-\sqrt{\omega^2-m^2}\hat{r_i^\prime}\cdot x_i)}\sim\delta^{(d-1)}(\hat{p_i}-\hat{r_i^\prime})\fft{\delta(|p_i|-\sqrt{\omega_i^2-m_i^2})}{|p_i|^{d-1}}\,.
\ee
Now we see the delta function mapping directions appear as in eq.~\eqref{smearsc}, and we can directly integrated it out. If we take the on-shell condition, we then have $\delta(0)$, giving the length of radius of our effective flat-space which is of the order $\ell$. On the other hand, the one-particle factor gives $p_i^2+m_i^2\propto 2|p_i|\ell$, we can then argue that one-particle factor and delta function of on-shell condition get canceled, leaving us kinematic factor $2|p_i|$ with some other things to be fixed by normalization. Including additional $|p_i|$ and $1/\Gamma(\Delta-d+1)$ in HKLL formula eq.~\eqref{global HKLL}, the kinematic factor $e^{i\omega_i t_i}\xi_{\omega_i\Delta_i}|\vec{p}_i|^{1-\Delta}$ in scattering smearing kernel eq.~\eqref{scatterkernel} is precisely produced! HKLL formula eq.~\eqref{global HKLL} also provides the Gamma function $\Gamma(\Delta-d/2+1)$, but we still miss some normalization factors, for example, correct scaling in $\ell$. The loss of correct normalization factors is resulted from our rough estimate of the integral where the delta function of on-shell condition arises. The on-shell condition is the saddle-point for $|p|$ at large $\ell$ limit, and a more careful analysis around this saddle-point may give rise to a function that cancels one-particle factor and includes the correct normalization. Nevertheless, we can fix the normalization by requiring tow-point S-matrix is canonically normalized, as we will show in appendix \ref{Normalization}
\be
S_{12}=\langle p_1|p_2\rangle=(2\pi)^d 2\omega\delta^{(d)}(p_1-p_2)\,.\label{2pt norm}
\ee

The Poincare smearing follows similarly. Except now we have
\bea
\int dT_id^{d-1}Y_i dx_d\exp[-i\omega_i T_i\ell+i \vec{k}_i\cdot Y_i\ell+i(k_d)_i (x_d)_i+(\Delta-d)\log\sigma_{{\rm Poincare}}]\,.
\eea
The saddle-points of $T_i$ and $Y_i$ are
\be
T_i-T_i^\prime=-\fft{i\omega_i(m_i+i(k_d)_i)}{|\bf{k_i}|}\,,\quad Y_i-Y_i^\prime=-\fft{ik_i(m_i+i(k_d)_i)}{|\bf{k_i}|}\,.
\ee
Let's only look into the important exponent. We find, after integrating out $T_i$ and $Y_i$
\bea
\int dx_d e^{-ip_i\cdot x_i+i(k_d-\sqrt{|{\bf k}|^2-m^2}x_d)}e^{-i\tilde{\alpha}_{k_d}}(\cdots)\,,
\eea
where $(\cdots)$ represents those not-so-essential factors that could be fixed by eq.~\eqref{2pt norm}. Note the Fourier-transform factor of Poincare smearing kernel \eqref{scattering kernel Poincare} already appears, while the further integration over $x_d$ gives, as in global case, the on-shell condition that is about to get canceled by one-particle factor.

In odd dimensions, the smearing function is modified by additional factor of $\log \sigma$. However, such logarithmic factor doesn't affect the exponent and the saddle-points. Thus it simply gives a constant $\log \sigma^{\ast}$ specified to saddle-points and can be absorbed in the normalization factor.

Now we understand the scattering smearing kernel as the flat-space limit of HKLL bulk reconstruction, the AdS subregion duality \cite{Bousso:2012sj,Czech:2012bh,Bousso:2012mh} then suggests that a local point (where the interactions happen) belongs to the overlap region of global and Poincare AdS can be reconstructed either from global smearing or Poincare smearing. It is thus not surprising that we can transform the Poincare scattering smearing to global scattering smearing, as we will show in section \ref{sec momentum-coordinate duality}.

\section{The flat-space limit from global smearing}
\label{Mellin etc}

\subsection{Known frameworks of the flat-space limit}
\label{sec known frameworks}
We begin with briefly reviewing the existed frameworks of flat-space limit, include Mellin space, coordinate space and partial-wave expansion, from historical point of view without providing very technical details. We will then show these frameworks are originated from global smearing kernel eq.~\eqref{scatterkernel} in the following subsections and dig in more physical details there. Our focus is always the flat-space limit $\ell\rightarrow\infty$, thus we may keep $\ell\rightarrow\infty$ implicit in the rest of this paper when there is no confusion.

\begin{itemize}
\item[] {\bf Mellin space}
\end{itemize}
\begin{itemize}
\item Massless

The Mellin formula (Mellin space will be reviewed shortly in the next subsection) describing the massless scattering in the flat-space limit was first proposed in \cite{Penedones:2010ue}, it gives
\bea
T(s_{ij})=\ell^{\fft{n(d-1)}{2}-d-1}\Gamma(\fft{\Delta_\Sigma-d}{2}) \int_{-i\infty}^{i\infty} \fft{d\alpha}{2\pi i}e^{\alpha}
\alpha^{\fft{d-\Delta_\Sigma}{2}}M(\delta_{ij}=-\fft{\ell^2}{4\alpha}s_{ij})\,,\label{massless Mellin}
\eea
where we use the shorthand notation $\Delta_\Sigma=\sum_{i=1}^n \Delta_i$. This formula was proved in \cite{Fitzpatrick:2011hu} by using the massless scattering smearing kernel (global AdS). We will actually follow the proof \cite{Fitzpatrick:2011hu} in appendix \ref{Mellin-derivation}. It also passes verification to work for supersymmetric theories, see e.g., \cite{Alday:2019nin,Alday:2020lbp,Aprile:2020mus,Alday:2021odx,Aprile:2020luw,1867320,Abl:2020dbx}.

\item Massive

The Mellin formula describing the massive scattering in the flat-space limit was conjectured in \cite{Paulos:2016fap}, and was recently rederived from massive formula in coordinate space \cite{Komatsu:2020sag}. In our conventions, it reads
\be
m_1^{\fft{n(d-1)}{2}-d-1}T(s_{ij})=\Delta_1^{\fft{n(d-1)}{2}-d-1} M\big(\delta_{ij}=\fft{\Delta_i\Delta_j}{\Delta_\Sigma}(1+\fft{\vec{p}_i\cdot\vec{p}_j}{m_im_j})\big)\,.\label{massive Mellin}
\ee
\end{itemize}

\begin{itemize}
\item[] {\bf Coordinate space}
\end{itemize}
\begin{itemize}
\item Massless

The massless scattering written in the coordinate space only has the version for four-point function, which first came out in \cite{Okuda:2010ym} and was rederived from Mellin descriptions in \cite{Penedones:2010ue}. Analysis of contact terms of Witten diagram also suggests the same expression \cite{Maldacena:2015iua}, which also phrases the name ``bulk-point limit''.
\bea
&& \langle \mathcal{O}_1\cdots\mathcal{O}_4\rangle=\prod_{i=1}^4 \fft{\mathcal{C}_{\Delta_i}}{\Gamma(\Delta_i)}\fft{i^{\Delta_\Sigma} \pi^{\fft{d+3}{2}}\ell^{\Delta_\Sigma-d}}{2^{\Delta_\Sigma}}\int ds (\fft{\sqrt{s}}{2})^{\Delta_\Sigma-\fft{d+7}{2}}\xi^{\fft{3-d}{2}}K_{\fft{d-3}{2}}(\sqrt{s}\xi)
\fft{iT(s,\sigma)}{2\sqrt{\sigma(1-\sigma)}}\,,\label{massless coordinate past}
\cr &&
\eea
where
\bea
&& \xi^2=-\lim_{{\rm det}P_{ij}\rightarrow 0} \fft{ \ell^2{\rm det}P_{ij}}{4P_{12}P_{34}\sqrt{P_{13}P_{24}P_{14}P_{23}}}\,,\quad \sigma =
\fft{P_{13}P_{24}}{P_{14}P_{23}}\,,
\eea
where ${\rm det}P_{ij}\sim (z-\bar{z})^2 \sim 0$ is called the bulk-point limit in \cite{Maldacena:2015iua}. One example of the development of this bulk-point is to start with boundary configuration where the Lorentzian time of $\mathcal{O}_{1,2}$ is $-\pi/2$ and the Lorentzian time of $\mathcal{O}_{3,4}$ is $\pi/2$ \cite{Maldacena:2015iua}, see Fig \ref{bulk point kinematics} (figure directly copied from \cite{Caron-Huot:2021kjy})

\begin{figure}
\centering \hspace{0mm}\def\svgwidth{50mm}
\begingroup%
  \makeatletter%
  \providecommand\color[2][]{%
    \errmessage{(Inkscape) Color is used for the text in Inkscape, but the package 'color.sty' is not loaded}%
    \renewcommand\color[2][]{}%
  }%
  \providecommand\transparent[1]{%
    \errmessage{(Inkscape) Transparency is used (non-zero) for the text in Inkscape, but the package 'transparent.sty' is not loaded}%
    \renewcommand\transparent[1]{}%
  }%
  \providecommand\rotatebox[2]{#2}%
  \newcommand*\fsize{\dimexpr\f@size pt\relax}%
  \newcommand*\lineheight[1]{\fontsize{\fsize}{#1\fsize}\selectfont}%
  \ifx\svgwidth\undefined%
    \setlength{\unitlength}{196.28793473bp}%
    \ifx\svgscale\undefined%
      \relax%
    \else%
      \setlength{\unitlength}{\unitlength * \real{\svgscale}}%
    \fi%
  \else%
    \setlength{\unitlength}{\svgwidth}%
  \fi%
  \global\let\svgwidth\undefined%
  \global\let\svgscale\undefined%
  \makeatother%
  \begin{picture}(1,1.21474999)%
    \lineheight{1}%
    \setlength\tabcolsep{0pt}%
    \put(0,0){\includegraphics[width=\unitlength,page=1]{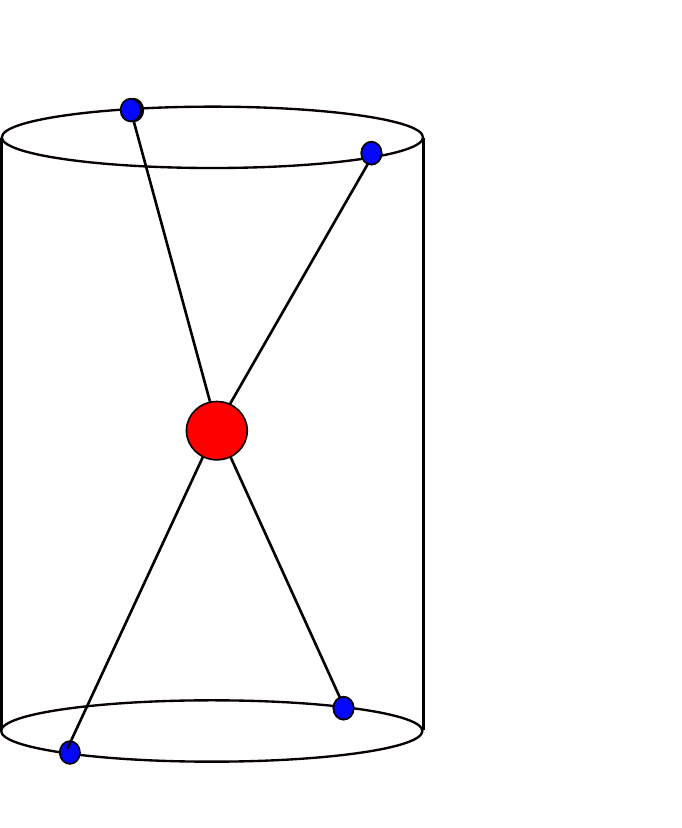}}%
    \put(0.02071316,0.02694037){\color[rgb]{0,0,0}\makebox(0,0)[lt]{\lineheight{1.25}\smash{\begin{tabular}[t]{l}$X_1$\end{tabular}}}}%
    \put(0.40568401,0.12290527){\color[rgb]{0,0,0}\makebox(0,0)[lt]{\lineheight{1.25}\smash{\begin{tabular}[t]{l}$X_2$\end{tabular}}}}%
    \put(0.09705314,1.09713775){\color[rgb]{0,0,0}\makebox(0,0)[lt]{\lineheight{1.25}\smash{\begin{tabular}[t]{l}$X_3$\end{tabular}}}}%
    \put(0.51723757,0.90006021){\color[rgb]{0,0,0}\makebox(0,0)[lt]{\lineheight{1.25}\smash{\begin{tabular}[t]{l}$X_4$\end{tabular}}}}%
    \put(0.38337339,0.59142934){\color[rgb]{0,0,0}\makebox(0,0)[lt]{\lineheight{1.25}\smash{\begin{tabular}[t]{l}$P$\end{tabular}}}}%
  \end{picture}%
\endgroup%

\caption{Bulk-point kinematics in Lorentzian cylinder of AdS. $X_1$ and $X_2$ are at Lorentzian time $-\pi/2$, $X_3$ and $X_4$ are at Lorentzian
time $\pi/2$, where particles are focused on the bulk-point $P$.}\label{bulk point kinematics}
\end{figure}
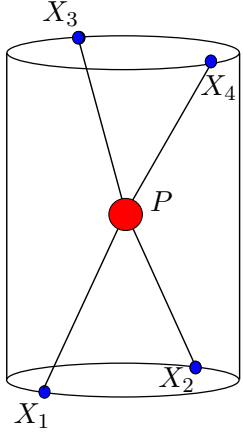

\item Massive

The flat-space limit for massive scattering was recently conjectured in \cite{Komatsu:2020sag} (the same parameterization was also obtained in \cite{Hijano:2019qmi}), rather straightforward by relating kinematics of flat-space scattering to embedding coordinate of CFT
\bea
P=(1,-\fft{\omega}{m},i\fft{\vec{p}}{m})\,,\quad T(s_{ij})=D \langle\mathcal{O}_1\cdots\mathcal{O}_n\rangle\,,
\label{coordinate massive balt}
\eea
where $D$ denotes the contact diagram in AdS, and it can represent the momentum conserving delta function being absorbed into $T$ to give the S-matrix conjecture \cite{Komatsu:2020sag}
\be
S=\langle\mathcal{O}_1\cdots\mathcal{O}_n\rangle\,.\label{coordinate massive S balt}
\ee
\end{itemize}
\begin{itemize}
\item[] {\bf Partial-wave expansion/phase-shift formula}

The description of flat-space limit in terms of partial-wave expansion only refers to four-point case (where $\Delta_2=\Delta_3, \Delta_4=\Delta_1$). The four-point function is expanded in terms of conformal blocks, and the flat-space amplitude is expanded in terms of the partial-waves (where the coefficients are usually named as scattering phase-shift), then one has a map for coefficients of expansions.
\end{itemize}
\begin{itemize}
\item Massless

Expand eq.~\eqref{massless coordinate past} in terms of conformal blocks and partial-waves for LHS and RHS respectively, one can have access to the formula \cite{Maldacena:2015iua}
\bea
e^{-i\pi\gamma_{n,\j}}\fft{c_{n,\j}}{c_{n,\j}^{(0)}}\Big|_{n\rightarrow\infty}=e^{2i\delta_{\j}}\,,\quad 4n^2=\ell^2 s\,.
\eea
One example of contact diagram at leading order was verified in \cite{Maldacena:2015iua}, see  \cite{Cardona:2019ywu} for examples of scalar and graviton tree-level exchange. It even passes checks at loop level \cite{Alday:2017vkk}. Surprisingly, this formula is recently verified to work for gluon scattering, without referring to explicit expression of conformal blocks and partial-waves \cite{Caron-Huot:2021kjy}.

\item Massive

The phase-shift formula for massive scattering was proposed in \cite{Paulos:2016fap}
\bea
\fft{1}{N_{\j}}\sum_{|\Delta-\sqrt{s}\ell|<\delta E}e^{-i\pi(\Delta-\Delta_1-\Delta_2-\j)}\fft{c_{\Delta,\j}}{c_{\Delta,\j}^{(0)}}\Big|_{\Delta\rightarrow\infty}=e^{2i\delta_{\j}}\,,
\eea
where $N_\j$ is the normalization factor to make sure $e^{2i\delta_{\j}}=1$ for free theory. Recently, by doing conformal blcok/partial-wave expansion for their flat-space limit in the coordinate space eq.~\eqref{coordinate massive balt}, \cite{Komatsu:2020sag} managed to derive the same phase-shift formula for identical particles.
\end{itemize}

It is not hard to see for each framework, the formulas for flat-space limit of massless scattering and massive scattering are quite different. For example, the massless Mellin formula is represented as integral over Mellin amplitudes, but massive Mellin formula doesn't have any integral to perform. We expect that massless scattering and massive scattering should be combined into one formula, as suggested by scattering smearing kernel. Meanwhile, as far as we know, some frameworks, for example, the coordinate framework for massive scattering, still remain as a conjecture with supportive examples \cite{Komatsu:2020sag}. In the following subsection, we will start with the global smearing eq.~\eqref{scatterkernel} and present how all those existed descriptions of flat-space limit naturally arise around the saddle-points of the scattering smearing kernel.

\subsection{Mellin space and saddle-points}
\label{subsec saddle-points}
For our purpose, we factorize out the time dependence of the scattering smearing kernel $K_s$ (\ref{scatterkernel}) and denote the remaining factor as kinematic factor
\be
\prod_i \ell^{\fft{d-1}{2}-\Delta_i}\xi_{\omega_i\Delta_i}|\vec{p}_i|^{1-\Delta}2^{\Delta_i}
 \pi^{\fft{d-2}{2}}\Gamma(1+\Delta_i-\fft{d}{2})
={\rm KI}\,.\label{KI factor}
\ee
Such a factor play its role when deriving the final formulae, but it is not relevant for saddle-points analysis.

The $n$-point function in CFT and thus the corresponding AdS amplitudes can be nicely and naturally represented in Mellin space \cite{Mack:2009mi,Mack:2009gy,Fitzpatrick:2011ia,Penedones:2010ue} (which is argued to be well-defined non-perturvatively \cite{Penedones:2019tng})
\bea
\langle\mathcal{O}_1\cdots\mathcal{O}_n\rangle&=&\fft{\mathcal{N}}{(2\pi i)^{\fft{n(n-3)}{2}}}
\int d\delta_{ij}\prod_{i<j}\Gamma(\delta_{ij})\big( P_{ij}\big)^{-\delta_{ij}}M(\delta_{ij})
\,,\quad \sum_{j\neq i}\delta_{ij}=\Delta_i\,,\label{Mellinrep}
\eea
where the normalization factor is
\be
\mathcal{N}=\fft{\pi^{\fft{d}{2}}}{2}\Gamma\big(\fft{\Delta_\Sigma-d}{2}\big)\prod_{i=1}^n \fft{\mathcal{C}_{\Delta_i}}{\Gamma\big(
\Delta_i\big)}\,,\quad \mathcal{C}_\Delta=\fft{\Gamma(\Delta)}{2\pi^{\fft{d}{2}}\Gamma\big(\Delta-\fft{d}{2}+1\big)}\,.
\ee
In Mellin space eq.~\eqref{Mellinrep}, $\delta_{ij}$ is called the Mellin variables, and their integral contours run parallel to the imaginary axis. Note in our coordinate (\ref{embedding}), we have
\be
-2P_i\cdot P_j:=P_{ij}=2(\cos\tau_{ij}-\hat{p}_i\cdot\hat{p}_j)\,,
\ee
where we have already used the fact that $\hat{r}\rightarrow \hat{p}$ due to the presence of delta function in smearing kernel (\ref{smearsc}). To play with the flat-space limit of Mellin amplitudes, we can redefine $\delta_{ij}=\ell^2 \sigma_{ij}$. This redefinition should not be understood as that Mellin variables in CFT depend on $\ell$, because a pure CFT correlator does not know about $\ell$; rather, as we will show shortly, this redefinition is taken for convenience, because the smearing kernel pushes $\delta_{ij}$ to regions that can be parameterized in terms of $\ell$. Moreover, to make order counting more obvious and straightforward, we make the following convention
\be
\tilde{P}_{ij}=\fft{|p_i||p_j|}{2\sum_k\Delta_k}P_{ij}\,,
\ee
such that $\tilde{P}_{ij}$ is well-defined with no subtlety for taking massless limit, and from now on we would shortly write $|\vec{p}|$ as $|p|$.
Such an redefinition is arbitrary and ambiguous as soon as the prefactor is factorized into sum of pair $i,j$, provided the constraints of $\delta_{ij}$, i.e.,
\be
\sum_{i<j}(b_{i}+b_{j})\delta_{ij}=\sum_i b_i\Delta_i\,.
\ee
Such a redefinition does nothing but provide additional prefactors that are not relevant to saddle-points analysis. We make our choice for latter convenience. In the flat-space limit, we can call Stirling approximation for $\Gamma(\delta_{ij})$ and rewrite $P_{ij}^{-\delta_{ij}}$ as an exponent. Then, we can further add the Lagrange multiplier, which is responsible for constraining $\delta_{ij}$, and we will have following exponent
\bea
\exp[\ell^2 \sum_{i<j}(-\sigma_{ij}+\sigma_{ij}\log \sigma_{ij}-\sigma_{ij}\log\tilde{P}_{ij})+i\ell\sum_i\omega_i\tau_i+i \sum_i \lambda_i (\sum_{j\neq i}\ell^2\sigma_{ij}-\Delta_i)]\,.
\eea
It is instructive to make variable change for $\lambda_i$
\be
e^{-i\lambda_i}=\beta_i\,,
\ee
which rewrites the exponent as
\bea
\exp\big[\ell^2 \sum_{i<j}(-\sigma_{ij}+\sigma_{ij}\log \sigma_{ij}-\sigma_{ij}\log\tilde{P}_{ij})+i\ell\sum_i\omega_i\tau_i- \ell^2\sum_{i<j}(\log\beta_i+\log\beta_j)\sigma_{ij}+\sum_i\Delta_i\log\beta_i \big]\,.
\cr &&\label{expstart}
\eea
We can actually immediately solve the saddle-points of above exponent for $\sigma_{ij}$
\be
\sigma_{ij}=\beta_i\beta_j \tilde{P}_{ij}\,.
\ee
Substitute this saddle-point into above exponent, we obtain
\be
\exp[\ell^2\sum_{i<j}\big(-\fft{\beta_i\beta_j}{\Delta_\Sigma}(\cos\tau_{ij}|p_i||p_j|-\vec{p}_i\cdot\vec{p}_j)\big)+ \sum_i\Delta_i
\log\beta_i +i\ell \sum_i\omega_i\tau_i]\,.
\ee
we can start from this exponent and go further to solve saddle-points for $\tau_i$ and $\beta_i$. We assume the momentum conservation, and we can then find a very simple solution to saddle-points equations. We can already notice the difference between massless formula and massive formula comes from the last two terms: they do not contribute for massless case but play their roles for massive case.
\begin{itemize}
\item All massless partiales

For the scattering where all particles are massless, $\Delta_i$ is order $1$, and thus we could neglect the last two terms to consider the saddle-point analysis. The equation gives below
\bea
&& \text{vary $\tau_i$} \rightarrow -\sum_{i\neq k}\fft{\beta_i\beta_k}{\Delta_\Sigma}\sin\tau_{ik} |p_i||p_k|=0\,,
\cr && \text{vary $\beta_i$} \rightarrow \sum_{i\neq k}\fft{\beta_i}{\Delta_\Sigma}(-\cos\tau_{ik}|p_i||p_k|+\vec{p}_i\cdot\vec{p}_k)=0\,.
\eea
It is not hard to find a very simple solution to above equation
\be
\sin\tau_{ij}=0\,,\quad \cos\tau_{ij}=\pm 1\,,\quad \beta_i=\beta\,,
\ee
where $\beta$ is arbitrary. There is $\pm$ sign, because we analytically continue the momentum such that all particles are in-going, which implies that energy $\omega$ of some particles are negative, i.e., $\omega=-|p|$, to guarantee the energy conservation. The saddle-points obtained above produce the known bulk-point configurations Fig \ref{bulk point kinematics} where massless scalars start around $\tau_i=\pm\pi/2$ \cite{Maldacena:2015iua}.

\item All massive particles

For scattering with all massive particles, we should scale $\Delta_i=m_i\ell$, and all terms in the exponent become the same order and participate in the saddle equations
\bea
&& \text{vary $\tau_i$} \rightarrow -\sum_{i\neq k}\fft{\beta_i\beta_k}{m_\Sigma}\sin\tau_{ik} |p_i||p_k|+i \omega_k=0\,,
\cr && \text{vary $\beta_i$} \rightarrow \sum_{i\neq k}\fft{\beta_i}{m_\Sigma}(-\cos\tau_{ik}|p_i||p_j|+\vec{p}_i\cdot\vec{p}_k)+\fft{m_k}{\beta_k}=0\,.\label{vary tau and beta}
\eea
The simple solution is
\be
\sin \tau_{ij}=i\fft{\omega_i m_j-\omega_j m_i}{|p_i||p_j|}\,,\quad \cos\tau_{ij}=\fft{-m_im_j+\omega_i\omega_j}{|p_i||p_j|}\,,\quad \beta_i=i\,.
\label{saddle massive}
\ee
We can easily verify that above solution for $\sin\tau_{ij}$ and $\cos\tau_{ij}$ is consistent on-shell, and there is a simple solution which we can take for convenience
\be
\tau_i=\pm\arcsin \fft{\omega_i}{|p_i|}\,.\label{tsaddle}
\ee
It is obvious that trivially shifting every $\tau_i$ above by the same constant still satisfies eq.~\eqref{saddle massive}. Choosing a convenient reference point can be understood as a sort of gauge choice or frame choice associated with $\tau$ translation symmetry (which is the constant scaling symmetry of a CFT) subject to saddle constraints eq.~\eqref{saddle massive} and the presumed range $\tau\in (-\pi/2-\delta,\pi/2-\delta)$. Amazing part is that the solution of $\tau_i$ is continuous without subtlety for massless limit, except that $\beta_i$ cannot be fully determined for massless case. We can easily show that the solution eq.~\eqref{tsaddle} is exactly what \cite{Komatsu:2020sag} suggests for writing the flat-space limit in coordinate space. We only need to scale $P$ by $\cos\tau$
\be
P\rightarrow (1,-i\tan\tau,\fft{\hat{r}}{\cos\tau})\,.
\ee
This scaling is allowed, because in embedding space, correlators are homogeneous in scaling $P$ weighted by conformal dimensions \cite{Costa:2011mg}. Compare with \cite{Komatsu:2020sag}, we can easily find
\be
n_0=-i\tan\tau\,,\quad n_i=\fft{\hat{r}}{\cos\tau}\,.
\ee
Taking the saddle-points (\ref{tsaddle}) (we choose the minus sign, i.e., $\sin\tau=-\omega/|p|$ and $\cos\tau=-im/|p|$), it is easy to find
\be
n_0=-\fft{\omega}{m}\,,\quad n_i=i\fft{\vec{p}}{m}\,,\label{kineiden}
\ee
which is exactly eq.~(2.9) in \cite{Komatsu:2020sag}! The scaling introduces $1/m$, making their parameterization \cite{Komatsu:2020sag} not suitable for addressing massless particles.

\item Mixing massless and massive particles

When external particles have both massless and massive particles, the situation makes no difference from scattering with all massive particles, thanks to analytic property at massless limit of saddle-points $\tau_i$. This fact is quite obvious but surprising: as soon as there is one massive particle, its contribution will make $\beta_i$ determined!

\end{itemize}

\begin{figure}[t]
\centering \hspace{0mm}\def\svgwidth{60mm}
\begingroup%
  \makeatletter%
  \providecommand\color[2][]{%
    \errmessage{(Inkscape) Color is used for the text in Inkscape, but the package 'color.sty' is not loaded}%
    \renewcommand\color[2][]{}%
  }%
  \providecommand\transparent[1]{%
    \errmessage{(Inkscape) Transparency is used (non-zero) for the text in Inkscape, but the package 'transparent.sty' is not loaded}%
    \renewcommand\transparent[1]{}%
  }%
  \providecommand\rotatebox[2]{#2}%
  \newcommand*\fsize{\dimexpr\f@size pt\relax}%
  \newcommand*\lineheight[1]{\fontsize{\fsize}{#1\fsize}\selectfont}%
  \ifx\svgwidth\undefined%
    \setlength{\unitlength}{234.64403447bp}%
    \ifx\svgscale\undefined%
      \relax%
    \else%
      \setlength{\unitlength}{\unitlength * \real{\svgscale}}%
    \fi%
  \else%
    \setlength{\unitlength}{\svgwidth}%
  \fi%
  \global\let\svgwidth\undefined%
  \global\let\svgscale\undefined%
  \makeatother%
  \begin{picture}(1,0.74561983)%
    \lineheight{1}%
    \setlength\tabcolsep{0pt}%
    \put(0,0){\includegraphics[width=\unitlength,page=1]{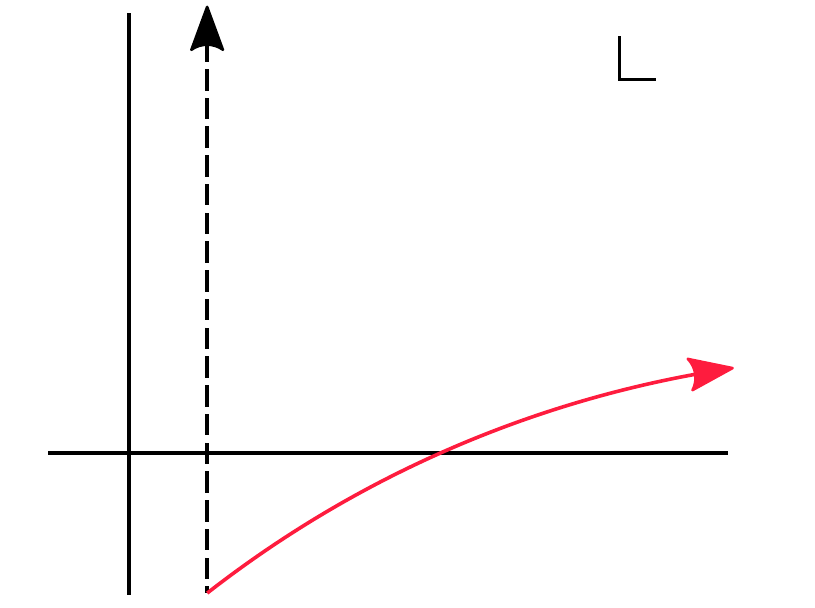}}%
    \put(0.77872118,0.67742081){\color[rgb]{0,0,0}\makebox(0,0)[lt]{\lineheight{1.25}\smash{\begin{tabular}[t]{l}$\alpha$\end{tabular}}}}%
    \put(-0.00381686,0.01247197){\color[rgb]{0,0,0}\makebox(0,0)[lt]{\lineheight{1.25}\smash{\begin{tabular}[t]{l}$-i\infty$\end{tabular}}}}%
    \put(0.03993244,0.69791151){\color[rgb]{0,0,0}\makebox(0,0)[lt]{\lineheight{1.25}\smash{\begin{tabular}[t]{l}$i\infty$\end{tabular}}}}%
    \put(0,0){\includegraphics[width=\unitlength,page=2]{contour.pdf}}%
    \put(0.55134882,0.11191398){\color[rgb]{0,0,0}\makebox(0,0)[lt]{\lineheight{1.25}\smash{\begin{tabular}[t]{l}$\alpha^{\ast}$\end{tabular}}}}%
  \end{picture}%
\endgroup%

\caption{We deform the contour of $\alpha$ to pass through along the steepest descent contour.}
\label{fig contour}
\end{figure}

Some more comments come in order. First, we have to note that above saddle-points analysis assume the energy and momentum conservation, which is, however, not guaranteed in AdS. When taking the flat-space limit, the dominant part of the spacetime is translational symmetric, giving rise to the conservation of the momentum. This fact can be made manifest when we are deriving the flat-space formula of Mellin amplitudes by using global scattering smearing kernel. The original scattering smearing kernel is constructed for the whole S-matrix eq.~\eqref{full global smearing}, and we can easily subtract the identity $\mathbb{I}$ (which represents the free field theory) to leave only scattering amplitudes $T$. It is obvious that the free QFT $\mathbb{I}$ corresponds to mean field theory (MFT) of CFT, because MFT factorizes CFT correlators into several pieces of two-point functions multiplied together
\be
\langle\mathcal{O}_1\cdots\mathcal{O}_{n}\rangle = \langle\mathcal{O}_1\mathcal{O}_2\rangle
\langle\mathcal{O}_3\mathcal{O}_4\rangle\cdots\langle\mathcal{O}_{n-1}\mathcal{O}_n\rangle +{\rm perm}\,,
\ee
which gives rise to a bunch of conserving factors $\delta(p_i+p_j)$
\be
\mathbb{I}= S_{12}S_{34}\cdots S_{n-1,n}+{\rm perm} =\delta(p_1+p_2)\delta(p_3+p_4)\cdots + {\rm perm}\,,
\ee
where $S_{ij}$ is defined in eq.~\eqref{norm state}. Writing in terms of scattering amplitudes, we have
\bea
T=-i\int d^{d+1} p_{\rm tot} \int \prod_{i=1}^n d\tau_i K_s \langle \mathcal{O}_1\cdots \mathcal{O}_n\rangle_{\rm c}\,,
\label{amplitude and correlator}
\eea
where the subscript ``c'' denotes the connected part of CFT correlator and we utilize an integral over $p_{\rm tot}$ to eliminate the momentum conservation delta function (without causing confusion, we will ignore the subscript for simplicity). In other words, the momentum conservation can be understood as saddle-points of integration over $p_{\rm tot}$. More precisely, we can define (we follow \cite{Fitzpatrick:2011hu})
\be
p_i=p'_i+q\,,\quad s_{ij}=s'_{ij}+\fft{2n}{n-2}q\cdot(p_i+p_j)-\fft{2n^2}{(n-1)(n-2)}q^2\,,\quad nq=p_{\rm tot}\,,
\ee
where $p'_i$ and $s'_{ij}$ are the saddle-points of $p_i$ and $s_{ij}$, satisfying
\be
\sum p'_i=0\,,\quad \sum_{j\neq i}s'_{ij}=(n-4)m_i^2+\sum_{j=1}^n m_j^2\,.
\ee
Then we could expand $\beta_i$, $\sigma_{ij}$ and $\tau_i$ around their saddle-points
\bea
\beta_i=\beta^{\ast}+\delta\beta_i\,,\quad \tilde{P}_{ij}=\fft{1}{2\Delta_{\Sigma}}\big(s_{ij}-(m_i+m_j)^2+\delta s_{ij}\big)\,,\quad \sigma_{ij}=\fft{\beta^2}{2\Delta_\Sigma}(s'_{ij}-(m_i+m_j)^2+\epsilon_{ij})\,,
\cr && \label{sadMellin}
\eea
where
\be
\delta s_{ij}=(-2\sin\tau_{ij}^{\ast}\delta\tau_{ij}
-\cos\tau_{ij}^{\ast} \delta\tau_{ij}^2+\cdots)|p_i||p_j|\,,
\ee
and perform the integral over the fluctuations $q\sim \delta\beta_i \sim \delta\tau_i\sim\epsilon_{ij}\ll 1$. For latter purpose of presenting the flat-space limit in coordinate space, we may do those integral separately. First, we integrate out $\epsilon_{ij}$ and $\delta\beta_i$, which is expected to take the form
\be
\langle \mathcal{O}_1\cdots \mathcal{O}_n\rangle=\fft{\mathcal{N}}{(2\pi i)^{\fft{n(n-3)}{2}}}\int d\beta \,\mathcal{D}(s_{ij},\beta)e^{S(q,\delta s_{ij},\beta)}
M\Big(\delta_{ij}=\fft{\ell^2 \beta^2}{2\Delta_\Sigma}\big(s_{ij}-(m_i+m_j)^2\big)\Big)\,.\label{correlator-Mellin}
\ee
We can then expand $S(q,\delta s_{ij},\beta)$ up to $\mathcal{O}(q^2)\sim\mathcal{O}(\delta\tau_i^2)$, and complete the Gaussian integral for $\delta\tau_i$ and $p_{\rm tot}$. The details are recorded in appendix \ref{Mellin-derivation}, and in the end we obtain a Mellin formula in flat-space limit that applies to arbitrary external scalar particles
\bea
T(s_{ij})=\fft{1}{\mathcal{N}_T} \int_{-i\infty}^{i\infty} \fft{d\alpha}{2i\pi} e^\alpha \alpha^{\fft{d-\Delta_\Sigma}{2}}M\Big(\delta_{ij}=-\fft{\ell^2}{4\alpha}\big(s_{ij}-(m_i+m_j)^2\big)\Big)\,,
\label{general Mellin}
\eea
where
\be
\mathcal{N}_T=\fft{\ell^{\fft{n(1-d)}{2}+d+1}}{\Gamma\big(\fft{\Delta_\Sigma-d}{2}\big)}\,.
\ee
Let us comment briefly on why this formula governs massless formula eq.~\eqref{massless Mellin} proposed in \cite{Penedones:2010ue} and massive formula eq.~\eqref{massless Mellin} proposed in \cite{Paulos:2016fap}. For massless scattering, due to $\Delta_i\ll \ell$, we can ignore $m_i$ in Mellin amplitudes and then the formula comes back to eq.~\eqref{massless Mellin}. On the other hand, if there exist one massive particle, then $\Delta_\Sigma$ become parametrically large, together with $e^\alpha$, $\alpha^{\fft{\Delta_\Sigma-d}{2}}$ exponentiates as $\alpha^{\fft{\Delta_\Sigma-d}{2}}=e^{\fft{\Delta_\Sigma-d}{2}\log \alpha}$ to locate the saddle-point of $\alpha$
\be
\alpha^\ast = \fft{\Delta_\Sigma}{2}\,.
\ee
Thus we can deform the contour of $\alpha$ to pass through $\alpha^\ast$, as shown in Fig \ref{fig contour}.
 Around this saddle-point $\alpha^\ast$, we have
\be
\int_{-i\infty}^{i\infty}  \fft{d\alpha}{2i\pi} e^{\alpha+\fft{\Delta_\Sigma-d}{2}\log \alpha}f(\alpha)\simeq \frac{e^{\frac{\Delta _{\Sigma }}{2}} 2^{\frac{1}{2} \left(-d+\Delta _{\Sigma }-2\right)} \Delta _{\Sigma }^{\frac{1}{2} \left(d-\Delta _{\Sigma }+1\right)}}{\sqrt{\pi }}f(\alpha^\ast)\,,
\ee
where the overall coefficient is precisely the large $\Delta_\Sigma$ limit of $1/\Gamma((\Delta_\Sigma-d)/2)$! Thus we are led to eq.~\eqref{massless Mellin}.

The inverse formula of eq.~\eqref{general Mellin} is straightforward and would be useful for going to formula in coordinate space
\be
M(\delta_{ij})=\mathcal{N}_T \int d\gamma e^{-\gamma}\gamma^{\fft{\Delta_\Sigma-d}{2}-1}T\Big(s_{ij}=-\fft{4\gamma \delta_{ij}}{\ell^2}+(m_i+m_j)^2\Big)\,.
\label{invMellin}
\ee

The second subtlety is about the effects of Mellin poles on saddle-points, which were posed recently in \cite{Komatsu:2020sag}. For some analytic regions of Maldastam variables, it turns out the deformation of integral contour to go through saddle-points along steepest descent contour would inevitably pick up poles of Mellin amplitudes, as result, the Mellin formula of the flat-space limit might have additional and isolated contribution from those Mellin poles. In terms of perturbative Witten diagram, this subtle phenomenon is corresponding to the existence of Landau pole \cite{Komatsu:2020sag}. A similar phenomenon is also observed in \cite{Caron-Huot:2021enk} where there exist saddle-points of AdS giving something different from flat-space S-matrix. We do not consider this subtlety in this paper, by appropriately assuming a nice analytic region of Maldastam variables and restricting the Maldastam variables to physical region. Nevertheless, we expect the global smearing kernel eq.~\eqref{full global smearing} always works since its construction does not have any subtlety. Thus we would like to think of eq.~\eqref{full global smearing} as a definition of a certain S-matrix in terms of a specific CFT correlator, where the underlying CFT theory should be supported with large $N$ limit and large gap $\Delta_{\rm gap}$. The details of the CFT correlator encode the interactions of the corresponding S-matrix, and universal properties of the CFT correlators would also have their landing point in S-matrices. Then we might be able to investigate the novel analytic region by directly studying analytic aspects of eq.~\eqref{full global smearing}, provided with axioms of CFT e.g., \cite{Kravchuk:2021kwe}. We leave this interesting question to future research.

\subsection{Conformal frame subject to saddle-points}

Before we move to other space, we would like to comment on the conformal frame subject to the saddle constraints eq.~\eqref{saddle massive}, which will benefit following subsections.

The saddle-points only constrain $\cos\tau_{ij}$ by eq.~\eqref{saddle massive}. We can shift $\tau_i$ by the same constant or shift $\tau_{ij}$ by $2\pi$ without changing the saddle-points and the physics. This reminds us the concept of frame choice. Nevertheless, it is quite trivial to shift a constant, which is nothing but choosing a specific starting time. Much more nontrivially, we notice that eq.~\eqref{saddle massive} only establishes a dictionary relating the conformal configurations to scattering kinematics. From point view of scattering process, we are allowed to choose different scattering frames which then have different $(\omega_i,\vec{p}_i)$ subject to on-shell condition and the momentum conservation. Constrained by saddle-points eq.~\eqref{saddle massive}, a choice of scattering frame then corresponds to a choice of conformal frame.

In our choice, we have explicitly
\be
P=-\fft{i}{|p|}(m,\omega,i\vec{p})\,.\label{P saddle-point}
\ee
The $i$ factor in front of spatial momentum $\vec{p}$ somehow wick rotates the spatial momentum to make $(\omega,i\vec{p})$ map precisely to momentum of scattering. Then straightforwardly, the frame choice of $p$ leads to the corresponding conformal frame $P$. For instance, we are allowed to take the rest frame where $\vec{p}=0$ for massive particles, even though $P$ seems to divergent, it can be scaled to give $P\sim(1,-1,0)$, representing the conformal position at $\infty$! Let's consider four-point case with $\Delta_3=\Delta_2, \Delta_4=\Delta_1$ to gain more insights about conformal frame constrained by eq.~\eqref{saddle massive} and prepare for discussions on the partial-wave expansion in subsection \ref{subsec partial-wave}.

\begin{figure}[t]
\centering \hspace{0mm}\def\svgwidth{70mm}
\begingroup%
  \makeatletter%
  \providecommand\color[2][]{%
    \errmessage{(Inkscape) Color is used for the text in Inkscape, but the package 'color.sty' is not loaded}%
    \renewcommand\color[2][]{}%
  }%
  \providecommand\transparent[1]{%
    \errmessage{(Inkscape) Transparency is used (non-zero) for the text in Inkscape, but the package 'transparent.sty' is not loaded}%
    \renewcommand\transparent[1]{}%
  }%
  \providecommand\rotatebox[2]{#2}%
  \newcommand*\fsize{\dimexpr\f@size pt\relax}%
  \newcommand*\lineheight[1]{\fontsize{\fsize}{#1\fsize}\selectfont}%
  \ifx\svgwidth\undefined%
    \setlength{\unitlength}{301.46392954bp}%
    \ifx\svgscale\undefined%
      \relax%
    \else%
      \setlength{\unitlength}{\unitlength * \real{\svgscale}}%
    \fi%
  \else%
    \setlength{\unitlength}{\svgwidth}%
  \fi%
  \global\let\svgwidth\undefined%
  \global\let\svgscale\undefined%
  \makeatother%
  \begin{picture}(1,0.61016025)%
    \lineheight{1}%
    \setlength\tabcolsep{0pt}%
    \put(0,0){\includegraphics[width=\unitlength,page=1]{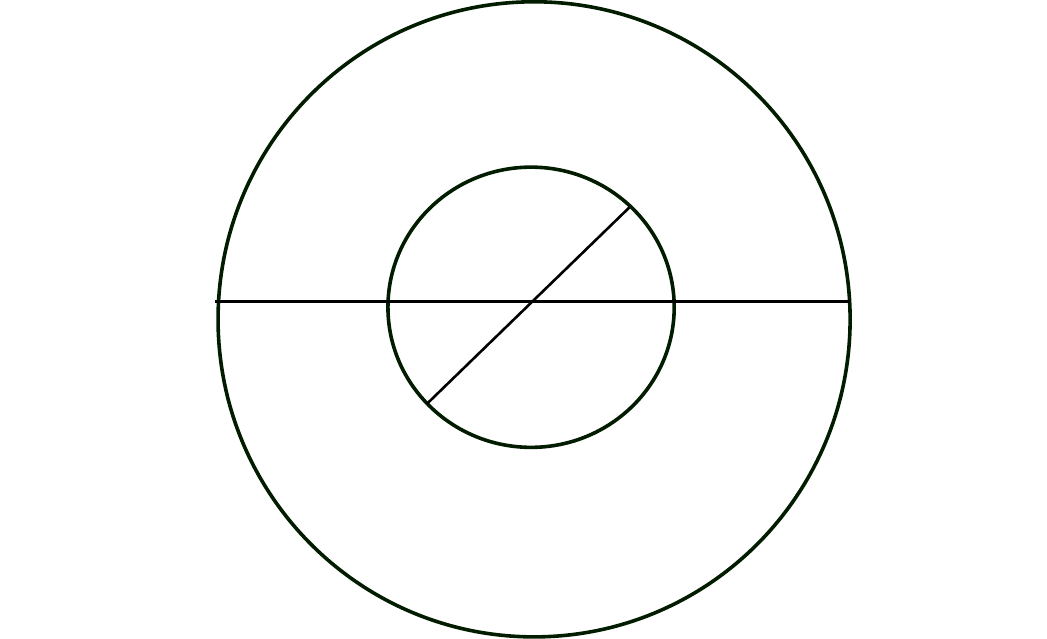}}%
    \put(0.57693008,0.43877012){\color[rgb]{0,0,0}\makebox(0,0)[lt]{\lineheight{1.25}\smash{\begin{tabular}[t]{l}$x_1=w$\end{tabular}}}}%
    \put(0.82334773,0.30812974){\color[rgb]{0,0,0}\makebox(0,0)[lt]{\lineheight{1.25}\smash{\begin{tabular}[t]{l}$x_3=1$\end{tabular}}}}%
    \put(0.27440316,0.14922472){\color[rgb]{0,0,0}\makebox(0,0)[lt]{\lineheight{1.25}\smash{\begin{tabular}[t]{l}$x_2=-w$\end{tabular}}}}%
    \put(-0.0029582,0.30812974){\color[rgb]{0,0,0}\makebox(0,0)[lt]{\lineheight{1.25}\smash{\begin{tabular}[t]{l}$x_4=-1$\end{tabular}}}}%
    \put(0,0){\includegraphics[width=\unitlength,page=2]{radial_frame.pdf}}%
  \end{picture}%
\endgroup%

\caption{Without restriction set by saddle-points, any four points of CFT can be brought to above conformal frame. Constrained by saddle-points of points in CFT, only massless or identical massive four-point function can have access to above conformal frame. Figure comes from \cite{Hogervorst:2013sma}.}
\label{fig radial frame}
\end{figure}

Consider four-point function in a CFT, it is especially useful to use the radial frame $(r,\theta)$ (or to write $w=re^{i\theta}$), which makes Caimir easy to keep track of series expansion of conformal block \cite{Hogervorst:2013sma} (see Fig \ref{fig radial frame} for illustration)
\bea
z\bar{z}=\fft{P_{12}P_{34}}{P_{13}P_{24}}=\fft{16r^2}{(1+r^2+2r\cos\theta)^2}\,,\quad (1-z)(1-\bar{z})=
\fft{P_{14}P_{23}}{P_{13}P_{24}}=\fft{(1+r^2+2r\cos\theta)^2}{(1+r^2-2r\cos\theta)^2}\,.
\cr &&\label{radial frame}
\eea
Constrained by eq.~\eqref{P saddle-point}, only massless scattering and identical massive scattering can have their CFT descriptions within the radial frame. Non-identical particles do not admit the radial frame! It would be very clear to observe these facts by using the center-of-mass frame for scattering amplitudes.
\begin{itemize}
\item {\bf Identical particles}
\end{itemize}

The center-of-mass frame for identical particles is especially simple
\be
p_1=(\omega,p \hat{n})\,,\quad p_2=(\omega,-p\hat{n})\,,\quad p_3=(-\omega,p \hat{n}^\prime)\,,\quad p_4=(-\omega,-p\hat{n}^\prime)\,.
\ee
These kinematic variables $(\omega,\theta)$ can be related to Maldastam variables
\be
\omega=\fft{\sqrt{s}}{2}\,,\quad \cos\theta=1+\fft{2t}{s-4m^2}\,.
\ee
Correspondingly we have
\be
P_{12}=P_{34}=4\,,\quad P_{23}=P_{14}=2\big(\fft{4m^2+s}{4m^2-s}+\cos\theta\big)\,,\quad P_{14}=P_{23}=
2\big(\fft{4m^2+s}{4m^2-s}-\cos\theta\big)\,.\label{Pij identical}
\ee
It is not hard to see this configuration allows the radial frame eq.~\eqref{radial frame} by identifying $\theta$ to scattering angle and
\be
s=\fft{4m^2(r-1)^2}{(r+1)^2}\,,\label{s and r identical}
\ee
where $r$ here can be defined by $r=e^{i\tau_{23}}$. For special case where $m=0$, it is obvious $r=-1=e^{-i\pi}$.

\begin{itemize}
\item {\bf Non-identical particles}
\end{itemize}

If $m_1\neq m_2$, it is then not possible to use the radial frame eq.~\eqref{radial frame}. We can still consider the center-of-mass frame, but now it is a bit more complicated in a sense that there must be different kinematic variables
\be
p_1=(\omega_1, p\hat{n})\,,\quad p_2=(\omega_2,-p\hat{n})\,,\quad p_3=(-\omega_2, p \hat{n}^\prime)\,,\quad p_4=(-\omega_1,-p\hat{n}^\prime)\,.
\label{amp frame}
\ee
Useful kinematic variables now take the form
\bea
&& \omega_1=\fft{s+m_{12}\bar{m}_{12}}{2\sqrt{2}}\,,\quad \omega_2=\fft{s-m_{12}\bar{m}_{12}}{2\sqrt{2}}\,,
\quad  p=\fft{1}{2}\sqrt{\fft{(s-m_{12}^2)
(s-\bar{m}_{12}^2)}{s}}\,,
\cr && \cos\theta=1+\fft{2st}{(s-m_{12}^2)(s-\bar{m}_{12}^2)}\,,\label{theta t nonidentical}
\eea
where $m_{12}=m_1-m_2$ and $\bar{m}_{12}=m_1+m_2$. There is no way to appropriately define $r$ in terms of above variables to reach eq.~\eqref{radial frame}. Nevertheless, we still have access to convenient conformal frame, which is particularly useful for solving conformal block at large conformal dimensions $\Delta,\Delta_i$ (appendix \ref{massive conformal block}) and then analyzing the partial-wave expansion for non-identical particles (subsection \ref{subsec partial-wave}). We only need to identify $\theta$ with scattering angle and then slightly generalize eq.~\eqref{s and r identical}
\be
s=\fft{\bar{m}_{12}^2 (r-1)^2}{(r+1)^2}\,,\quad \cos\theta=1+\fft{2st}{(s-m_{12}^2)(s-\bar{m}_{12}^2)}\,.\label{s and r nonidentical}
\ee
For $m_1=m_2$, eq.~\eqref{s and r nonidentical} reduces to eq.~\eqref{s and r identical}. In this case we have
\bea
&& P_{12}=P_{34}=\fft{4s}{s-m_{12}^2}\,,\quad P_{13}=P_{24}=\fft{4s(s+t-m_{12}^2)}{(m_{12}^2-s)(s-\bar{m}_{12}^2)}\,,
\cr && P_{23}=\fft{4s(4m_2^2-t)}{(m_{12}^2-s)(s-\bar{m}_{12}^2)}\,,\quad P_{14}=\fft{4s(4m_1^2-t)}{(m_{12}^2-s)(s-\bar{m}_{12}^2)}\,.
\eea
The frame now reads (in terms of $(s,t)$)
\bea
z\bar{z}=\fft{(s-\bar{m}_{12}^2)^2}{(s+t-\bar{m}_{12}^2)^2}\,,\quad (1-z)(1-\bar{z})=\fft{m_{12}^4+(\bar{m}_{12}^2-t)^2-
2m_{12}^2(\bar{m}_{12}^2+t)}{(s+t-\bar{m}_{12}^2)^2}\,.\label{frame nonidentical}
\eea
We can use eq.~\eqref{theta t nonidentical} and eq.~\eqref{s and r nonidentical} to explicitly write eq.~\eqref{frame nonidentical} in terms of $r$ and $\cos\theta$, the final expression cannot be simplified to the radial frame eq.~\eqref{radial frame} unless $m_1=m_2$.

\subsection{From Mellin space to coordinate space}

Recently, \cite{Komatsu:2020sag} proposed two conjectures for the (massive) flat-space limit in coordinate space, as we reviewed in subsection \ref{sec known frameworks}, see eqs.~\eqref{coordinate massive balt} and \eqref{coordinate massive S balt}. The key point is the kinematic identification (\ref{kineiden}) that we derived. We could now find a way to derive the flat-space limit in coordinate space by using the inverse Mellin formula (\ref{invMellin}). The idea is to start from Mellin representation of $n$-point function in CFT (\ref{Mellinrep}) subject to kinematic identification (\ref{tsaddle}) and $\hat{r}=\hat{p}$, and work out the integral by picking up the saddle-points $\sigma_{ij}=\beta_i\beta_j \tilde{P}_{ij}$, which can establish a formula relating CFT $n$-point function to Mellin amplitudes specified to those saddle-points. Next, we use the inverse Mellin formula (\ref{invMellin}) to produce the formula directly relating $n$-point function in coordinate space to flat-space scattering amplitudes or S-matrix.

Let us start with (\ref{correlator-Mellin}) and specify to saddle-points, we now have
\be
\langle \mathcal{O}_1\cdots \mathcal{O}_n\rangle=\fft{\mathcal{N}}{(2\pi i)^{\fft{n(n-3)}{2}}}\int d\beta \,\mathcal{D}(s_{ij},\beta)e^{S(0,\delta s_{ij},\beta)}
M\Big(\delta_{ij}=\fft{\ell^2 \beta^2}{2\Delta_\Sigma}\big(s_{ij}-(m_i+m_j)^2\big)\Big)\,,
\ee
where we keep $\delta s_{ij}$ nonzero up to sub-leading order to regulate the integral. We will see later that this regulation is exactly corresponding to bulk-point singularity \cite{Maldacena:2015iua}. Using (\ref{invMellin}) yields
\bea
&& \langle \mathcal{O}_1\cdots \mathcal{O}_n\rangle=\fft{\mathcal{N}\mathcal{N}_T}{(2\pi i)^{\fft{n(n-3)}{2}}}\int d\beta \,\mathcal{D}(s_{ij},\beta)e^{S(0,\delta s_{ij},\beta)}
\int d\gamma e^{-\gamma}\gamma^{\fft{\Delta_\Sigma-d}{2}-1}
\cr && \times T\Big(\fft{2\gamma \beta^2}{\Delta_\Sigma}\big(-s_{ij}+(m_i+m_j)^2\big)+(m_i+m_j)^2
\Big)\,.\label{coor-formula}
\eea
We shall explain in details on this formula for massive case and massless case separately.

\subsubsection{All massless particles: bulk-point singularity}

For all external particles are massless, we have
\bea
 \langle \mathcal{O}_1\cdots \mathcal{O}_n\rangle &=& \fft{\mathcal{N}\mathcal{N}_T}{(2\pi i)^{\fft{n(n-3)}{2}}}\int d\beta \,\mathcal{D}(s_{ij},\beta)e^{S(0,\delta s_{ij},\beta)}
\int d\gamma e^{-\gamma}\gamma^{\fft{\Delta_\Sigma-d}{2}-1}
\cr && \times T\Big(-\fft{2\gamma \beta^2}{\Delta_\Sigma}s_{ij}
\Big)\,.
\eea
We can redefine $\gamma$ by
\be
\tilde{\gamma}=-\fft{2\gamma \beta^2}{\Delta_\Sigma}s_{12}\,,
\ee
which gives
\bea
 \langle \mathcal{O}_1\cdots \mathcal{O}_n\rangle &=& \fft{\mathcal{N}\mathcal{N}_T}{(2\pi i)^{\fft{n(n-3)}{2}}}\int d\tilde{\gamma} \int d\beta \,\mathcal{D}(s_{ij},\beta)e^{S(0,\delta s_{ij},\beta)+\fft{\Delta_\Sigma \tilde{\gamma}}{2\beta^2 s_{12}}}\big(-\fft{\Delta_\Sigma}{2\beta^2
s_{12}}\big)^{\fft{\Delta_\Sigma-d}{2}}
\cr && \times \tilde{\gamma}^{\fft{\Delta_\Sigma-d}{2}-1} T(\tilde{\gamma},\fft{s_{ij}}{s_{12}})\,.
\eea
Now $\tilde{\gamma}$ in the amplitudes play exactly the role as scattering energy $s$. From appendix \ref{Mellin-derivation}, we have
\bea
&& \mathcal{D}(s_{ij},\beta)=(-1)^{\fft{1}{4}n(n+1)}(\fft{\ell^2}{2\Delta_\Sigma})^{\fft{1}{2}\Delta_\Sigma}(2\pi)^{\fft{1}{2}n(n-1)}\beta^{\Delta_\Sigma-n}
\prod_i \omega_i^{\Delta_i}\sqrt{\fft{(2\pi)^{n+1}}{{\rm det}A_\beta}}\,,
\cr &&  S(0,\delta s_{ij},\beta)=-\fft{\ell^2\beta^2}{2\Delta_\Sigma}(\sum_i\omega_i\delta\tau_i)^2\,,
\eea
where $A_\beta$ can be found in eq.~\eqref{Abeta}. We can integrate out $\beta$ to have a Bessel function. We obtain
\bea
\langle \mathcal{O}_1\cdots \mathcal{O}_n\rangle &=& \fft{\mathcal{N}\mathcal{N}_T}{(2\pi i)^{\fft{n(n-3)}{2}}}\int d\tilde{\gamma} D\big(s_{ij},\omega_i\big) \delta^{\fft{1}{2}(n-d-1)} \big(\fft{i\ell\sqrt{\tilde{\gamma}}}{\sqrt{s_{12}}}\big)^{\fft{1}{2}(1+d-n)+\Delta_\Sigma-d-2}
\cr && \times K_{\fft{d+1-n}{2}}\big(\fft{i\ell\sqrt{\tilde{\gamma}}\delta}{\sqrt{s_{12}}}\big) T(\tilde{\gamma},\fft{s_{ij}}{s_{12}})\,,
\label{massless coor formula}
\eea
where
\bea
D\big(s_{ij},\omega_i\big) &=& \fft{(-1)^{\fft{1}{4}(n^2+n+2)}
2^{\fft{1}{2}((n^2-3n-2)+d-\Delta_\Sigma)}\Delta_\Sigma^{\fft{1-n}{2}}\ell^{n+1}\pi^{\fft{1}{2}(n^2-3n-2)}}{s_{12}} \prod_i \omega_i^{\Delta_i}\sqrt{\fft{(2\pi)^{n+1}}{{\rm det}A_\beta}}\,.
\cr &&
\eea
Note we use a shorthand notation $\delta=\sum_i\omega_i\delta\tau_i$. Taking $n=4$, above formula reduces to known massless flat-space limit formula first proposed in \cite{Okuda:2010ym}. More specifically, we can neaten up eq.~\eqref{massless coor formula}
\bea
&& \langle \mathcal{O}_1\cdots\mathcal{O}_4\rangle=\prod_{i=1}^4 \fft{\mathcal{C}_{\Delta_i}}{\Gamma(\Delta_i)}\fft{i \pi^{\fft{d+3}{2}}\ell^{\Delta_\Sigma-\fft{3}{2}(d-1)}}{2^{\Delta_\Sigma+1}}\int ds (\fft{i\sqrt{s}}{2})^{\Delta_\Sigma-\fft{d+7}{2}}\epsilon^{\fft{3-d}{2}}K_{\fft{d-3}{2}}(i\ell\sqrt{s}\epsilon)
\fft{iT(s,\theta)}{|\sin\theta|}\,,\label{4pt massless form}
\cr &&
\eea
where $\epsilon=\delta/\sqrt{s_{12}}$. We use the standard notation for scattering energy i.e., $s=\tilde{\gamma}$, and $\theta$ is the scattering angle $\cos\theta=1+2t/s$. We can see eq.~\eqref{4pt massless form} precisely give eq.~\eqref{massless coordinate past} that is proposed in \cite{Okuda:2010ym}, provided with $i\epsilon=\xi$ and eq.~\eqref{Pij identical} (where $m=0$). The same formula was also understood as bulk-point singularity in CFT \cite{Maldacena:2015iua}, because integrating over $\tilde{\gamma}$ leads to divergence in $\delta=0$, and this is also the reason we keep $\delta\neq 0$ to regulate the answer. In terms of cross-ratio $(z,\bar{z})$, the singularity $\epsilon\rightarrow0$ is actually $z-\bar{z}^{\circlearrowleft}\rightarrow0$ where $\circlearrowleft$ represents the analytic continuation which is automatically done in our discussion.

\subsubsection{Include massive particles}

\begin{figure}
\centering \hspace{0mm}\def\svgwidth{50mm}
\begingroup%
  \makeatletter%
  \providecommand\color[2][]{%
    \errmessage{(Inkscape) Color is used for the text in Inkscape, but the package 'color.sty' is not loaded}%
    \renewcommand\color[2][]{}%
  }%
  \providecommand\transparent[1]{%
    \errmessage{(Inkscape) Transparency is used (non-zero) for the text in Inkscape, but the package 'transparent.sty' is not loaded}%
    \renewcommand\transparent[1]{}%
  }%
  \providecommand\rotatebox[2]{#2}%
  \newcommand*\fsize{\dimexpr\f@size pt\relax}%
  \newcommand*\lineheight[1]{\fontsize{\fsize}{#1\fsize}\selectfont}%
  \ifx\svgwidth\undefined%
    \setlength{\unitlength}{100.43828441bp}%
    \ifx\svgscale\undefined%
      \relax%
    \else%
      \setlength{\unitlength}{\unitlength * \real{\svgscale}}%
    \fi%
  \else%
    \setlength{\unitlength}{\svgwidth}%
  \fi%
  \global\let\svgwidth\undefined%
  \global\let\svgscale\undefined%
  \makeatother%
  \begin{picture}(1,0.71037149)%
    \lineheight{1}%
    \setlength\tabcolsep{0pt}%
    \put(0,0){\includegraphics[width=\unitlength,page=1]{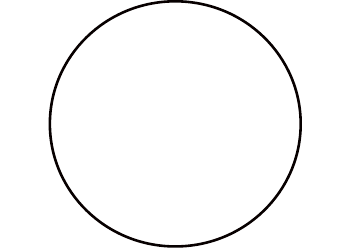}}%
    \put(0.03994888,0.57228614){\color[rgb]{0,0,0}\makebox(0,0)[lt]{\lineheight{1.25}\smash{\begin{tabular}[t]{l}$1$\end{tabular}}}}%
    \put(-0.01162349,0.35574092){\color[rgb]{0,0,0}\makebox(0,0)[lt]{\lineheight{1.25}\smash{\begin{tabular}[t]{l}$2$\end{tabular}}}}%
    \put(0.03754107,0.12958388){\color[rgb]{0,0,0}\makebox(0,0)[lt]{\lineheight{1.25}\smash{\begin{tabular}[t]{l}$3$\end{tabular}}}}%
    \put(0.7430525,0.26423162){\color[rgb]{0,0,0}\makebox(0,0)[lt]{\lineheight{1.25}\smash{\begin{tabular}[t]{l}$n$\end{tabular}}}}%
    \put(0.29861353,0.08991087){\color[rgb]{0,0,0}\makebox(0,0)[lt]{\lineheight{1.25}\smash{\begin{tabular}[t]{l}$\cdots$\end{tabular}}}}%
    \put(0,0){\includegraphics[width=\unitlength,page=2]{AdSCFT_contact.pdf}}%
  \end{picture}%
\endgroup%

\caption{The contact Witten diagram. The dots represents other legs.}\label{Contact Witten diagram}
\end{figure}

As we explain in the previous subsection, if at least one external particle is massive, $\beta$ and $\gamma$ pick their saddle-points up
\be
\beta^\ast=i\,,\quad \gamma^\ast=\fft{\Delta_\Sigma}{2}\,.
\ee
So the formula (\ref{coor-formula}) simply becomes
\bea
&& \langle \mathcal{O}_1\cdots \mathcal{O}_n\rangle=\fft{\mathcal{N}\ell^{\fft{n(1-d)}{2}+d+1}}{(2\pi i)^{\fft{n(n-3)}{2}}} \mathcal{F}(s_{ij}) T(s_{ij})\,,\label{massiveform}
\eea
where $\mathcal{F}(s_{ij})$ is the determinant factor and the rest exponents from picking up saddle-points of $\beta$ and $\gamma$
\be
\mathcal{F}=(-1)^{\fft{1}{4}n(n+1)}(\fft{\ell^2}{2\Delta_\Sigma})^{\fft{1}{2}\Delta_\Sigma}(2\pi)^{\fft{1}{2}(n^2-3n-2)}\prod_i|p_i|^{\Delta_i} {\rm i}^{\Delta_\Sigma-n}e^{-\fft{1}{2}\Delta_\Sigma} \sqrt{\fft{(2\pi)^{n}}{{\rm det}(A_{ij})}}\Big|_{\beta=i}\,,
\ee
Let us explain this factor $\mathcal{F}(s_{ij})$ together with the normalization. Assume we consider the simplest contact interaction with no any derivatives
\be
\mathcal{L}_{\rm int}=\phi_1\phi_2\cdots\phi_n\,.\label{contact Lagrangian}
\ee
This contact interaction is illustrated using Witten diagram in Fig \ref{Contact Witten diagram}.
In flat-space, this kind of contact interaction simply gives $T(s_{ij})=1$, which indicates that the factor $\mathcal{F}(s_{ij})$ is nothing more than contact Witten diagram at large AdS radius limit $\ell\rightarrow\infty$. This fact was verified for $n=4$ identical particles in \cite{Komatsu:2020sag} and for non-identical particles in appendix \ref{app contact witten diagram}. Now we can see that the formula (\ref{massiveform}) is exactly the amplitudes conjecture of the flat-space limit in coordinate space \cite{Komatsu:2020sag}. Moreover, \cite{Komatsu:2020sag} shows that the contact Witten diagram can actually give rise to momentum conservation delta function, see also appendix \ref{app contact witten diagram} for a more general case. Since the contact Witten diagram can be understood as delta function of momentum conservation, multiplying it with amplitudes will then be interpreted as S-matrix which equates CFT correlator.

\subsection{From coordinate space to partial-waves}
\label{subsec partial-wave}
To consider partial-waves, we focus on four-point amplitudes. It is natural to start with the flat-space limit in the coordinate space and then expand CFT and amplitudes in terms of conformal blocks and partial-waves respectively. As consequence, a dictionary map between phase-shift and the OPE coefficients (together with anomalous dimensions) can be established. At tree-level, such a dictionary relates the partial-wave amplitudes to the anomalous-dimensions at leading order.

Represented by partial-waves, the massless scattering and massive scattering is sharply distinguished. The origin of this sharp difference results from the spectra of exchanged operators in four-point function of a CFT $\langle\mathcal{O}_1\mathcal{O}_2\mathcal{O}_2\mathcal{O}_1\rangle$, which can be approximately represented as the double-twist family \cite{Fitzpatrick:2011dm}
\be
[\mathcal{O}_1\mathcal{O}_2]_{n,\j}=\mathcal{O}_1\partial^{2n}\partial_{\mu_1}\cdots\partial_{\mu_\j}\mathcal{O}_2\,,\quad \Delta=\Delta_1+\Delta_2
+\j+2n+\gamma_{n,\j}\,,
\ee
where $\gamma_{n,\j}$ is the anomalous dimension. For external massless particles where $\Delta_1\sim \Delta_2\sim \mathcal{O}(1)$, the four-point function is dominated by massive exchanged particles $\Delta\sim 2n\rightarrow\infty$, effectively making $n$ continuous. On the other hand, for massive $\mathcal{O}_1$ or $\mathcal{O}_2$, double-twist dimension $\Delta$ is already large, and thus we should include all integer $n$.

We will need the conformal block in a limit that the exchanged operator is heavy, i.e., large $\Delta$ limit \cite{Kos:2013tga}
\bea
G_{\Delta,\j}(r,\theta)|_{\Delta\rightarrow\infty}=\fft{\j!}{(d-2)_\j}\fft{(4r)^\Delta C_\j^{\fft{d}{2}-1}(\cos\theta)}{(1-r^2)^{\fft{d}{2}-1}
\sqrt{(1+r^2)^2-4r^2\cos^2\theta}}\,.\label{large delta limit}
\eea
Nonetheless, we should not take eq.~\eqref{large delta limit} for granted. This conformal block eq.~\eqref{large delta limit} assumes $\Delta_i\ll\Delta$ and thus is only applicable for massless scattering in principle. For massive scattering, we have additional large parameters $\Delta_{i}\sim\Delta$, which may modify eq.~\eqref{large delta limit}. Fortunately, as we will see in appendix \ref{massive conformal block}, only $\Delta_{12}$ can appear in the Casimir equation eq.~\eqref{Casi}. Thus eq.~\eqref{large delta limit} is still valid for identical masses. \cite{Komatsu:2020sag} considers identical particles and apply eq.~\eqref{large delta limit} to study partial-wave/phase-shift formula. A worse situation is the scattering with non-identical massive particles, where a standard $(r,\theta)$ frame breaks down, thus we have to be careful about the conformal block eq.~\eqref{large delta limit}. In appendix \ref{massive conformal block}, we focus on non-identical operators and adopt a new conformal frame (see eq.~\eqref{ztorho}) which reduces to eq.~\eqref{s and r nonidentical} and \eqref{frame nonidentical} when $\Delta=\sqrt{s}\ell$. We solve the conformal block, and surprisingly, the expression eq.~\eqref{large delta limit} is still valid, but with slightly modified normalization and $(r,\theta)$ defined differently!

The dictionary are nicely presented in the literature for both massless amplitudes and massive amplitudes, here we will derive them in a hopefully original way.

\subsubsection{Massless phase-shift}
For massless case, the conformal block eq.~\eqref{large delta limit} can be further modified. Notice there is bulk-point singularity $\epsilon\rightarrow 0$ (according to eq.~\eqref{Pij identical}, we should then have $r=e^{-i\epsilon-i\pi}$), which could be served as UV cut-off of spectrum $\Delta$. Thus a more physical limit is taking $\Delta\rightarrow\infty, r\rightarrow 1$ but keeping $\Delta\epsilon $ fixed. The conformal block with this limit (analytically continued to Lorentzian signature) is \cite{Maldacena:2015iua}
\be
G_{\Delta,\j}(e^{-i\epsilon-i \pi},\theta)=\fft{2^{\fft{1-d}{2}+2\Delta}\j!e^{-i\pi\Delta}}{\sqrt{\pi}(d-2)_\j}\sqrt{\Delta}(i\epsilon)^{\fft{3-d}{2}}
K_{\fft{d-3}{2}}(i\Delta\epsilon)\fft{C_\j^{\fft{d}{2}-1}(\cos\theta)}{|\sin\theta|}\,.
\ee
The four-point function can be expanded in terms of this conformal block, namely
\bea
\langle\mathcal{O}_1\cdots\mathcal{O}_4\rangle_c=4^{-(\Delta_1+\Delta_2)}\sum_{n,\j}a_{n,\j}G_{\Delta,\j}(e^{-i\epsilon-i \pi},\theta)\,.
\label{block massless}
\eea
On the other hand, the amplitudes $T$ can take the partial-wave expansion
\bea
T=\sum_{\j} \fft{2^{d+1}(2\j+d-2)\pi^{\fft{d-1}{2}}\Gamma(d-2)}{\Gamma(\fft{1}{2}(d-1))}\fft{1}{s^{\fft{d-3}{2}}}a_{\j}C_{\j}^{\fft{d}{2}-1}(\cos\theta)\,,\quad a_{\j}=i(1-e^{2i\delta_{\j}})\,,\label{amp massless expansion}
\eea
where $a_{\j}$ is called the partial-wave amplitudes and $\delta_{\j}$ is the scattering phase-shift. Comparing eq.~\eqref{4pt massless form} with the conformal block expansion eq.~\eqref{block massless}, it is not hard to find perfect match with the following dictionary, which is expected to be valid to any loop order and even nonperturbatively \cite{Alday:2017vkk}
\bea
e^{-i\pi\gamma_{n,\j}}\fft{c_{n,\j}}{c_{n,\j}^{(0)}}\Big|_{n\rightarrow\infty}=e^{2i\delta_{\j}}\,,\quad 4n^2=\ell^2 s\,,
\label{phase-shift massless}
\eea
where $c_{n,\j}^{(0)}$ is the OPE coefficients in MFT that can sum to disconnected contribution \cite{Fitzpatrick:2011dm}
\bea
&& c_{n,\j}^{(0)}=\frac{\sqrt{\pi } (d+2 J-2) \Gamma (d+J-2) 2^{-d+3}}{\Gamma \left(\frac{d}{2}-\frac{1}{2}\right) \Gamma (J+1) \Gamma (n+1) \Gamma \left(\frac{d}{2}+J+n\right)}\times
\cr &&
 \cr && \frac{\left(-\frac{d}{2}+\Delta _1+1\right)_n \left(-\frac{d}{2}+\Delta _2+1\right)_n \left(\Delta _1\right)_{J+n} \left(\Delta _2\right)_{J+n}}{\left(-d+n+\Delta _1+\Delta _2+1\right)_n \left(J+2 n+\Delta _1+\Delta _2-1\right)_J \left(-\frac{d}{2}+J+n+\Delta _1+\Delta _2\right)_n}\,.
\eea
At tree-level (i.e, $1/N^2$ order), it reduces to a more familiar formula $\gamma_{n,\j}|_{n\rightarrow\infty}=-1/\pi a_{\j}$ \cite{Maldacena:2015iua}, which is verified to be valid even for gluons \cite{Caron-Huot:2021kjy}.

\subsubsection{Massive phase-shift}

We work with $n=4$ for eq.~\eqref{massiveform}
\bea
\langle \mathcal{O}_1\cdots\mathcal{O}_4\rangle_c &=& \mathcal{N}\fft{2^{\fft{3}{2}-4\bar{\Delta}_{12}}\ell^{1-d+\bar{\Delta}_{12}}\pi^{\fft{d}{2}+1}
e^{-\bar{\Delta}_{12}+i\pi\bar{\Delta}_{12}}\bar{m}_{12}^{1-\bar{\Delta}_{12}}(s-m_{12}^2)^{\bar{\Delta}_{12}}(s-\bar{m}_{12}^2)
^{\bar{\Delta}_{12}}}{\sqrt{(s-\bar{m}_{12}^2)(4m_1m_2-t)(s+t-m_{12}^2)}}\times
\cr && iT(s,t)\,.\label{coordinate massive n=4}
\eea
Similar to massless scattering, we should then do conformal block and partial-wave expansion. The partial-wave expansion of amplitudes is rather straightforward, slightly generalizing eq.~\eqref{amp massless expansion} to account for massive phase-space volume (see appendix \ref{app contact witten diagram})
\bea
T=\sum_{\j} \fft{2^{d+1}(2\j+d-2)\pi^{\fft{d-1}{2}}\Gamma(d-2)}{\Gamma(\fft{1}{2}(d-1))}\fft{s^{\fft{d-1}{2}}}
{(s-m_{12}^2)^{\fft{d-2}{2}}(s-\bar{m}_{12}^2)^{\fft{d-2}{2}}}a_{\j}C_{\j}^{\fft{d}{2}-1}(\cos\theta)\,,\label{amp mass expansion}
\eea

On the other hand, expanding the conformal correlator in terms of conformal block is a bit technically subtle. We use the conformal block eq.~\eqref{limit of block general} we solve in appendix \ref{massive conformal block}. Carefully include all relevant factor, we have conformal block expansion
\bea
&& \langle \mathcal{O}_1\cdots\mathcal{O}_4\rangle=\fft{(s-m_{12}^2)^{\bar{\Delta}_{12}}(s+t-m_{12}^2)^{\Delta_{12}}}{4^{\bar{\Delta}_{12}}
s^{\bar{\Delta}_{12}}(4m_1^2-t)^{\Delta_{12}}}\times
\cr && \sum_{\Delta,\j} c_{\Delta,\j} \Big(\fft{m_{12}^2(1+r_\Delta^2+2r_\Delta \eta_\Delta)+m^2(1+r_{\Delta}^2-2r_\Delta\eta_\Delta)+2m_{12}m(1-r_{\Delta}^2)}{(m^2-m_{12}^2(1+r_\Delta^2+2r_\Delta \eta_\Delta)^2)}\Big)^{\fft{\Delta_{12}}{2}}
 g_{\Delta,\j}(r_\Delta,\eta_\Delta)\,,\label{block expansion original}
 \cr &&
\eea
where $(r_\Delta,\eta_\Delta)$ is defined by $(w,\bar{w})$ in eq.~\eqref{ztorho}. We emphasize here that $(r_\Delta,\eta_\Delta)$ is not $(r,\eta=\cos\theta)$ defined via $(s,t)$ in eq.~\eqref{s and r nonidentical}. They only match when $\Delta=\sqrt{s}\ell$. More general, when $\Delta$ deviates from $\sqrt{s}\ell$, we find
\bea
&& r_{\Delta}=\fft{\bar{m}_{12}-\sqrt{s}}{\bar{m}_{12}+\sqrt{s}}-\fft{2m_{12}^2 \bar{m}_{12}(\bar{m}_{12}-\sqrt{s})}{(\sqrt{s}+\bar{m}_{12})
(\bar{m}_{12}^2(m_{12}^2-s)+st)}\delta +\cdots\,,
\cr &&
\cr && \eta_{\Delta}=\frac{s (\bar{m}_{12}^2-s-2 t)+m_{12}^2 (s-\bar{m}_{12}^2)}{(m_{12}^2-s)(s-\bar{m}_{12}^2)}+\frac{4 m_{12}^2 \sqrt{s} t (m_{12}^2 (s-\bar{m}_{12}^2)-s (-\bar{m}_{12}^2+s+t))}{(m_{12}^2-s)^2 (s-\bar{m}_{12}^2)(s (t-\bar{m}_{12}^2)+m_{12}^2 \bar{m}_{12}^2)}\delta
+\cdots\,,
\cr && \label{r eta expand}
\eea
where $\delta=m-\sqrt{s}$. On the other hand, we can factorize MFT OPE $c^{(0)}_{\Delta,\j}$ out, which exponentiates
\bea
&& c^{(0)}_{\Delta,\j}=\frac{2^{d+2} \ell^{-\frac{d}{2}} (d+2 J-2)\Gamma (d+J-2)}{\sqrt{\pi } \Gamma (\frac{d-1}{2}) \Gamma (J+1)}
m^{\frac{3 d}{2}-2 \Delta} (m-m_{12})^{\Delta-\Delta_{12}-\frac{d}{2}} (m+m_{12})^{\Delta+\Delta_{12}-\frac{d}{2}} \times
\cr && (m-\bar{m}_{12})^{\bar{\Delta}_{12}-\Delta-\frac{d}{2}} (\bar{m}_{12}+m)^{\bar{\Delta}_{12}+\Delta-\frac{3 d}{2}} (\bar{m}_{12}-m_{12})^{2\Delta_2+\frac{d}{2}} (\bar{m}_{12}+m_{12})^{\frac{d}{2}-2\Delta_1}\,.
\eea
We assume $c_{\Delta,\j}/c^{(0)}_{\Delta,\j}$ does not have further exponentially large factor, then we can use this MFT OPE and single out $\Delta$ dependence of $(r_{\Delta},\eta_\Delta)$ (i.e, use eq.~\eqref{r eta expand}) to estimate the weighted sum of eq.~\eqref{block expansion original}. ultimately, we find an exponential factor
\bea
\mathcal{E}_\delta=\exp\big[-\frac{\ell \delta ^2 s \bar{m}_{12} \left(-\bar{m}_{12}^2+m_{12}^2+t\right)}{\left(\bar{m}_{12}-\sqrt{s}\right) \left(\bar{m}_{12}+\sqrt{s}\right) \left(s \left(\bar{m}_{12}^2-t\right)-m_{12}^2 \bar{m}_{12}^2\right)}\big]\,.
\eea
The appearance of this exponential factor extends the finding in \cite{Komatsu:2020sag} to non-identical particles. This exponential factor decays if $\Delta-\sqrt{s}\ell$ is large enough to go beyond $\mathcal{O}(\sqrt{\ell})$, which then effectively creates a spectra window together with additional factor that measures the width of the Gaussian distribution
\bea
\sum_{\Delta,\j} (\cdots) \mathcal{E}_{\delta} \simeq \sum_\j \fft{1}{N_\j}\sum_{|\Delta-\sqrt{s}\ell|<\delta E} (\cdots) \times
\Big(\frac{\pi  \ell  \left(\bar{m}_{12}-\sqrt{s}\right) \left(\bar{m}_{12}+\sqrt{s}\right) \left(s \left(\bar{m}_{12}^2-t\right)-m_{12}^2 \bar{m}_{12}^2\right)}{s \bar{m}_{12} \left(-\bar{m}_{12}^2+m_{12}^2+t\right)}\Big)^{\fft{1}{2}}\,,
\cr &&
\eea
where $\delta E\preceq \mathcal{O}(\sqrt{\ell})$. Usually, include the exponential Gaussian factor, we could ignore the sum or integral and evaluate everything at the origin of Gaussian distribution multiplied by Gaussian width factor. However, we will see $(\cdots)$ contains phase factor $e^{-i\pi\Delta}$ which is then sensitive to finite change of $\Delta$. Thus we keep the sum here but now the sum runs over a small window. $1/N_J$ appears to compensate for the remained sum and keep the normalization. The form of this window sum is exactly the one in \cite{Paulos:2016fap}. Gather all factors, we find
\bea
\fft{\langle\mathcal{O}_1\cdots\mathcal{O}_4\rangle}{D_c} &=& -i \sum_\j \fft{1}{N_\j}\sum_{|\Delta-\sqrt{s}\ell|<\delta E} e^{-i\pi(\Delta-\Delta_1-\Delta_2)}
\fft{c_{\Delta,\j}}{c^{(0)}_{\Delta,\j}}\times
\fft{2^{d+1}(2\j+d-2)\pi^{\fft{d-1}{2}}\Gamma(d-2)}{\Gamma(\fft{1}{2}(d-1))}\times
\cr && \fft{s^{\fft{d-1}{2}}}
{(s-m_{12}^2)^{\fft{d-2}{2}}(s-\bar{m}_{12}^2)^{\fft{d-2}{2}}}C_\j^{\fft{d}{2}-1}(\eta)\,,
\eea
where
\be
D_c=i\mathcal{N}\fft{2^{\fft{3}{2}-4\bar{\Delta}_{12}}\ell^{1-d+\bar{\Delta}_{12}}\pi^{\fft{d}{2}+1}
e^{-\bar{\Delta}_{12}}\bar{m}_{12}^{1-\bar{\Delta}_{12}}(s-m_{12}^2)^{\bar{\Delta}_{12}}(s-\bar{m}_{12}^2)
^{\bar{\Delta}_{12}}}{\sqrt{(s-\bar{m}_{12}^2)(4m_1m_2-t)(s+t-m_{12}^2)}}\,.\label{contact Dc}
\ee
Use eq.~\eqref{coordinate massive n=4} (subtract the MFT part) and compare to eq.~\eqref{amp mass expansion}, we conclude
\be
e^{2i\delta_\j}=\fft{1}{N_\j}\sum_{|\Delta-\sqrt{s}\ell|<\delta E} e^{-i\pi(\Delta-\Delta_1-\Delta_2)}\fft{c_{\Delta,\j}}{c^{(0)}_{\Delta,\j}}\,.
\ee
For MFT, we can estimate $N_\j$
\be
N_\j\simeq 2\delta E\,,
\ee
which is also consistent with what found in \cite{Komatsu:2020sag}. It is pointed out that there are some bound states below $\Delta=\Delta_1+\Delta_2$, we refer \cite{Komatsu:2020sag} for more discussions.

\section{Momentum-coordinate duality}
\label{sec momentum-coordinate duality}
The last section is devoted to discussions of variants stemming from the global scattering smearing. In addition to those flat-space limits discussed in the last section, we can also construct the flat-space amplitudes from momentum space of a CFT, as originally suggested by \cite{Raju:2012zr}. The origin of this momentum space prescription is Poincare AdS reconstruction. Naturally, we should ask, can we also establish connections between global scattering smearing and Poincare scattering smearing?

The answer is positive. Intuitively, when the AdS radius is large enough, the wave packets propagate freely in the bulk until they scatter through each other around a bulk region which is extremely local compared to the AdS radius. This region is where the flat-space S-matrix can be defined and we may call it the scattering region \cite{May:2019odp}. Physically, the scattering smearing kernel describes the bulk reconstruction of scattering region. The scattering region we are going to reconstruct must fall in one subregion $A$ of AdS, then according to the subregion duality, this scattering region can be reconstructed from smearing over the subregion of boundary $A_b$ spanned by $A$. For example, applying to one Poincare patch, we can reconstruct any scattering region inside the patch by the full $M^{d}$ plane (which can be wick rotated to $R^d$), which is exactly what we find in eq.~\eqref{scattering kernel Poincare}: reconstruct the scattering in terms of the CFT correlator in the momentum space. Meanwhile, it is also possible to find another AdS subregion $B$ which has overlap with $A$, and the overlap includes the same scattering region. If $B$'s spanned boundary region $B_b$ is different from $A_b$, then we can reconstruct the same S-matrix by two different CFT prescriptions. In a very robust way, since the S-matrix is the same one defined in the same scattering region, the two prescriptions of CFT correlators should be identified.

\begin{figure}[t]
\centering \hspace{0mm}\def\svgwidth{145mm}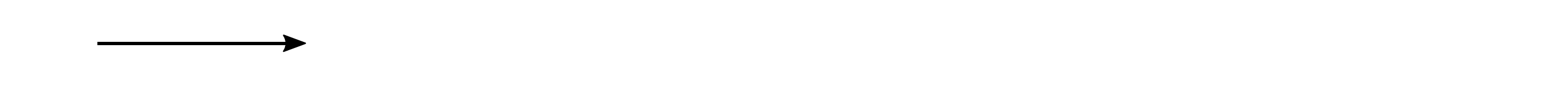
\caption{The analytic operations taking CFT on $M^d$ to CFT on Lorentzian $R\times S^{d-1}$.}\label{analytic operation}
\end{figure}

A bit trivial use of the idea suggested by subregion duality described above is to take $A$ a certain Poincare patch and $B$ the global AdS, as we study in this paper. Then we should be able to equate the global scattering smearing and Poincare scattering smearing, giving
\bea
&& (\prod_i\sqrt{k_{id}}|{\bf k}_i|^{\Delta_i-\fft{d}{2}}e^{-i\tilde{\alpha}_{k_{id}}})\langle\mathcal{O}_{1}(\omega_1\ell,{\bf k}_1\ell)\cdots \mathcal{O}_n(\omega_n\ell,{\bf k}_n\ell)\rangle_{L}
\cr &&
\cr && =\int \big(\prod_i d\tau_i  e^{i\omega_i \tau_i\ell}\ell^{-\fft{d-1}{2}}\xi_{\omega_i\Delta_i}|\vec{p}_i|^{1-\Delta_i}
2^{\fft{d}{2}-1} \pi^{\fft{d-1}{2}}\big)\langle \mathcal{O}_1(\tau_1,\hat{p}_1)\cdots\mathcal{O}_n(\tau_n,\hat{p}_n)\rangle\,,
\cr &&\label{momentum-coordinate dual Lorentzian}
\eea
where we eliminate Gamma functions by assuming large $\Delta_i$. For those finite $\Delta_i$, the normalization depending on only Gamma functions can be easily restored. This equation \eqref{momentum-coordinate dual Lorentzian} establishes a relation representing the Lorentzian CFT in the momentum space (with large momentum) by the CFT on Lorentzian $R\times S^{d-1}$. We call this relation the momentum-coordinate duality of a CFT. Such a duality is highly nontrivial, it connects two very different space of CFT, which can not be simply transformed via conformal map but via tricky operations as shown in Fig \ref{analytic operation}.

However, the momentum space in the Lorentzian signature is quite hard to keep track of, thus we may use a mild version of momentum-coordinate duality, starting with the middle of Fig \ref{analytic operation} where the momentum space is already analytically continued to Euclidean space
\bea
\langle\mathcal{O}_{1}(p_1\ell)\cdots \mathcal{O}_n(p_n\ell)\rangle_{E} &=& \int \big(\prod_i d\tau_i  e^{i\omega_i \tau_i\ell}\ell^{-\fft{d-1}{2}}\fft{\xi_{\omega_i\Delta_i}|\vec{p}_i|^{1+\fft{d}{2}}}{\sqrt{
\omega_i}}
 e^{i\tilde{\alpha}_{\omega_i}}2^{\fft{d}{2}-1}\pi^{\fft{d-1}{2}}\big)\times
 \cr && \langle \mathcal{O}_1(\tau_1,\hat{p}_1)\cdots\mathcal{O}_n(\tau_n,\hat{p}_n)\rangle\,.
\cr &&\label{momentum-coordinate dual Euclidean}
\eea

How is this momentum-coordinate duality possible? Note that the momentum of CFT is parametrically large, scaling as $\ell$. This fact implies that the Fourier-transform can be approximately evaluated by some saddle-points. Let's play with single Fourier transform of one operator
\be
\int d^dX e^{ip\cdot X\ell}\mathcal{O}_{\rm flat}(X)\,.
\ee
To make contact with LHS of eq,~\eqref{momentum-coordinate dual Euclidean}, we make a conformal transformation, mapping $\mathcal{O}_{\rm flat}$ to $\mathcal{O}_{\rm cyl}$ (see \cite{Simmons-Duffin:2016gjk} eq.~$(93)$ for this map)
\be
\int d^dX e^{ip\cdot X\ell}\mathcal{O}_{\rm flat}(X)\rightarrow \int d\tau d\Omega_{d-1}e^{-i\ell p e^\tau \Omega_p\cdot\Omega-(\Delta-d) \tau}\mathcal{O}_{\rm cyl}(\tau,\hat{n})\,,
\ee
where we have used
\be
r=\sqrt{X^2}=e^\tau\,.
\ee
Then we just wick rotate $\tau\rightarrow i\tau$ and play with
\be
\int d\tau d\Omega_{d-1}e^{i\ell p e^{i\tau} \Omega_p\cdot\Omega-i(\Delta-d) \tau}\mathcal{O}(\tau,\hat{n})\,.
\ee
Since it is not possible for CFT correlators to develop exponentially growing factors of $\hat{n}$, we can then approximate the integral of $\Omega_{d-1}$ by the saddle-points of $\hat{n}$ in the Fourier factor. The saddle-points are precisely those directions along the momentum, i.e., $\hat{n}=\hat{p}\,$ !
\be
\int d\tau d\Omega_{d-1}e^{i\ell p e^{i\tau} \Omega_p\cdot\Omega-i(\Delta-d) \tau}\mathcal{O}(\tau,\hat{n})
=\int d\tau \big(\fft{2\pi}{p \ell}\big)^{\fft{d-1}{2}}e^{i\ell p e^{-i\tau}-i(\Delta-\fft{1}{2})\tau}\mathcal{O}(\tau,\hat{p})\,.
\label{middle of go to global}
\ee
It comes close to the LHS of eq.~\eqref{momentum-coordinate dual Euclidean}, but we still have to figure out how Fourier factor depending on $\tau$ can be identical. Note the extremum of the remaining exponents in eq.~\eqref{middle of go to global} is not giving the correct saddle-points of $\tau$, because CFT correlators develop further exponential growing terms involving $\tau$. As we show in the last section \ref{Mellin etc}, the global smearing kernel is not the end of the story, the $\tau$ integral can actually be dominated by saddle-points eq.~\eqref{tsaddle}. We can see, if we use eq.~\eqref{middle of go to global} rather than the global smearing kernel eq.~\eqref{scatterkernel}, we only need to slightly change the first line of eq.~\eqref{vary tau and beta}
\be
-\sum_{i\neq k}\fft{\beta_i\beta_k}{m_\Sigma}\sin\tau_{ik} |p_i||p_k|+i (e^{-i\tau_i}p-m\tau_i)=0\,,
\ee
which gives rise to the exactly same saddle-points eq.~\eqref{tsaddle}! Thus we can simply estimate $e^{-i\tau}$ around these saddle-points just for showing eq.~\eqref{middle of go to global} can be identified to global smearing,
\be
e^{i\tau}\simeq e^{i\tau^{\ast}}(1+i(\tau-\tau^\ast))\,.
\ee
Picking up the linear $\tau$ term, it explicitly gives
\be
\exp[i\ell p e^{i\tau^\ast}-i\Delta \tau^\ast] = e^{i\omega \tau}\,.
\ee
Other terms with $\tau^\ast$ simply gives
\be
\xi_{\omega,\Delta}e^{i\tilde{\alpha}_{\omega}}\,,
\ee
both giving rise to $\xi$ factor and cancelling $e^{-i\tilde{\alpha}_{\omega}}$. Till now we basically show
\be
\int d^dX e^{ip\cdot X\ell}\mathcal{O}_{\rm flat}(X)\sim \int d\tau e^{i\omega\tau} \mathcal{O}(\tau,\hat{p})\,.
\ee
However, we have to note that using the described trick is not possible to exactly determine the correct normalization, because we partially use the saddle-points approximation, which completely ruin the information of normalization \footnote{One can convince himself about this fact by a simple example $\int dx e^{\ell a^3 \log x-1/3 \ell x^3}f(x)$ where $f(x)$ has no large exponential terms. If we directly evaluate it by saddle-point approach, we obtain $\sqrt{2\pi/(3\ell)}a^{a^3\ell-1/2}e^{-a^3\ell/3}f(a)$. However, if we first linearize $\log x$ around $x=a$, and then evaluate the integral using saddle-point, we find $\sqrt{\pi/\ell}a^{a^3\ell-1/2}e^{-a^3\ell/3}f(a)$, which is basically the same answer but losing a numerical factor of normalization $\sqrt{2/3}$.}. Nevertheless, as the form $e^{i\omega\tau}$ is established, we can easily normalize it as shown in Appendix \ref{Normalization}.

As summary, we use the notation of subregion duality to relate the global scattering smearing and Poincare scattering smearing, which indicates the momentum-coordinate duality. Although the examples of global AdS and Poincare AdS are a bit trivial, this notion of duality has its potential to be more general. The scattering region, as shown in \cite{May:2019odp} recently, must lie in the connected entanglement wedge of boundary subregion where CFT correlators are defined. We may find different entanglement wedges contain the same scattering region, and then it is possible to connect different CFT prescriptions by saddle-points approximation. We leave this idea for future work.

\section{Fun with spinning flat-space limit}
\label{sec spinning flat-space}
In this section, we aim to gain some insights about the flat-space limit for spinning operators/particles. We do not have a much rigorous way to present a convincing formula for flat-space limit of spinning operators, but it is quite natural to state that the saddle-points of embedding coordinate should not change even for spinning particles. A new building block for spinning operators is the embedding polarization $Z$, which is subject to null conditions
\be
Z^2=0\,,\quad Z\cdot P=0\,,
\ee
and the redundancy $Z\simeq Z+\# P$. Constrained by these conditions, we conjecture the following parameterization
\bea
P=-\fft{i}{|p|}(m,\omega,i\vec{p})\,,\quad Z=(\fft{\vec{p}\cdot \vec{\epsilon}}{\omega-m},\fft{\vec{p}\cdot \vec{\epsilon}}{\omega-m},i\vec{\epsilon})\,,\label{parameterization spin}
\eea
where $\vec{\epsilon}$ represents the spatial polarization and is null $\vec{\epsilon}\cdot\vec{\epsilon}=0$. Since we have no way to fix appropriate overall factor for $Z$, we will not give ourselves a hard time on normalization throughout this section. Not exactly similar to $P$ where $(\omega,i\vec{p})$ in $P$ is the wick rotated momentum $p$, $(\vec{p}\cdot \vec{\epsilon}/(\omega-m),i\vec{\epsilon})$ in $Z$ is not the wick rotated polarization $\epsilon$ except for massless case.

We will play with photon-photon-massive three-point function $\langle VV\mathcal{O}\rangle$ using eq.~\eqref{parameterization spin}. We will verify that the flat-space limit indeed gives rise to correct three-point amplitudes in QFT.

In \cite{Caron-Huot:2021kjy}, the authors construct the helicity basis for $d=3$ CFT. The helicity basis resembles the helicity states in QFT and is found to diagonalize three-point pairing, shadow matrix, OPE matrix and parity-conserving anomalous dimensions of gluon scattering at tree level, where the partial-wave expansion is also found to satisfy bulk-point phase-shift formula eq.~\eqref{phase-shift massless} compared to flat-space gluon amplitudes \cite{Caron-Huot:2021kjy}. It is then of interest to ask: does three-point function in helicity basis already match with three-point amplitude?

The construction of helicity basis starts with working in the conformal frame $(0,x,\infty)$ and then Fourier-transform $x$ to $p$, though the concept of helicity is naturally conformal invariant \cite{Caron-Huot:2021kjy}. The trick is to use ${\rm SO}(2)$ which stablize $p$ to label the helicity, separating the indices that are perpendicular or along $p$. The constructed structure is then automatically orthogonal with respect to contracting $p$. As discussed in \cite{Caron-Huot:2021kjy}, this trick is easily to extend to higher dimensions, where one organize the structures by ${\rm SO}(d-1)$ subgroup that fixes $p$. One can perform the dimension reduction of ${\rm SO}(d)$ group to ${\rm SO}(d-1)$, which lists perpendicular indices $\j'<\j$ for spin $\j$ operator. The following differential operator help single out the perpendicular indices
\be
\mathcal{P}^{(k)}_\epsilon=\big(1-\fft{2^k (p\cdot\epsilon)^k}{p^{2k}(d-2-k-2+2n)_k}p^\mu
\mathcal{D}_\mu^{\epsilon}\big)\mathcal{P}^{(k-1)}_\epsilon\,,\quad \mathcal{P}^{(0)}_\epsilon=1\,,
\ee
where the differential operator $\mathcal{D}_\mu^\epsilon$ is used to restore the indices from $\epsilon$ \cite{Dobrev:1975ru}
\be
\mathcal{D}_\mu^\epsilon=(\fft{d}{2}-1+\epsilon\cdot \fft{\partial}{\partial\epsilon})\fft{\partial}{\partial\epsilon^\mu}-
\fft{1}{2}\epsilon_\mu \fft{\partial^2}{\partial\epsilon\cdot\partial\epsilon}\,.\label{Tod}
\ee
The parity-even three-point structures can then be constructed \footnote{We constructed these structures with Simon Caron-Huot during the preparation of \cite{Caron-Huot:2021kjy}. \cite{Caron-Huot:2021kjy} only presents $d=3$ case, where these structures reduce to parity-even helicity basis.}
\bea
&& T_{123}^{i_1, i_2, i_3}(p)\propto (p\cdot \epsilon_1)^{J_1-i_1}(p\cdot \epsilon_2)^{J_2-i_2}(p\cdot \epsilon_3)^{J_3-i_3}
p^{\alpha} \times
\cr &&
 \mathcal{P}^{(i_1)}_{\epsilon_1}\mathcal{P}^{(i_2)}_{\epsilon_2}\mathcal{P}^{(i_3)}_{\epsilon_3}(\epsilon_1\cdot\epsilon_2)^{
\fft{i_{123}}{2}}(\epsilon_1\cdot\epsilon_3)^{\fft{i_{132}}{2}}(\epsilon_2\cdot\epsilon_3)^{\fft{i_{231}}{2}}\,,\label{constrmom}
\eea
where $i_{abc}=i_a+i_b-i_c$ and $\alpha=\Delta_{123}-(J_1-i_1)-(J_2-i_2)-(J_3-i_3)$ (we also denote $\Delta_{123}=\Delta_1+\Delta_2-\Delta_3$). By taking different integers from $0$ to $J_1$, $J_2$ for $i_1$, $i_2$ respectively followed by taking $i_3$ among $|i_1-i_2|, |i_1-i_2|+2, \cdots i_1+i_2$, different structures that are orthogonal in $p$ can thus be produced. The overall normalization is not relevant to our purpose. This construction follows the same spirit of construction of scattering amplitudes using center-of-mass frame, ensuring a counting map to flat-space \cite{Kravchuk:2016qvl}.

We will be focusing on conserved spin-$1$ operator, which is dual to photon or more general gluon (the difference is the color structure encoded in OPE). There are two parity-even structures \cite{Caron-Huot:2021kjy}
\bea
&& T_p=\Big\{\fft{\left[p^2 (\epsilon_1\cdot\epsilon_3)-(p\cdot\epsilon_1)(p\cdot\epsilon_3)\right]\left[p^2 (\epsilon_2\cdot\epsilon_3)-(p\cdot\epsilon_2)(p\cdot\epsilon_3)\right]}
{(p\cdot\epsilon_3)^2}
- \frac{p^2 (\epsilon_1\cdot\epsilon_2)-(p\cdot\epsilon_1)(p\cdot\epsilon_2)}{d-1}
\,,
\cr &&
\cr && p^2 (\epsilon_1\cdot\epsilon_2)-(p\cdot\epsilon_1)(p\cdot\epsilon_2)\Big\}(p\cdot\epsilon_3)^{J_3}p^{d-4-\Delta_3-\j_3}\,.
\label{VVmom}
\eea
We can Fourier-transform these structures back to coordinate space and rewrite in terms of embedding formalism
\be
T_x=M_V.B_V\,,
\ee
where $B_V$ is the basis constructed in embedding space
\bea
&& B_V=\fft{1}{P_{12}^{\frac{1}{2} \left(2 d-\Delta _3-J_3\right)}P_{13}^{\frac{1}{2} \left(\Delta _3+J_3\right)} P_{23}^{\frac{1}{2} \left(\Delta _3+J_3\right)}}\times
\cr && \Big\{-H_{12}(-V_3)^{\j_3}\,,H_{31}H_{23}(-V_3)^{\j_3-2}\,,V_1V_2 (-V_3)^{\j_3}\,,H_{31}V_2(-V_3)^{\j_3-1}\,,H_{23}V_1(-V_3)^{\j_3-1}\Big\}\,,
\cr &&
\eea
in which we follow \cite{Costa:2011mg} to define
\bea
&& H_{ij}=-2 \left(P_i\cdot P_j Z_i\cdot Z_j-P_i\cdot Z_j P_j\cdot Z_i\right)\,,\quad V_i:=V_{i,jk}=\frac{P_i\cdot P_k P_j\cdot Z_i-P_i\cdot P_j P_k\cdot Z_i}{P_j\cdot P_k}\,.
\cr &&
\eea
The $2\times 5$ matrix $M_V$ is given below
\bea
&& \left(
\begin{array}{ccccc}
 \ft{2 n \beta }{1-d} & \ft{-\mathcal{J}_3+(d-1) (2-\tilde{\Delta}_3)^2}{d-1} & \ft{2 n (4-2d(n+1)-4J_3)}{1-d} & \ft{2 n (J_3-(d-1)(\tilde{\Delta}_3-2))}{1-d} & \ft{2 n (J_3-(d-1)(\tilde{\Delta}_3-2))}{1-d} \\
 2 n (d-\beta -1) & (1-J_3) J_3 & 2 n (2J_3+\beta) & -2 n J_3 & -2 n J_3 \\
\end{array}
\right)\,,
\cr &&
\eea
where we have defined $\mathcal{J}_3=\j_3(\j_3+d-2)$, $\Delta_3-\j_3=2(d-2+n)$ and $\tilde{\Delta}_3=d-\Delta_3$ to simplify the expression. We use our parameterization eq.~\eqref{parameterization spin} with center-of-mass frame
\bea
p_1=(\omega,\vec{p})\,,\quad p_2=(\omega,-\vec{p})\,,\quad p_3=(-2\omega,0)\,,\label{center-of-mass frame gluon}
\eea
where we set $|p_3|=0$ by scaling $P_3$. Since $\mathcal{O}_3$ is massive, we should scale it $\Delta_3\sim m_3 \ell$ and only keep the leading term that dominates at $\ell\rightarrow\infty$. In the end, by identifying $\epsilon=\vec{\epsilon}, p=|\vec{p}|$ we find
\bea
T_x \propto \Delta_3^2 T_p\,.
\eea
This is a spinning version of momentum-coordinate duality we discuss in the previous section!

They are also equal to three-point amplitudes in flat-space, where the corresponding vertex is \cite{Chakraborty:2020rxf} (for simplicity, we consider photon, while gluon follows similarly)
\bea
\Big\{\partial_{\mu_1}\cdots\partial_{\mu_{\j_3-2}}F_{\mu_{\j_3-1}\nu}F_{\mu_{\j_3}}\,^\nu \mathcal{O}^{\mu_1\cdots\mu_{\j_3}}\,, \partial_{\mu_1}\cdots
\partial_{\mu_{\j_3}}(F_{\mu\nu}F^{\mu\nu})^2 \mathcal{O}^{\mu_1\cdots\mu_{\j_3}}\Big\}\,.
\eea
By Feynman rule, we can easily read off the three-point amplitudes. We still adopt the center-of-mass frame eq.~\eqref{center-of-mass frame gluon}. After making orthogonal combination of these vertices, we indeed verify
\be
T_{\rm amp} \propto \int d^{d}x e^{ip\cdot x}\langle V(0)V(x)\mathcal{O}(\infty)\rangle \propto \langle V(0)V(x)\mathcal{O}(\infty)\rangle\,.
\ee

We verify that the structures eq.~\eqref{VVmom} are indeed corresponding to nicely orthogonal structures of amplitude, however, there is a puzzle. Using eq.~\eqref{VVmom}, \cite{Caron-Huot:2021kjy} find a messily non-diagonal shadow and OPE matrices except for $d=3$ even for MFT, which is counterintuitive comparing to amplitude. The resolution is simple. We have to notice that the OPE matrix contains ratio of rational function of $\Delta$ where $\Delta$ is the conformal dimension of exchanged operator that is massive. To match with flat-space, we should really take $\Delta\rightarrow\infty$ and keep the leading term. The leading term is perfectly diagonal (the OPE matrix remains diagonal up to $\mathcal{O}(1/\Delta^2)$)
\be
c^{\rm MFT}(\Delta,\j)=\fft{1}{2(d-2)^2(d-1)^3}\left(
\begin{array}{cc}
 1 & 0 \\
 0 & \frac{(J-1) J}{(d-2) (d+J-2) (d+J-1)} \\
\end{array}
\right)\,,
\ee
which readily generalizes $d=3$ diagonal OPE matrix obtained in \cite{Caron-Huot:2021kjy}.

\section{Conclusion}

In this paper, we constructed the scattering smearing kernels for both global AdS (eq.~\eqref{full global smearing}) and Poincare AdS (eq.~\eqref{scattering kernel Poincare}), which represent flat-space S-matrix in $d+1$ in terms of CFT correlator in $d$. We found that the scattering smearing kernel from Poincare AdS is a simple Fourier factor that brings the CFT correlator to momentum space. The scattering smearing kernel from global AdS is more nontrivial, and we found that it is served as the unified origin of other known frameworks of flat-space limit: Mellin space, coordinate space, and partial-waves.

We focused on global AdS and employed the Mellin representation of CFT correlators. We found that the scattering smearing kernel is dominated by specific configurations of CFT embedding coordinate, which is the coordinate parameterization conjectured in \cite{Komatsu:2020sag}. These kinematic saddle-points are valid regardless of mass, but we found that one more saddle-point regarding Mellin constraints is developed for massive scattering. According to this crucial observation, we found a Mellin formula that unifies massless formula and massive formula, see eq.~\eqref{general Mellin}. We used the unified Mellin formula to readily derive a unified formula describing the flat-space limit in coordinate space eq.~\eqref{coor-formula}, which reduces to the bulk-point limit \cite{Maldacena:2015iua} for massless scattering and also gives rise to both amplitude and S-matrix conjecture proposed in \cite{Komatsu:2020sag}. We readily derived the phase-shift formula for massless scattering by doing the partial-wave expansion. As the positions of CFT operators are restricted by kinematic saddle-points, we introduced a new conformal frame, which solves the conformal block at the heavy limit of both internal and external conformal dimensions. This conformal block was then used to derive a phase-shift formula for non-identical massive scattering, proving the proposal of \cite{Paulos:2016fap}.

The notion of subregion duality suggests that the Poincare scattering smearing kernel eq.~\eqref{scattering kernel Poincare} should be transformed to the global scattering smearing kernel eq.~\eqref{scattering kernel Poincare}. We thus came up with a momentum-coordinate duality, which establishes a bridge for the large momentum limit of CFT correlator and smeared CFT correlator in the coordinate space eq.~\eqref{momentum-coordinate dual Lorentzian}. By analyzing the saddle-points of Fourier-transform, we verified this duality and thus connected the flat-space limit in momentum space with other frameworks of flat-space limit. As this final gap was filled, the main result of this paper is to show that all existed frameworks of the flat-space limit of AdS/CFT are equivalent.

The final part of this paper is to play with the flat-space limit for spinning operators. We proposed a reasonable parameterization of embedding polarizations and then verified that the coordinate space and the momentum space of three-point function $\langle VV\mathcal{O}\rangle$ in the flat-space limit are indeed equivalent to each other, and they are equivalent to photon-photon-massive three-point amplitudes. We also quoted the MFT OPE matrix of conserved current four-point function, which becomes diagonal by taking the flat-space limit of intermediate operators $\Delta\rightarrow\infty$.

There are some interesting questions that we do not explore in this paper. Since OPE and anomalous dimensions in CFT can be identified to the phase-shift in QFT, it is then natural to ask, does taking the flat-space limit of Lorentzian inversion formula \cite{Caron-Huot:2017vep,Simmons-Duffin:2017nub} yield the Froissart-Gribov formula (see \cite{Correia:2020xtr} for a review)? A related question is that does the flat-space limit of CFT dispersive sum rule \cite{Carmi:2019cub,Caron-Huot:2020adz} give rise to dispersion relation in QFT? These questions are all relevant to analytic and unitary properties of AdS/CFT \cite{Meltzer:2019nbs,Meltzer:2020qbr,Meltzer:2021bmb,Ponomarev:2019ofr} under the flat-space limit and the investigations of them are in active progress \cite{Caron-Huot:2021enk,Balt:flatspace}. Regarding the analytic analysis, the AdS impact parameter space \cite{Cornalba:2006xk} can serve as an important tool (e.g., probe the conformal Regge limit \cite{Costa:2012cb}), and its flat-space limit (see, e.g., \cite{Antunes:2020pof}) could potentially cover large spin regime where $s\sim \Delta^2-\j^2$ \cite{Caron-Huot:2021enk}. These aspects could shed light on constraining AdS EFT (e.g., \cite{Camanho:2014apa,Kundu:2021qpi}) by recently developed techniques of numerically obtaining EFT bounds \cite{Caron-Huot:2020cmc,Caron-Huot:2021rmr}.

It is also of great importance to derive complete formulas of flat-space limit for spinning correlators, or at least do more examples at four-point level in terms of Mellin space, coordinate space or partial-wave expansion, see e.g., \cite{Hijano:2020szl,Chandorkar:2021viw} for recent nice trying. This could shed light on color-kinematic duality and double-copy relation (see \cite{Kawai:1985xq,Bern:2008qj}) in CFT (see \cite{Armstrong:2020woi,Albayrak:2020fyp,Jain:2021wyn,Jain:2021qcl} for insightful studies in momentum space of AdS/CFT).

Another interesting topic is to investigate the relation to celestial amplitude. Flat-space massless four-point amplitudes, as projected to celestial sphere, develop two lower-dimensional CFT structures with bulk-point delta function $\delta(z-\bar{z})$ \cite{Pasterski:2017ylz}, it is then interesting to clarify its relation to bulk-point limit, as was done in four dimensions \cite{Lam:2017ofc}.

\begin{acknowledgments}
We would like to thank for Simon Caron-Huot for valuable discussions and support. We would also like to thank for Balt C. van Rees, Anh-Khoi Trinh, Zahra Zahraee and Xiang Zhao for useful conversations. We are also grateful to referee's comments and suggestions.
Work of Y.-Z.L. is supported in parts by the Fonds de Recherche du Qu\'ebec - Nature et Technologies and by the Simons Collaboration on the Nonperturbative Bootstrap.
\end{acknowledgments}

\begin{appendix}

\section{Momentum space for Euclidean CFT}
\label{app Euclidean momentum}

In subsection \ref{Poincare}, we construct the scattering smearing kernel from Poincare AdS, which Fourier transform Lorentzian CFT correlators, giving rise to the flat-space limit in the momentum space eq.~\eqref{scattering kernel Poincare}. However, Lorentzian CFTs admit more subtle analytic structures (see \cite{Kravchuk:2018htv} for fun), making it not easy to perform Fourier transform. It is better to represent S-matrix in terms of Euclidean CFT, where the Fourier transform is much straightforward. This is the flat-space limit proposed in \cite{Raju:2012zr}. In this appendix, we demonstrate how, in a direct way, to rewrite eq.~\eqref{scattering kernel Poincare} in terms of Euclidean CFT, which, as the massless condition is turned on, reduces to \cite{Raju:2012zr}.

Of course we should wick rotate Lorentzian CFT to Euclidean CFT, i.e., $T\rightarrow i T$. Correspondingly, we have $E\rightarrow i E$ where $E$ now is spatial momentum rather than energy. However, this procedure causes some troubles for modes expansion eq.~\eqref{poincare}, as we discussed there. A simple resolution is to wick rotates $z\rightarrow i z$, and consequently the Bessel function of the first kind $J_\nu$ remains valid as mode functions. Importantly, we should also retain the spacetime in the flat-space limit eq.~\eqref{scattering kernel Poincare} as a Minkowski space. We can formally do this by taking $\ell\rightarrow i \ell$ and $x_d\rightarrow ix_d$. To be more clear, we do wick rotations as follows
\bea
T\rightarrow i T\,,\quad z\rightarrow i z\,,\quad \ell\rightarrow i\ell\,,\quad x_d\rightarrow i x_d\,,\quad t\rightarrow -t\,,\quad
x_{i<d} \rightarrow i x_{i<d}\,.
\eea
It is easy to see that after doing these analytic continuations, AdS becomes dS and the flat-space limit remains as Minkowski. It is then readily to find the remanning parts of analyzing the flat-space limit still follow subsection \ref{Poincare}, but with the momentum continued correspondingly
\bea
\omega\rightarrow \omega\,,\quad k_{i<d} \rightarrow -ik_{i<d}\,,\quad k_{d}\rightarrow i k_{d} =i\sqrt{|{\bf{k}}|^2+m^2}\,,
\eea
where $|{\bf{k}}|=\sqrt{\omega^2+k_{i<d}^2}$. Now it is easy to see that $\omega$ is no longer the energy but one component of spatial momentum, and the additional momentum coming from bulk $k_d$ is the actual energy as the proposal in \cite{Raju:2012zr}. We may stick to the usual notation calling energy $\omega$, then the scattering smearing kernel eq.~\eqref{scattering kernel Poincare} basically remains the same but replacing $k_d\rightarrow i\omega$ since $k_d$ now is energy
\be
S=\int \big(\prod_id^dx_i 2^{1-\fft{d}{2}+\Delta_i}\ell^{-\Delta_i}\sqrt{\fft{\Gamma(1+\Delta_i-\fft{d}{2})}{\Gamma(\fft{d}{2}-\Delta_i)}}\fft{\omega^{\fft{1}{2}}}
{|p_i|^{\Delta_i-\fft{d}{2}}}e^{-i\tilde{\alpha}_{\omega}}e^{ip_i\cdot x_i}\big)\langle\mathcal{O}_1\cdots \mathcal{O}_n\rangle_{\rm E}\,,\,.
\ee

\section{Normalizing scattering smearing kernel}
\label{Normalization}
The scattering smearing kernels we construct in section \ref{scattering smearing} are already normalized. We show in subsection \ref{HKLL LSZ sec} that using HKLL formula and LSZ can somehow determine the scattering smearing kernels up to normalization. Here we demonstrate we can fix the normalization by requiring the canonical condition
\be
S_{12}=\langle p_1|p_2\rangle=(2\pi)^d 2\omega\delta^{(d)}(p_1-p_2)\,.\label{norm state}
\ee

\subsection{Global smearing}

For global smearing, we start with a smearing kernel with momentum dependence unknown for
$S_{12}$
\be
S_{12}=\int dt_1 dt_2 e^{i(\omega_2t_2-\omega_1t_1)}A_g(p_1) A_g(p_2) \langle\mathcal{O}_1(\tau_1,\hat{p}_1)\mathcal{O}_2(\tau_2,\hat{p}_2)\rangle\,,
\label{S12 todo}
\ee
where $A_g(p)$ is the yet-to-be-determined normalization. We use the following representation of two-point function, basically constructed from quantization eq.~\eqref{quantize operator global}
\bea
&& \langle\mathcal{O}_1(\tau_1,\hat{p}_1)\mathcal{O}_2(\tau_2,\hat{p}_2)\rangle= \fft{\mathcal{C}_\Delta}{2^\Delta (\cos\tau_{12}-\hat{p}_1\cdot\hat{p}_2)^\Delta}=\sum_{n,\j}(N_{\Delta,n,\j}^{\mathcal{O}})^2e^{iE_{n,\j}(\tau_1-\tau_2)}Y_{\j m_i}(\hat{p}_1)
Y_{\j m_i}(\hat{p}_2)\,.
\cr &&
\eea
As we show in subsection \ref{global}, taking $\ell\rightarrow\infty$ yields
\bea
&& \langle\mathcal{O}_1(\tau_1,\hat{p}_1)\mathcal{O}_2(\tau_2,\hat{p}_2)\rangle=\int d\omega \fft{2^{d-2\Delta-1}\ell^{2\Delta-d+1}p^{2\Delta-d}}
{\xi_{\omega \Delta}^2 \Gamma(\Delta+1-\fft{d}{2})^2}e^{i\omega(t_1-t_2)}\delta^{(d-1)}(\hat{p}_1-\hat{p}_2)\,.
\eea
Plugging into eq.~\eqref{S12 todo}, we can perform the integral of $t_{1,2}$ to have $(2\pi)^2\delta(\omega-\omega_1)\delta(\omega-\omega_2)$. Then we can integrate out $\omega$, leaving only one delta function $\delta(\omega_1-\omega_2)$. We have
\bea
S_{12} &=& \fft{2^{d-2\Delta+1}\ell^{2\Delta-d+1}p_1^{2\Delta-d}\pi^2}
{\xi_{\omega_1 \Delta}^2\Gamma(\Delta+1-\fft{d}{2})^2} A_g(p_1)^2 \delta(\omega_1-\omega_2)\delta^{(d-1)}(\hat{p}_1-\hat{p}_2)\,,
\cr &&
\cr &=& \fft{2^{d-2\Delta+1}\ell^{2\Delta-d+1}p_1^{2(\Delta-1)}\pi^2}
{\xi_{\omega_1 \Delta}^2\Gamma(\Delta+1-\fft{d}{2})^2} A_g(p_1)^2 \omega_1 \delta^{(d)}(p_1-p_2)\,,
\eea
where we have used the on-shell condition to rewrite the delta functions
\be
\delta(\omega_1-\omega_2)\delta^{(d-1)}(\hat{p}_1-\hat{p}_2)=\omega_1 p_1^{d-1}\delta^{(d)}(p_1-p_2)\,.
\ee
Equating to eq.~\eqref{norm state}, we obtain correctly
\bea
A_g(p)=2^\Delta \ell^{\fft{d-1}{2}-\Delta}p^{1-\Delta}\pi^{\fft{d-2}{2}}\xi_{\omega\Delta}\Gamma(\Delta+1-\fft{d}{2})\,.
\eea
\subsection{Poincare smearing}

Similarly, we consider $S_{12}$ with normalization factor $A_p$ to be fixed
\be
S_{12}=\int d^dx_1d^dx_2 e^{i(p_1\cdot x_1-p_2 \cdot x_2)}A_p(p_1)A_p(p_2)\langle \mathcal{O}_1(T_1,Y_1)\mathcal{O}_2(T_2,Y_2)\rangle\,,
\ee
where
\be
\langle \mathcal{O}_1(T_1,Y_1)\mathcal{O}_2(T_2,Y_2)\rangle=\fft{\mathcal{C}_\Delta}{|-(T_1-T_2)^2+(Y_1-Y_2)^2|^{\Delta}}\,.
\ee
It is more convenient to work with Euclidean CFT, and we can also work with variables $x_{12}$ and $x_2$
\be
S_{12}=\int d^dx_{12}d^dx_2 e^{i p_1\cdot x_{12} +i p_{12}\cdot x_2}A_p(p_1)A_p(p_2) \fft{\mathcal{C}_{\Delta}\ell^{2\Delta}}
{x_{12}^{2\Delta}}\,.
\ee
The integral of $x_{12}$ performs the Fourier transform for $p^{(d)}_1$, and the integral of $x_2$ simply gives delta function $(2\pi)^d \delta(p_1-p_2)$
\be
S_{12}=2^{d-2\Delta-1} p_1^{2\Delta-d}\fft{\Gamma(\fft{d}{2}-\Delta)}{\Gamma(1+\Delta-\fft{d}{2})}\ell^{2\Delta}\times A_p(p_1)^2 (2\pi)^d \delta^{(d)}(p_1-p_2)\,.
\ee
Compare with eq.~\eqref{norm state}, and then analytically continue back to Lorentzian signature, we find
\be
A_p(p)=2^{1-\fft{d}{2}+\Delta}\ell^{-\Delta}\sqrt{\fft{\Gamma(1+\Delta-\fft{d}{2})}{\Gamma(\fft{d}{2}-\Delta)}}\fft{k_d^{\fft{1}{2}}}
{|{\bf k}|^{\Delta-\fft{d}{2}}}\,.
\ee
\section{Derivation of formulas in Mellin space}
\label{Mellin-derivation}
We break our derivation of Mellin space formula into two steps. First, we approximate four-point function in terms of Mellin amplitudes at saddle-points of $\delta_{ij}$ and then we recall scattering kernel and perform integration over time around its saddle-point for massless case and massive case separately.

\subsection{Limit of Mellin representation and massive formula}

Start with Mellin representation of four-point functions eq.~\eqref{Mellinrep}, we scale $\delta_{ij}=\ell^2\sigma_{ij}$ and exponentiate all integrands as we describe in subsection \ref{subsec saddle-points}, include the explicit prefactor we have
\be
\langle\mathcal{O}_1\cdots \mathcal{O}_n\rangle=\fft{\mathcal{N}}{(2\pi i)^{\fft{n(n-3)}{2}}}\int \prod_{i=1}^n\fft{d\beta_i}{2\pi} [d\sigma_{ij}]
\ell^{n(n-1)+\Delta_\Sigma}\prod_{i=1}^n\fft{{\rm i}}{\beta_i} (\fft{2\pi}{\ell^2})^{\fft{n(n-1)}{4}}\fft{\prod_{i=1}^n |p_i|^{\Delta_i}}{(2\Delta_{\Sigma})
^{\fft{1}{2}\Delta_\Sigma}}\prod_{i<j}(\sigma_{ij}^\ast)^{-\fft{1}{2}} \exp[\cdots]\,,
\ee
where the exponent is exactly eq.~\eqref{expstart}. To be general, we expand the exponent around saddle-points as recorded in eq.~\eqref{sadMellin}, which works for both massless and massive situation. In general, $\beta$ is not determined unless further saddle-points are dominated as for massive particles. We may take a gauge choice that sets $\beta_1=\beta$ to keep track of $\beta$, which introduces additional integration
\be
\int \fft{d\delta\beta_0}{2\pi} \exp[i\delta\beta_0\delta\beta_1]\,.
\ee
We can make further simplification by following \cite{Fitzpatrick:2011hu} to redefine $\epsilon_{ij}$
\be
u_{ij}=\epsilon_{ij}-\fft{2n}{n-2}q\cdot(p_i+p_j)-\delta s_{ij}\,,
\ee
and we obtain
\bea
&& \langle\mathcal{O}_1\cdots \mathcal{O}_n\rangle=\fft{\mathcal{N}}{(2\pi i)^{\fft{n(n-3)}{2}}}\times
\cr && \int d\beta \prod_{a=0}^{n}\fft{\delta\beta_a}{2\pi}[du_{ij}]
(\fft{\ell^2 \beta^2}{2\Delta_\Sigma})^{\fft{n(n-1)}{2}}(\fft{i}{\beta})^n
(\fft{2\pi}{\ell^2})^{\fft{n(n-1)}{4}}\prod_{i<j}(\sigma_{ij}^\ast)^{-\fft{1}{2}}(\fft{\ell^2 \beta^2}{2\Delta_\Sigma})^{\fft{1}{2}\Delta_\Sigma}\prod_i|p_i|^{\Delta_i}\exp[\cdots]\,,
\cr &&
\eea
where the exponent here is
\bea
&& \exp\Big[
i\delta\beta_0
\delta\beta_1-\fft{\ell^2\beta}{2\Delta_\Sigma}\sum_{i<j}\big((\delta\beta_i+\delta\beta_j)
(s'_{ij}-(m_i+m_j)^2)
 +\beta\delta s_{ij}\big)-\fft{\ell^2 n^2\beta^2}{2\Delta_\Sigma}q^2
 \cr && +\fft{\ell^2\beta}{\Delta_\Sigma}\sum_{i<j}\big(\fft{\beta u_{ij}^2}{
4(s'_{ij}-(m_i+m_j)^2)}-\fft{(\delta\beta_i+\delta\beta_j)}{2}(u_{ij}+2n\fft{q\cdot(p_i+p_j)}{n-2}+\delta s_{ij})\big)
\cr && +\fft{\ell^2}{4\Delta_\Sigma}\sum_{i<j}\big((s'_{ij}-(m_i+m_j)^2)(\delta\beta_i+\delta\beta_j)^2\big)+\fft{1}{\beta}\sum_i
\Delta_i\delta\beta_i(1-\fft{\delta\beta_i}{2\beta})+\fft{1}{2}\beta^2\Delta_\Sigma\Big]\,.
\cr &&
\eea
Integrating out $u_{ij}$ gives an overall factor
\be
(2\pi)^{\fft{n(n-1)}{4}}\big(\prod_{i<j}(s'_{ij}-(m_i+m_j)^2)\big)^{\fft{1}{2}}\big(-\fft{2\Delta_\Sigma}{\ell^2\beta^2}\big)
^{\fft{n(n-1)}{4}}\,,
\ee
accompanied with an exponent
\be
\exp\big[-\fft{\ell^2}{4\Delta_\Sigma}\sum_{i<j}(s'_{ij}-(m_i+m_j)^2)(\delta\beta_i+\delta\beta_j)^2\big]\,.
\ee
We should then integrate out $\delta\beta_i$. The exponent relevant to $\delta\beta_a$ can be concisely written in terms of matrices
\be
\exp[-\fft{1}{2}\delta\beta.A_\beta.\delta\beta^{\rm T}+B_\beta.\delta\beta^{\rm T}]\,,\quad \delta\beta=(\delta\beta_0,\cdots,\delta\beta_n)\,,
\ee
where
\bea
&& 
(A_\beta)_{0i}=
(A_\beta)_{i0}=-{\rm i}\,\delta_{i1}\,,\quad (A_\beta)_{ij}=\fft{1}{\beta^2}\Delta_i\delta_{ij}+\fft{\ell^2}{2\Delta_\Sigma}(s'_{ij}-(m_i+m_j)^2)\,,
\cr &&
\cr && (B_\beta)_0=0\,,\quad (B_\beta)_i=\fft{\Delta_i}{\beta}-\fft{\ell^2\beta}{2\Delta_\Sigma}\sum_{j\neq i}(s'_{ij}-(m_i+m_j)^2+\delta s_{ij}+
\fft{2n}{n-2}q\cdot(p_i+p_j))\,.
\cr &&\label{Abeta}
\eea
Integrating out $\delta\beta_a$ thus simply gives
\be
\sqrt{\fft{(2\pi)^{n+1}}{{\rm det}A_\beta}} \exp[\fft{1}{2}\sum_{i,j}(A^{-1}_\beta)_{ij}(B_\beta)_i(B_\beta)_j]\,.
\ee
${\rm det}A$ is difficult to be evaluated for general $n$, nevertheless we can find its pattern follows
\bea
&& {\rm det}A_\beta=\fft{\ell^{2(n-1)}{\rm det}^\prime(s_{ij}-(m_i+m_j)^2)}{(2\Delta_\Sigma)^{n-1}}+
 \fft{\prod_{i=2}^n\Delta_i}{\beta^{2(n-1)}}
 \cr &&
 \cr && +\sum_{m=2}^{n-2}(-1)^{m+1}\sum_{\{i_m\}\neq 1}\big(\prod_{i=2,i\neq \{i_m\}}^n \Delta_i\big)
\big(\prod_{(k,l)>1,(k,l)\neq \{\bar{i_m}\}}^n (s_{kl}-(m_k+m_l)^2)\big)\fft{\ell^{2m}}{4\beta^{2(n-1-m)}}\,,
\cr &&
\eea
where ${\rm det}^\prime$ denotes the determinant with discarding the first raw and column. We should explain more on the notation. $\{i_m\}$ denotes a length $m$ list of numbers and $\{\bar{i_m}\}$ denotes the complementary of $\{i_m\}$ through $i>1$. For massless case, all the followed terms are subdominate compare to the first term, thus the expression reduces to
\be
{\rm det} A_\beta\simeq \fft{\ell^{2(n-1)}{\rm det}^\prime(s_{ij})}{(2\Delta_\Sigma)^{n-1}}\,.
\ee
Including all pieces, we obtain
\be
\langle \mathcal{O}_1\cdots \mathcal{O}_n\rangle=\fft{\mathcal{N}}{(2\pi i)^{\fft{n(n-3)}{2}}}\int d\beta \,\mathcal{D}(s_{ij},\beta)e^{S(q,\delta s_{ij},\beta)}
M\Big(\delta_{ij}=\fft{\ell^2 \beta^2}{2\Delta_\Sigma}\big(s_{ij}-(m_i+m_j)^2\big)\Big)\,,\label{limit of Mellin}
\ee
where
\bea
&& \mathcal{D}(s_{ij},\beta)=(-1)^{\fft{1}{4}n(n+1)}(\fft{\ell^2}{2\Delta_\Sigma})^{\fft{1}{2}\Delta_\Sigma}(2\pi)^{\fft{1}{2}(n^2-3n-2)}
\beta^{\Delta_\Sigma-n}
\prod_i |p_i|^{\Delta_i}\sqrt{\fft{(2\pi)^{n+1}}{{\rm det}A_\beta}}\,,
\cr && S(q,\delta s_{ij},\beta)=-\fft{\ell^2\beta^2}{2\Delta_\Sigma}\sum_{i<j}\delta s_{ij}-\fft{\ell^2\beta^2 n^2}{2\Delta_\Sigma}q^2+
\fft{1}{2}\sum_{i,j}(A^{-1}_\beta)_{ij}(B_\beta)_i(B_\beta)_j+\fft{1}{2}\beta^2\Delta_\Sigma\,.
\eea

The second step is then integrating time and $q$. Generally evaluating this two integrals analytically is technically difficult, fortunately we can discuss massive case and massless case separately, which can largely simplify the problem. For formula involving massive external particles, the situation is much more trivial and it is actually not necessary to really do the derivation. In this case, eq.~\eqref{limit of Mellin} can be further simplified by assigning $\beta=i$ to integrands and dropping integral of $\beta$. Performing integral over $\tau_i$ and $q$, we simply obtain a formula that equates flat-space amplitudes to Mellin amplitudes with $\delta_{ij}=-\ell^2 \beta^2/(2\Delta_\Sigma)\big(s_{ij}-(m_i+m_j)^2$ up to an overall normalization, namely
\bea
T(s_{ij})\propto M\Big(\delta_{ij}=-\fft{\ell^2}{2\Delta_\Sigma}\big(s_{ij}-(m_i+m_j)^2\big)\Big)\,.
\eea
The proportional factor is universal, since it is originated from universal kinematic factor ${\rm KI}$ in eq.~\eqref{KI factor} and universal factor $\mathcal{D}(s_{ij},i)e^{S(q,\delta s_{ij},i)}$ in eq.~\eqref{limit of Mellin}. Thus we can determine the proportional factor by simply considering a contact example eq.~\eqref{contact Lagrangian}. Both flat-space amplitude and Mellin amplitude of such contact interaction are simply coupling constant, thus the proportional factor of above formula is simply $1$!

\subsection{Derivation of massless formula}
When all external particles are massless, the derivation becomes highly nontrivial. The expected form of the formula is
\be
T(s_{ij})\sim\int d\beta f(\beta) M\Big(\delta_{ij}=\fft{\ell^2 \beta^2}{2\Delta_\Sigma}\big(s_{ij}-(m_i+m_j)^2\big)\Big)\,,\label{expect massless}
\ee
however, the existence of integral over $\beta$ makes it impossible to simply determine the proportional function $f(\beta)$ by contact interaction, unless we know $f(\beta)$. A nice derivation is available in \cite{Fitzpatrick:2011hu}, and we review their derivation here but with a different gauge.

Let's first describe how our gauge choice can be transformed to the one used in \cite{Fitzpatrick:2011hu}. The gauge choice in \cite{Fitzpatrick:2011hu} is $\epsilon_{12}=0$ rather than $\beta_1=\beta$ we use. To transform the gauge to $\epsilon_{12}=0$, we only need to redefine $\beta$ by $\beta\rightarrow \beta-\delta\beta_1$ with a specific $\delta\beta_1$ rendering $\epsilon_{12}=0$
\be
\delta\beta_1\simeq\fft{\beta\epsilon_{12}}{2s_{12}}\,.
\ee
Then we have
\be
\exp[i\delta\beta_0\delta\beta_1]\rightarrow \exp[i\delta\beta_0 \fft{\epsilon_{12}\beta}{2s'_{12}}]\,.
\ee
We can then change some variables by
\be
\fft{\beta\ell^2}{2\Delta_\Sigma}\delta\beta_i=i\lambda_i\,,\quad \fft{\delta\beta_0\beta}{2s'_{12}}=\lambda_0\,,
\ee
which provide the following prefactors
\be
(-\fft{2i\Delta_\Sigma}{\beta\ell^2})^n \fft{2s'_{12}}{\beta}\,.
\ee
Then trivially changing the variable $\beta$ by $\beta=i\sqrt{\Delta_\Sigma/(2\alpha)}$ (which will also be used with our gauge anyway) makes the integrand become
\be
\prod_{i=1}d\delta\tau_id\alpha\prod_{a=0}\lambda_a[d\epsilon_{ij}](-\fft{\ell^2}{4\alpha})^{\fft{n(n-3)}{2}}(-\fft{s'_{12}}{\alpha})
(\fft{2\pi}{\ell^2})^{\fft{n(n-1)}{4}}\prod_{i<j}(\sigma_{ij}^\ast)^{-\fft{1}{2}}(-\fft{\ell^2 }{4\alpha})^{\fft{1}{2}\Delta_\Sigma}\prod_i\omega_i^{\Delta_i}\,,
\ee
where the exponent is exactly eq.~(107) in \cite{Fitzpatrick:2011hu} by simply noting $\ell\big|_{\rm here}=R\big|_{\rm there}$ and $\delta\tau_{ij}\big|_{\rm here}=t_{ij}/R\big|_{\rm there}$. It is also easy to check that the prefactors also match with \cite{Fitzpatrick:2011hu}.

With our gauge, we now should start with eq.~\eqref{limit of Mellin} and integrate both $\delta\tau_i$ and $q$ over. The massless limit simplifies the exponent in eq.~\eqref{limit of Mellin}
\be
\fft{1}{2}\sum_{i,j}(A^{-1}_\beta)_{ij}(B_\beta)_i(B_\beta)_j=2n^2(\fft{\ell^2\beta}{2\Delta_\Sigma})^2\sum_{i,j}(q\cdot p_i)(q\cdot p_i)(A^{-1}_\beta)_{ij}-\fft{1}{2}\sum_{l,m}(A^1_{\tau})_{lm}\delta\tau_l\delta\tau_m\,,
\ee
where
\bea
(A^1_{\tau})_{lm}&=&-4\big(\fft{\ell^2\beta}{2\Delta_\Sigma}\big)^2 n\Big(\sum_i q\cdot p_i \big((A^{-1}_\beta)_{il}+(A^{-1}_\beta)_{im}\big)\omega_m
\omega_l (1-\delta_{lm})
\cr && -\sum_i \sum_{k\neq m}q\cdot p_i\big((A^{-1}_\beta)_{im}+(A^{-1}_\beta)_{ik}\big)\omega_k\omega_m \delta_{lm} \Big)\,.
\eea
Now let us first take a look at $\delta\tau_i$. We follow \cite{Fitzpatrick:2011hu} to introduce an exponent $\exp[-\sum_i\fft{\delta\tau_i^2}{2T^2}]$ with cut-off $T\rightarrow\infty$, which benefits the derivation. Then we can write the time relevant exponent as
\be
\exp[-\fft{1}{2}\delta\tau.A_\tau.\delta\tau^{\rm T}]\,,\quad \delta\tau=(\delta\tau_1,\cdots,\delta\tau_n)\,.
\ee
The linear term is suppressed by large AdS radius $\ell$ and the matrix $A_\tau$ can be organized as
\be
(A_\tau)_{lm}=(A^{0}_\tau)_{lm}+(A^q_\tau)_{lm}\,,\quad (A^q_\tau)_{lm}=(A^1_\tau)_{lm}
+(A^2_\tau)_{lm}\,,\quad (A^{0}_\tau)_{lm}=\fft{1}{T^2}\delta_{lm}+\fft{\beta^2\ell^2}{\Delta_\Sigma}\omega_l\omega_m\,,
\ee
where
\be
A^2_\tau=-\fft{\beta^2\ell^2}{\Delta_\Sigma}nq_0\omega_l\delta_{lm}\,,\quad A^3_\tau=\fft{\beta^2\ell^2}{\Delta_\Sigma} \omega_l\omega_m\,.
\ee
The inverse of $A_\tau$ can be evaluated as \cite{Fitzpatrick:2011hu}
\bea
&& ((A^0_\tau)^{-1})_{lm}=T^2\delta_{lm}+\omega_l\omega_m(-\fft{T^2}{\sum\omega_i^2}+\fft{\Delta_\Sigma}{\beta^2\ell^2
(\sum_i\omega_i^2)^2})+\mathcal{O}(T^{-2})\,,
\cr &&
\cr && (A_\tau)^{-1}=(A^0_\tau))^{-1}(1-A^q_\tau (A^0_\tau)^{-1}+(A^q_\tau (A^0_\tau)^{-1})^2)\,.
\eea
Then performing the integral over $\delta\tau_i$, the following prefactor is obtained
\be
{\rm pref}_\tau=\big(\fft{\Delta_\Sigma}{\sum\omega_k^2 \beta^2}\big)^{\fft{1}{2}}\fft{T^{n-1}(2\pi)^{\fft{n}{2}}}{\ell}\,,
\ee
which comes with the following exponent
\be
\exp[-\fft{1}{2}\sum_{ij}(A^{-1})_{ij}\omega_i\omega_j\ell^2]\,.
\ee
The remaining exponent is recorded below
\be
\exp[-\fft{1}{2}\fft{\Delta_\Sigma}{\beta^2}+Q(q_\mu)]\,,\label{beta exp}
\ee
where $Q(q_\mu)$ can be organized as
\bea
Q(q)&=&-\fft{\ell^2 n^2\beta^2}{2\Delta_\Sigma}q^2+2n^2(\fft{\ell^2\beta}{2\Delta_\Sigma})^2\sum_{i,j}(q\cdot p_i)(q\cdot p_i)(A^{-1}_\beta)_{ij}
\cr &&
\cr &&-\fft{1}{2}\big(\fft{\Delta_\Sigma T}{\sum\omega_k^2\ell}\big)^2\sum_{kl}(\delta_{kl}-\fft{\omega_k\omega_l}{\sum\omega_i^2})\tilde{A}^q
_{k}\tilde{A}^q_{l}\,,
\eea
where
\bea
\tilde{A}^q_m=\sum\omega_i A^q_{im}=4(\fft{\ell^2}{2\Delta_\Sigma})^2 n\sum_{i,k}q\cdot p_i\big((A^{-1}_\beta)_{im}+(A^{-1}_\beta)_{ik}\big)\omega_k\omega_m(\omega_m-\omega_k)-\fft{\ell^2}{\Delta_\Sigma}nq_0\omega_k^2\,.
\cr &&
\eea
Finally we are in the right position to integrate over $q$ to get
\be
\sqrt{\fft{(2\pi)^{d+1}}{{\rm det}Q_{qq}}}\,,\quad Q_{qq}=-\fft{\partial}{\partial q^\mu}\fft{\partial}{\partial q^\nu}Q\,,
\ee
where explicitly we obtain
\bea
Q_{qq}&=&\fft{\ell^2 n^2\beta^2}{\Delta_\Sigma}\delta_{\mu\nu}-2n^2(\fft{\ell^2\beta}{2\Delta_\Sigma})^2\sum_{i,j}(p^\mu_ip^\nu_j+p^\mu_jp^\nu_i)(q\cdot p_i)(A^{-1}_\beta)_{ij}
\cr &&
\cr &&+\fft{1}{2}\big(\fft{\Delta_\Sigma T}{\sum\omega_k^2\ell}\big)^2\sum_{kl}(\delta_{kl}-\fft{\omega_k\omega_l}{\sum\omega_i^2})\big(\fft{\partial}{\partial q^\mu}\tilde{A}^q
_{k}\fft{\partial}{\partial q^\nu}\tilde{A}^q_{l}+\fft{\partial}{\partial q^\nu}\tilde{A}^q
_{k}\fft{\partial}{\partial q^\mu}\tilde{A}^q_{l}\big)\,.
\eea
It is not hard to find that the second and the third term in $Q_{qq}$ is only rank-$(n-1)$ up to $\mathcal{O}(q)$, thus by taking $T\rightarrow\infty$, the whole determinant of $Q_{qq}$ can be evaluated by multiplying the rank-$(n-1)$ determinant of the last term with the rank-$(d-n+2)$ determinant of the first term \cite{Fitzpatrick:2011hu}. Using this trick, we can pull out $\beta$ and $T$, which is crucial for determining $f(\beta)$. Pulling out $T$ cancels $T^{n-1}$ in ${\rm pref}_\tau$, leaving the final answer independent of cut-off $T$. On the other hand, it contributes $\beta^{-(d-n+2)}$. Together with eq.~\eqref{limit of Mellin} (also note eq.~\eqref{beta exp}), one can readily find the $\beta$ (or $\alpha$) dependence $f(\beta)\sim \beta^{\Delta_\Sigma-d}$
\be
d\beta\beta^{\Delta_\Sigma-d}e^{-\fft{1}{2}\fft{\Delta_\Sigma}{\beta^2}}\sim d\alpha \alpha^{\fft{d-\Delta_\Sigma}{2}}e^{\alpha}\,.
\ee
The remaining part is technically difficult to evaluate, but nevertheless it is not necessary to evaluate it. The form of $f(\beta)$ in \eqref{expect massless} is now fixed, and the remaining factor serves simply as normalization factor and should be determined by contact interaction.

\section{$n=4$ Contact Witten diagram}
\label{app contact witten diagram}

We consider Witten diagram given by contact interaction
\be
\mathcal{L}=\phi_1^2\phi_2^2\,.
\ee
The AdS amplitude is simply
\be
A=\int d^{d+2}X \prod_{i=1}^4 G_{b\partial}(X,P_i)\,,
\ee
where $G_{b\partial}$ is the bulk-to-boundary propagator
\bea
G_{b\partial}(X,P_i)=\fft{\mathcal{C}_{\Delta_i}}{\ell^{\fft{d-1}{2}}(-2P_i\cdot X/\ell)^{\Delta_i}}\,.
\eea
The contact Witten diagram can be represented by D-function \cite{DHoker:1999kzh}
\bea
 A=\ell^{3-d}\mathcal{C}_{\Delta_1}^2\mathcal{C}_{\Delta_2}^2 D_{\Delta_1\Delta_2\Delta_2\Delta_1}(P_i)\,,
\eea
where
\bea
D_{\Delta_1\Delta_2\Delta_2\Delta_1}(P_i)=\fft{1}{\ell\Gamma(\Delta_1)^2\Gamma(\Delta_2)^2}
\int_0^\infty (\prod_i dt_i t_i^{\Delta_i-1})\int dX e^{-2\sum_{i=1}^4 t_i \fft{P_i\cdot X}{\ell}}\,.
\eea
Integrate out the bulk coordinate $X$, one found a simple representation of this Witten diagram \cite{Penedones:2010ue}
\bea
A=\ell^{3-d}\pi^{\fft{d}{2}}\Gamma(\fft{\Delta_\Sigma-d}{2})\prod_{i=1}^4 \fft{\mathcal{C}_{\Delta_i}}{\Gamma(\Delta_i)}
\int_0^\infty (\prod_{i=1}^4 dt_i t_i^{\Delta_i-1})e^{-\sum_{i<j}t_i t_j P_{ij}}\,.
\eea
This representation can be straightforwardly transformed into Mellin amplitudes. We start with this representation, it is then not surprise it gives rise to the same answer as Mellin space provides. We find saddle-points of $t_i$ are
\bea
&& t_1=-\fft{i\sqrt{\ell}(m+m_{12})(m+\bar{m}_{12})}{4\sqrt{\bar{m}_{12}}m}\,,\quad t_2=-\fft{i\sqrt{\ell}(m-m_{12})(m+\bar{m}_{12})}{4\sqrt{\bar{m}_{12}}m}\,,
\cr && t_3=\fft{i\sqrt{\ell}(m+m_{12})(m-\bar{m}_{12})}{4\sqrt{\bar{m}_{12}}m}\,,\quad t_2=\fft{i\sqrt{\ell}(m-m_{12})(m-\bar{m}_{12})}{4\sqrt{\bar{m}_{12}}m}\,.\label{saddle ts}
\eea
Picking up these saddle-points and including all reasonable normalization, we find it indeed gives rise to $D_c$ in eq.~\eqref{contact Dc} for $s=m^2$.

We can also follow the routine of \cite{Komatsu:2020sag} to verify that the contact Witten diagram is equivalent to momentum conservation delta function. To show this, we evaluate
\be
\int \fft{d|p_3|}{2\omega_3}\fft{d^d p_4}{2\omega_4} |p_3|^{d-1} A\,.\label{integral contact Witten}
\ee
In flat-space, this evaluates the phase-space volume
\be
\int \fft{d|p_3|}{2\omega_3}\fft{d^d p_4}{2\omega_4}|p_3|^{d-1}\delta^{(d+1)}(p_1+p_2+p_3+p_4)=\fft{(s-m_{12}^2)^{\fft{d-2}{2}}(s-\bar{m}_{12}^2)^{\fft{d-2}{2}}}{2^ds^{\fft{d-1}{2}}}\,,
\ee
which is the factor appear in partial-wave expansion of amplitudes eq.~\eqref{amp mass expansion}. To show the match, we still use saddle-points of $P_i$ eq.~\eqref{P saddle-point} but setting $p_3, p_4$ off-shell in frame eq.~\eqref{amp frame}
\be
p_1=(\omega_1, p\hat{n})\,,\quad p_2=(\omega_2,-p\hat{n})\,,\quad p_3=(-|\omega_3|, |p_3| \hat{n}^\prime)\,,\quad p_4=(-|\omega_4|,-|p_4|\hat{n}'')\,,
\label{amp frame off-shell}
\ee
where $|\omega_i|=\sqrt{m_i^2+|p_i|^2}$. Then we find the saddle-points of eq.~\eqref{integral contact Witten} are eq.~\eqref{saddle ts} together with
\be
|p_3|=|p_4|=p\,,\quad \hat{n}''=\hat{n}^\prime\,.
\ee
Include all relevant factors, it is equivalent to momentum conservation delta function.

\section{Conformal blocks with large $\Delta$ and $\Delta_{1,2}$}
\label{massive conformal block}
\subsection{From Casimir equation}

\begin{figure}[t]
\centering \hspace{0mm}\def\svgwidth{78mm}
\begingroup%
  \makeatletter%
  \providecommand\color[2][]{%
    \errmessage{(Inkscape) Color is used for the text in Inkscape, but the package 'color.sty' is not loaded}%
    \renewcommand\color[2][]{}%
  }%
  \providecommand\transparent[1]{%
    \errmessage{(Inkscape) Transparency is used (non-zero) for the text in Inkscape, but the package 'transparent.sty' is not loaded}%
    \renewcommand\transparent[1]{}%
  }%
  \providecommand\rotatebox[2]{#2}%
  \newcommand*\fsize{\dimexpr\f@size pt\relax}%
  \newcommand*\lineheight[1]{\fontsize{\fsize}{#1\fsize}\selectfont}%
  \ifx\svgwidth\undefined%
    \setlength{\unitlength}{298.24325368bp}%
    \ifx\svgscale\undefined%
      \relax%
    \else%
      \setlength{\unitlength}{\unitlength * \real{\svgscale}}%
    \fi%
  \else%
    \setlength{\unitlength}{\svgwidth}%
  \fi%
  \global\let\svgwidth\undefined%
  \global\let\svgscale\undefined%
  \makeatother%
  \begin{picture}(1,0.61674927)%
    \lineheight{1}%
    \setlength\tabcolsep{0pt}%
    \put(0,0){\includegraphics[width=\unitlength,page=1]{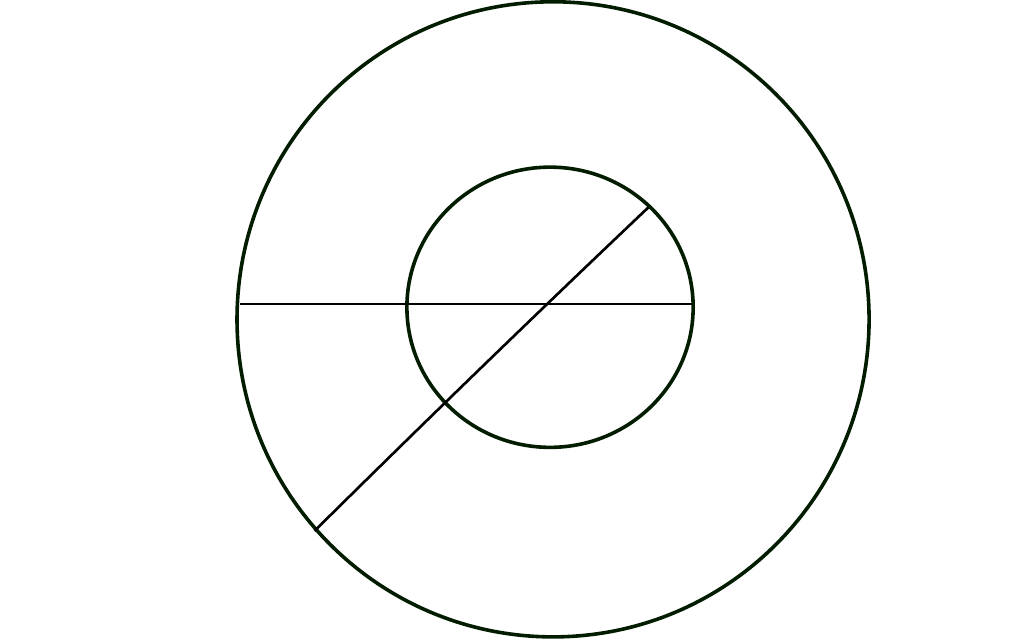}}%
    \put(0.58926498,0.45266525){\color[rgb]{0,0,0}\makebox(0,0)[lt]{\lineheight{1.25}\smash{\begin{tabular}[t]{l}$x_1=\alpha_1w$\end{tabular}}}}%
    \put(0.67046669,0.28398638){\color[rgb]{0,0,0}\makebox(0,0)[lt]{\lineheight{1.25}\smash{\begin{tabular}[t]{l}$x_3=\alpha_1$\end{tabular}}}}%
    \put(0.03623399,0.05621451){\color[rgb]{0,0,0}\makebox(0,0)[lt]{\lineheight{1.25}\smash{\begin{tabular}[t]{l}$x_2=-\alpha_2w$\end{tabular}}}}%
    \put(-0.00299015,0.31145718){\color[rgb]{0,0,0}\makebox(0,0)[lt]{\lineheight{1.25}\smash{\begin{tabular}[t]{l}$x_4=-\alpha_2$\end{tabular}}}}%
    \put(0,0){\includegraphics[width=\unitlength,page=2]{massive_frame.pdf}}%
  \end{picture}%
\endgroup%

\caption{A convenient conformal frame for solving conformal block at $\Delta,\Delta_i\rightarrow\infty$. Non-identical operators subject to flat-space saddle-points can also have access to above conformal frame. In general, $\alpha_1\neq\alpha_2$.}
\label{fig massive frame}
\end{figure}

We consider four-point function expanded in terms of conformal block
\be
\langle\mathcal{O}_1\cdots\mathcal{O}_4\rangle=\fft{1}{(P_{12}P_{34})^{\fft{\Delta_1+\Delta_2}{2}}}
\Big(\fft{P_{24}}{P_{14}}\Big)^{\fft{\Delta_{12}}{2}}\Big(\fft{P_{14}}{P_{13}}\Big)^{\fft{\Delta_{21}}{2}}
\sum_{\Delta,\j}c_{\Delta,\j}G_{\Delta,\j}(z,\bar{z})\,.
\ee
Acting with Casimir operator yields the Casimir equation \cite{Dolan:2003hv}
\bea
\mathcal{D}G_{\Delta,\j}=(\Delta(\Delta-d)+\j(\j+d-2))G_{\Delta,\j}\,, \label{Casi}
\eea
where
\bea
&& \mathcal{D}=\mathcal{D}_z+\mathcal{D}_{\bar{z}}+2(d-2)\fft{z\bar{z}}{z-\bar{z}}((1-z)\partial_z-(1-\bar{z})\partial_{\bar{z}})\,,
\cr && \mathcal{D}_z=2(z^2(1-z)\partial_z^2-(1+a+b)z^2\partial_z-abz)\,.
\eea
$(z,\bar{z})$ is the usual cross-ratios, and note $a=b=\Delta_{21}/2$. For $\Delta_{1}=\Delta_2$ or $\Delta_i\ll \Delta$, the Casimir equation simplifies and easily gives eq.~\eqref{large delta limit}. For $\Delta_1\neq \Delta_2$, the term with $\Delta_{12}$ is very important. Inspired by eq.~\eqref{s and r nonidentical} and \eqref{frame nonidentical}, we now adopt the following conformal frame
\be
z=\fft{4m^2w}{(m^2-m_{12}^2)(1+w)^2}\,,\quad \bar{z}=\fft{4m^2\bar{w}}{(m^2-m_{12}^2)(1+\bar{w})^2}\,,\label{ztorho}
\ee
where $w=r e^{i\theta}$, which is depicted in Fig \ref{fig massive frame}.
For $m_{12}=0$, the parameterization eq.~\eqref{ztorho} reduces to the usual radial frame \cite{Hogervorst:2013sma}. The Casimir equation eq.~\eqref{Casi} now reads
\bea
&& \mathcal{A} G(r,\eta) + \mathcal{B}_1 \partial_r G(r,\eta) +\mathcal{B}_2 \partial_\eta G(r,\eta) +\mathcal{C}_1 \partial_r^2 G(r,\eta) +\mathcal{C}_2 \partial_\eta^2 G(r,\eta)+\mathcal{C}_3 \partial_r\partial_\eta G(r,\eta)=0
\,, \cr &&
\eea
where
\bea
&& \mathcal{A}=(r^2-1)(r^2-2 \eta  r+1)^3 ((d-2) J (m_{12}-m) (m+m_{12})(r^2+2 \eta  r+1)^2
\cr && +\Delta(-d (m_{12}-m) (m+m_{12}) (r^2+2 \eta  r+1)^2-m^3 \ell (r^2+2 \eta  r+1)^2
\cr && +m_{12}^2 \Delta(r^4+(4 \eta ^2-6) r^2+1))+J^2(m_{12}-m) (m+m_{12})(r^2+2 \eta  r+1)^2)\,,
\cr &&
\cr && \mathcal{B}_1=-r (r^2+2 \eta  r+1) (-m^2 (r^2-2 \eta  r+1)^2 (d (r^2+1) (r^4+(2-4 \eta ^2) r^2+1)
\cr && +r^2 (r^4+4 \eta ^2 (r^2+3) -7 r^2-9)-1)+m_{12}^2 (r^2+2 \eta  r+1) (d (r^4-2 \eta  (r^2+1) r
\cr && -6 r^2+1) (r^2-2 \eta  r+1)^2+r (2 \eta  +r (r^6+18 r^4+8 \eta ^3 (r^2+3) r-16 r^2-4 \eta ^2 (5 r^4
\cr && -4 r^2+7)-2 \eta  (r^4+r^2+15) r+30))-1) +8 \eta  m^2 m_{12} r (r^2-1)^2 \ell  (r^2-2 \eta  r+1)^2)\,,
\cr &&
\cr && \mathcal{B}_2=(1-r^2) (r^2+2 \eta  r+1) (\eta  (-m^2) (r^2-2 \eta  r+1)^2 (d r^4-4 (d+1) \eta ^2 r^2+2 d r^2
\cr && +d-r^4+6 r^2-1)+m_{12}^2 (r^2+2 \eta  r+1) (-8 (d+1) \eta ^4 r^3+4 (3 d+1) \eta ^3 r^2 (r^2+1)
\cr && +2 \eta ^2 r ((5-3 d) r^4+2 (d-3) r^2-3 d+5) +\eta  (r^2+1) ((d-1) r^4-2 (7 d+1) r^2+d-1)
\cr && +4 r ((d-2) r^4+2 (d+2) r^2+d-2)) -8 (\eta ^2-1) m^2 m_{12} r (r^2+1) \ell  (r^2-2 \eta  r+1)^2)\,,
\cr &&
\cr && \mathcal{C}_1=-r^2 (r^2-1) (r^2-2 \eta  r+1) (r^2+2 \eta  r+1)^2 (m_{12}^2 (r^4+(4 \eta ^2-6) r^2+1)
\cr && -m^2 (r^2 -2 \eta  r+1)^2)\,,
\cr &&
\cr && \mathcal{C}_2=(\eta ^2-1) (1-r^2) (r^2-2 \eta  r+1) (r^2+2 \eta  r+1)^2 (m_{12}^2 (r^4+(4 \eta ^2-6) r^2+1)
\cr && -m^2 (r^2-2 \eta  r+1)^2)\,,
\cr &&
\cr && \mathcal{C}_3=8 \left(\eta ^2-1\right) m_{12}^2 r^2 \left(r^2-1\right)^2 \left(r^2-2 \eta  r+1\right) \left(r^2+2 \eta  r+1\right)^2\,.
\cr &&
\eea
We denote $\eta=\cos\theta$. The expression looks horrible, but we find it is especially useful to define
\be
g_{\Delta,\j}(z,\bar{z})=\Big(\fft{P_{24}}{P_{14}}\Big)^{\fft{\Delta_{12}}{2}}\Big(\fft{P_{14}}{P_{13}}\Big)^{\fft{\Delta_{21}}{2}}
G_{\Delta,\j}(z,\bar{z})\,.
\ee
Then the leading order of Casimir equation is trivially satisfied by scaling $r^{\Delta}f(r,\eta)$ and the sub-leading order of the equation reads
\bea
&& r (d ((r^2+1)^2-4 \eta ^2 r^2)+4 (\eta ^2-1) (r^2+1))f+
(r^2-1) (r^4+(2-4 \eta ^2) r^2+1) \partial_r f=0\,.
\cr &&
\eea
Finally, we end up with a simple solution (include reasonable normalization)
\bea
g_{\Delta,\j}(r,\theta)|_{\Delta,\Delta_i\rightarrow\infty}
&=& \fft{\j! N_{\Delta}}{(d-2)_\j}
\fft{(4r)^\Delta C_\j^{\fft{d}{2}-1}(\cos\theta)}{(1-r^2)^{\fft{d}{2}-1}
\sqrt{(1+r^2)^2-4r^2\cos^2\theta}}\,,\label{limit of block general}
\eea
where
\be
N_{\Delta}=\fft{\Delta^{2\Delta}}{(\Delta-\Delta_{12})^{\Delta-\Delta_{12}}(\Delta+\Delta_{12})^{\Delta+\Delta_{12}}}\,.
\ee
The expression looks the same as eq.~\eqref{large delta limit} up to additional normalization factor, but the definition of $(r,\theta)$ is no longer the same, besides, $g_{\Delta,\j}$ is defined by including appropriate prefactors. It is also worth noting that here $(r,\theta)$ depend on $\Delta$, so when we sum over conformal blocks, we should be careful about addressing conformal block itself. To avoid confusion, we may denote $(r_{\Delta},\theta_{\Delta})$ in the main text. In the next subsection, we verify our solution by working specifically in $d=2,4$.

\subsection{Explicit check in $d=2,4$}

In $d=2,4$, the conformal block can be exactly solved \cite{Dolan:2003hv,Dolan:2000ut}
\bea
&& d=2\,,\qquad G_{\Delta,\j}= k_{\Delta+\j}^{a,b}(z) k_{\Delta-\j}^{a,b}(\bar{z}) +k_{\Delta+\j}^{a,b}(\bar{z}) k_{\Delta-\j}^{a,b}(z)\,,
\cr &&
\cr && d=4\,,\qquad G_{\Delta,\j}=\fft{z\bar{z}}{z-\bar{z}}\big( k_{\Delta+\j}^{a,b}(z) k_{\Delta-\j-2}^{a,b}(\bar{z})-k_{\Delta+\j}^{a,b}(\bar{z}) k_{\Delta-\j-2}^{a,b}(z)\big)\,,\label{d=2,4 block}
\eea
where
\be
k_{\beta}^{a,b}(z)=z^{\fft{\beta}{2}}\,_2F_1(a+\fft{\beta}{2},b+\fft{\beta}{2},\beta,z)
\ee

We can find $k_{\beta}^{a,a}(z)|_{\beta,a\rightarrow\infty}$ by using the Barnes representation
\be
\,_2F_1(a,b,c,z)=\fft{\Gamma(c)}{\Gamma(a)\Gamma(b)} \int_{-i\infty}^{i\infty} \fft{ds}{2\pi i} \fft{\Gamma(a+s)\Gamma(b+s)\Gamma(-s)}{\Gamma(c+s)}(-z)^s\,.
\ee
We deform the contour to right and find there is a saddle-point of $s$
\be
s^\ast=\fft{w\beta(\beta+2a)}{(1-w)\beta-2a(1+w)}\,.
\ee
Then by performing the integral dominated by this saddle-point, we obtain
\bea
k_{\beta}^{a,a}&=&\frac{\beta ^{\beta }}{\sqrt{1-w }} (a+\frac{\beta }{2})^{-2 a-\beta } (w +1)^{2 a-\frac{1}{2}} w ^{\beta /2} (\beta -2 a)^{\frac{\beta  (\beta -2 a)}{2 a (w +1)+\beta  (w -1)}} (2 a+\beta )^{\frac{\beta  w  (2 a+\beta )}{2 a (w +1)+\beta  (w -1)}} \times
\cr && (\beta ^2-4 a^2)^{\frac{(w +1) (4 a^2-\beta ^2)}{2 a (w +1)+\beta  (w -1)}-\frac{\beta  (2 a (w -1)+\beta  (w +1))}{2 \beta  (1-w )-4 a (w +1)}} (2 \beta  (1-w )-4 a (w +1))^{-2 a}\label{rep of k}
\eea
This expression looks tough, but it turns out those transcendental factors exactly give rise to the wanted prefactor. Plug eq.~\eqref{rep of k} in eq.~\eqref{d=2,4 block} and absorb the prefactors, we find
\bea
&& d=2\,,\qquad g_{\Delta,\j}= N_{\Delta} \fft{(4r)^{\Delta}}{\sqrt{1+r^2-2r^2\cos(2\theta)}}\times 2\cos(\j\theta)\,,
\cr && d=4\,,\qquad g_{\Delta,\j}=N_{\Delta} \fft{(4r)^{\Delta}}{(1-r^2)\sqrt{1+r^2-2r^2\cos(2\theta)}}\times \fft{\sin((\j+1)\theta)}{\sin\theta}\,,
\eea
which precisely match with the general result eq.~\eqref{limit of block general}.

\end{appendix}

\bibliographystyle{JHEP}
\bibliography{flatspace}
\end{document}